\documentclass[12pt,a4paper]{article}
\usepackage{latexsym}
\usepackage{epsf}
\font\mybb=msbm10 at 12pt
\def\bb#1{\hbox{\mybb#1}}
\def\Z {\bb{Z}}
\def\R {\bb{R}}

\makeatletter
\@addtoreset{equation}{section}
\makeatother

\def\unit{\hbox to 3.3pt{\hskip1.3pt \vrule height 7pt width .4pt \hskip.7pt
\vrule height 7.85pt width .4pt \kern-2.4pt
\hrulefill \kern-3pt
\raise 4pt\hbox{\char'40}}}

\def\x{\times}

\begin{document}

\begin{flushright}
\footnotesize
UG-8/97\\
QMW-PH-97-28 \\
CERN--TH/97--229\\
IFT-UAM/CSIC-97-2\\
{\bf hep-th/9712115}\\
December  $12$th, $1997$
\normalsize
\end{flushright}

\begin{center}


{\LARGE {\bf Massive Branes}}

\vspace{.9cm}

{
{\bf Eric Bergshoeff}
\footnote{E-mail address: {\tt E.Bergshoeff@phys.rug.nl}}\\
{\it Institute for Theoretical Physics\\
University of Groningen\\
Nijenborgh 4, 9747 AG Groningen, The Netherlands}
}

\vspace{.5cm}

{
{\bf Yolanda Lozano}
\footnote{E-mail address: {\tt Y.Lozano@qmw.ac.uk}}\\
{\it Physics Department\\
Queen Mary \& Westfield College\\
Mile End Road, London E1 4NS, U.K.}
}

\vspace{.2cm}

{and}

\vspace{.2cm}

{
{\bf Tom\'as Ort\'{\i}n}
\footnote{E-mail addresses: {\tt Tomas.Ortin@cern.ch, 
tomas@leonidas.imaff.csic.es}}
\\
{\it C.E.R.N.~Theory Division}\\
{\it CH--1211, Gen\`eve 23, Switzerland}\\
{\it and}\\
{\it Instituto de F\'{\i}sica Te\'orica, C-XVI}\\
{\it Universidad Aut\'onoma de Madrid, C.U.~Cantoblanco}\\
{\it E-28049-Madrid, Spain}\\
{\it and}\\
{\it I.M.A.F.F., C.S.I.C.,}\\
{\it Calle de Serrano 113, E-28006-Madrid, Spain.}
}

\vspace{.8cm}

\newpage

{\bf Abstract}

\end{center}

\begin{quotation}

\small

We investigate the effective worldvolume theories of branes in a
background given by (the bosonic sector of) 10-di\-men\-sio\-nal
massive IIA supergravity (``massive branes'') and their M-theoretic
origin. In the case of the solitonic 5-brane of type~IIA superstring
theory the construction of the Wess-Zumino term in the worldvolume
action requires a dualization of the {\sl massive}
Neveu-Schwarz/Neveu-Schwarz target space 2-form field.  We find that,
in general, the effective worldvolume theory of massive branes
contains new worldvolume fields that are absent in the massless case,
i.e.~when the mass parameter $m$ of massive IIA supergravity is set to
zero.  We show how these new worldvolume fields can be introduced in a
systematic way.

In particular, we find new couplings between the massive solitonic
5-brane and the target space background, involving an additional
worldvolume 1-form and 6-form.  These new couplings have implications
for the anomalous creation of branes. In particular, when a massive
solitonic 5-brane passes through a D8-brane a stretched D6-brane is
created.  Similarly, in M-theory we find that when an M5-brane passes
through an M9-brane a stretched Kaluza-Klein monopole is created.

We show that pairs of massive branes of type~IIA string theory can be
viewed as the direct and double dimensional reduction of a single
``massive M-brane'' whose worldvolume theory is described by a gauged
sigma model. For D-branes, the worldvolume gauge vector field becomes
the Born-Infeld field of the 10-dimensional brane. The construction of
the gauged sigma model requires that the 11-dimensional background has
a Killing isometry.  This background can be viewed as an
11-dimensional rewriting of the 10-dimensional massive IIA
supergravity theory. We present the explicit form and discuss the
interpretation of (the bosonic sector of) this so-called ``massive
11-dimensional supergravity theory''.

\end{quotation}

\vspace{1cm}

\begin{flushleft}
\footnotesize
CERN--TH/97--229\\
\normalsize
\end{flushleft}

\newpage

\pagestyle{plain}


\section*{Introduction}

In order to get a better understanding of the dynamics of branes in
string theory, it is important to know precisely what the effective
worldvolume theory is that describes the dynamics of these objects.
This worldvolume theory contains a great deal of information.  For
instance, fields propagating on a brane's worldvolume describe the
dynamics of the intersections with other branes ending on it
\cite{kn:T2,kn:S}.  In particular, the Born-Infeld (BI) vector field
present in the worldvolume of all D-p-branes describes the $U(1)$
field whose sources are open string endpoints. The anti-selfdual
2-form potential living on the M-5-brane worldvolume describes the
dynamics of an M-5-brane intersecting with an M-2-brane over a 1-brane.
The dynamics of these worldvolume fields has recently received some
attention \cite{kn:CM,kn:HLW2,kn:G}.

At present the structure of the worldvolume theory is quite well
understood for the case in which the brane propagates in a
supergravity background without a cosmological constant, the so-called
``massless supergravity theories''.  As is well known, in the case of
IIA supergravity, an extension is possible with a non-zero
cosmological constant proportional to $m^2$ with $m$ a mass parameter
\cite{kn:R}.  Such backgrounds are essential for the existence of
D-8-branes whose charge is proportional to $m$ \cite{kn:PW,kn:BRGPT}.
It is the purpose of this paper to investigate the worldvolume theory
of branes that propagate in such a massive background.  We will call
such branes ``massive branes''\footnote{In this paper we reserve the
  name ``massive branes'' for branes that propagate in a massive IIA
  supergravity background as opposed to branes that propagate in a
  background with zero mass parameter. Of course, all branes are
  massive in the sense that their physical mass is nonzero.}.  We will
in turn use the worldvolume theories to investigate the 11-dimensional
origin of these massive branes.

The basic branes of string theory are the fundamental string
(p-1-brane), the solitonic 5-brane (p-5-brane), the Brinkmann wave,
the Kaluza-Klein (KK) monopole and the D-p-branes ($0\le p\le 9$).
The charge of the first two objects is carried by a
Neveu-Schwarz/Neveu-Schwarz (NS/NS) field whereas the D-p-branes carry
a Ramond-Ramond (R-R) charge. For objects that carry a NS/NS charge
there are three possibilities: the brane may move in a Heterotic, IIA
or IIB supergravity background. Each of these three possibilities is
described by a different worldvolume theory. We distinguish between
these cases using an obvious notation, e.g.~for the different
p-5-branes

\begin{displaymath}
\mbox{p-5-brane} \rightarrow \left\{ 
                                \begin{array}{c}
                                \mbox{ p-5H-brane}\\
                                 \\
                                \mbox{ p-5A-brane}\\
                                 \\
                                \mbox{ p-5B-brane}
                                \end{array}
\right.
\end{displaymath}

\noindent On the other hand the D-p-branes carry a R-R charge and 
therefore always propagate in a IIA (p even) or IIB (p odd)
background.

It is instructive to first remind the worldvolume theories that
describe the dynamics of massless branes, i.e.~branes that move in a
background with zero cosmological constant.

The $(0+1)$-dimensional worldvolume theory corresponding to the
Brink\-mann wave is given by a kinetic term corresponding to a
massless particle. The difference between the worldvolume theories of
the Heterotic wave (WH), IIA-wave (WA) and IIB-wave (WB) is in the
fermionic terms and in the coupling to the dilaton.

In the case of the p-1-brane (NS/NS string) there is no difference
between the worldvolume theories of the p-1H-, p-1A- and p-1B-branes
at the bosonic level.  These actions are given by a Nambu-Goto kinetic
term and a WZ term that describes the coupling of the brane to the
NS/NS 2-form $B$.  The dilaton does not appear in any of these two
terms, since the tension of fundamental objects is independent of the
string coupling constant $g$.

In the case of the p-5-brane the kinetic term is again of the
Nambu-Goto form but now there is an extra dilaton factor $e^{-2\phi}$
in front of it showing that the p-5-brane is a solitonic object whose
mass is proportional to $1/g^2$.  Concerning the WZ term there is a
distinction between the three different cases already at the bosonic
level. The reason for this is that the p-5-brane couples to the 6-form
dual of the NS/NS 2-form $B$. The definition of this dual 6-form
depends on the background involved because either the field strengths
or the couplings of $B$ in the action are different. There are then
three different dual 6-forms denoted by ${\tilde B}_{\rm H}, {\tilde
  B}_{\rm IIA}$ and ${\tilde B}_{\rm IIB}$.

The kinetic term of the (5+1)-dimensional worldvolume theory
corresponding to the KK-monopole has only been recently constructed
\cite{kn:BJO2}.  It is given by a gauged sigma model involving an
auxiliary worldvolume 1-form. The kinetic term carries a factor
$e^{-2\phi}k^{2}$ indicating that the effective tension of this object
is proportional to $1/g^{2}$ (and so it is solitonic) and to $R^{2}$,
$R$ being the radius of the dimension to be compactified. This
behavior is characteristic of KK monopoles. There is also a difference
in the bosonic worldvolume content of the KK-H, KK-10A and KK-10B
monopoles \cite{Hull}.  The WZ terms for these monopoles have not been
constructed explicitly so far (see, however, \cite{kn:BJO3}).

Finally, the worldvolume theory of the D-p-branes is given by a
Dirac-Born-Infeld (DBI) kinetic term with a $e^{-\phi}$ dilaton
coupling in front of it and a WZ term whose leading term is the R-R
(p+1)-form field. The dilaton coupling shows that the mass of these
objects is proportional to $1/g$.

In this work we will consider the massive extension of the worldvolume
theory corresponding to the p-1A-brane, p-5A-brane and the D-p-branes
(p even).  In the latter case, the result has already been given in
the literature \cite{kn:BR,kn:GHT}. We find that the (bosonic part of
the) worldvolume theory of the massive p-1A-brane is identical to that
of the massless p-1A-brane. In the case of the p-5A-brane, however,
there are striking differences.  It turns out that there are extra
couplings, involving an additional 1-form and 6-form worldvolume
field, that are proportional to the mass parameter $m$ of massive IIA
supergravity. Both fields are auxiliary and do not contribute to the
worldvolume degrees of freedom.

The reason for the presence of the worldvolume 6-form is that in order
to build the WZ term one has to dualize the {\sl massive} NS/NS target
space 2-form field along the lines recently discussed in \cite{kn:Q}.
An interesting feature is that, whereas in the usual formulation of
IIA supergravity the R-R 1-form $C^{(1)}$ is a Stueckelberg field that
gets ``eaten up'' by the Neveu-Schwarz/Neveu-Schwarz (NS/NS) 2-form
$B$ which becomes massive:

\begin{displaymath}
\left\{
\begin{array}{rcl}
C^{(1)} &\rightarrow& {\rm Stueckelberg\ field}\, ,\\
& &  \\
B       &\rightarrow& {\rm massive\ field}\, ,\\
\end{array}
\right.
\end{displaymath}

\noindent in the dual formulation the situation is
reversed: the dual NS/NS 6-form ${\tilde B}_{\rm IIA}$ becomes a
Stueckelberg field giving mass to the dual R-R 7-form $C^{(7)}$:

\begin{displaymath}
\left\{
\begin{array}{rcl}
C^{(7)} &\rightarrow& {\rm massive\ field}\, ,\\
& & \\
{\tilde B}_{\rm IIA} &\rightarrow& {\rm Stueckelberg\ field}\, .\\
\end{array}
\right.
\end{displaymath}

An immediate consequence of the above observation is that, since
${\tilde B}_{\rm IIA}$ occurs as the leading term of the WZ term of
the p-5A-brane, we must introduce an independent auxiliary 6-form
worldvolume field in order to cancel the Stueckelberg transformations
of ${\tilde B}_{\rm IIA}$.  This is on top of an auxiliary worldvolume
1-form which must be introduced in order to construct a WZ term that
is invariant under the ``massive gauge transformations'' of massive
IIA supergravity.  The massive p-5A-brane therefore contains extra
couplings to a worldvolume 1-form and 6-form that are absent in the
massless case.  The situation is similar to that of a D-0-brane in a
massive background. In that case the Stueckelberg variation of the
leading term $C^{(1)}$ in the WZ term is canceled by an auxiliary BI
1-form field that couples to the D-0-brane with a strength
proportional to the mass parameter $m$.

As mentioned at the beginning, the second goal of this work is to
investigate a possible ``M-theoretic origin'' of the massive branes
considered above, thereby generalizing the massless case considered in
\cite{kn:T}.  This is a nontrivial problem in view of the fact that
the massive IIA supergravity background fields have no known
11-dimensional interpretation.

We will deal with the above problem as follows: first, we will show
that the massive brane worldvolume actions with 10-dimensional target
space can be quite naturally rewritten as worldvolume actions of
objects (``massive M-branes'') moving in an 11-dimensional target
space.  These massive M-brane actions give the known massive brane
actions of string theory upon direct or double dimensional reduction
exactly as it happens in the massless case\footnote{As we shall see,
  the situation for the M-5-brane is more subtle.}.  Therefore, a
single massive M-brane unifies two different massive branes of string
theory.  The situation is depicted in Figure~\ref{fig:reduction}
(Section~\ref{sec-mbraneactions}).

We find that the worldvolume theory of a general massive M-brane is
given by a gauged sigma model\footnote{The massive M-2-brane has
  already been considered in \cite{kn:L,kn:O} and the present work
  should be viewed as an extension of these works to the other
  M-branes.  }. To preserve the (p+1)-form potential gauge invariance
in the gauging some modifications of the standard 11-dimensional
supergravity background transformations have to be introduced.  These
modifications are the same for all massive M-branes and, thus, we are
led to the construction of a ``massive 11-dimensional supergravity
theory'', the natural background in which massive M-branes move.

The construction of this theory requires the existence of a Killing
isometry (the one which is gauged in the sigma models), i.e.~in
adapted coordinates the fields of 11-dimensional supergravity do not
depend on one of the coordinates, say $y$.  In the limit in which the
mass parameter is set to zero, the dependence on $y$ can be restored
and the action is given by ordinary (massless) 11-dimensional
supergravity. Furthermore, the theory has the sought-after property
that it gives type~IIA supergravity upon dimensional reduction in the
direction $y$.

Of course this so-called ``massive 11-dimensional supergravity
theory'' is not a proper 11-dimensional theory in the sense that the
fields do not depend on the special coordinate $y$\footnote{It is due
  to this feature that the no-go theorem of \cite{kn:BDHS} can be
  circumvented.}.  Nevertheless, we believe that the massive
11-dimensional supergravity theory we propose should lead to a better
understanding of a possible M-theoretic origin of the mass parameter
$m$.  An interesting feature is that the massive 11-dimensional
supergravity theory we consider in this work is on a rather similar
footing with a recent proposal \cite{kn:HLW} for an 11-dimensional
supergravity theory that, upon reduction, leads to a new
10-dimensional massive supergravity theory. A more detailed comparison
between the two theories can be found in the Conclusion Section.

Finally, as a by-product of our work in which we were obliged to
introduce additional worldvolume fields, we have found a systematic
description for worldvolume p-forms, their gauge transformations and
field strengths.  These worldvolume p-forms can be defined as the dual
of the BI field in the worldvolume action of the D-(p+2)-brane. The
integrals of their (p+1)-form field strengths can be identified with
the WZ terms of D-p-branes. Our description is valid both for type~IIA
and type~IIB object worldvolumes and we believe that the structure of
these fields has a deep significance.

The organization of the paper is as follows. The construction of the
worldvolume theories describing the massive branes of string theory
requires certain results both on the structure of the supergravity
background fields, in particular their dual formulations, as well as
on the structure of the different p-form worldvolume fields that will
be needed. Therefore, as a preliminary, we first discuss in
Section~\ref{sec-masssugras} the target space supergravity fields and
in Section~\ref{sec-worldvolume} the p-form worldvolume fields.  Next,
these results are applied in Section~\ref{sec-mbraneactions} to
construct the sigma models describing the massive branes.

More specifically, in Section~\ref{sec-masssugras} we discuss massive
supergravities in 10 and 11 dimensions together with their dual
formulations.  The massless supergravity theories are included as a
special case by taking the mass parameter equal to zero. The different
cases we consider are: (i) massive 11-dimensional supergravity
(Section~\ref{sec-massiveM}); (ii) the dual formulation of massless
d=11 supergravity (Section~\ref{sec-dual11}); (iii) the dual
formulation of massive 10-dimensional IIA supergravity
(Section~\ref{sec-dual10}) and finally (iv) the dual formulation of
massive d=11 supergravity (Section~\ref{sec-dualmassive11}).  Next, in
Section~\ref{sec-worldvolume} we introduce the different p-form
worldvolume fields. Some of them will play the role of describing
dynamical degrees of freedom, others will be auxiliary fields.  All
these different supergravity backgrounds and worldvolume fields are
needed when we construct in Section~\ref{sec-mbraneactions} the
worldvolume effective actions of the massive branes. The strategy we
follow in this section will be to first construct the sigma models
corresponding to the massive M-branes.  In a second stage we perform
the direct and double dimensional reduction of the different massive
M-branes and show that they give rise to pairs of massive branes of
string theory. This is represented in Figure~\ref{fig:reduction}. Our
conclusions are given in Section~\ref{sec-conclusion}.  Finally, there
are three appendices.  Appendix~\ref{sec-massdual} gives a general
self-contained discussion on the duality of massive $k$-form fields;
Appendix~\ref{sec-fields} (Appendix~\ref{sec-wvfields}) gives our
notation for all the different target space (worldvolume) fields
occurring in this paper and summarizes some useful formulae needed in
the text.  


\section{Target Space Fields}
\label{sec-masssugras}


\subsection{Massive 11-Dimensional Supergravity}
\label{sec-massiveM}

In this Section we present, using an 11-dimensional notation, an
action that upon dimensional reduction gives massive 10-dimensional
IIA supergravity. This theory provides the 11-dimensional background
for the worldvolume actions of the {\it massive} M-branes (see
Section~\ref{sec-mbraneactions}).  As discussed in the introduction
the construction of the action requires that the 11-dimensional fields
exhibit a Killing isometry.  The action we will give below has the
property that when the mass parameter $m$ is set to zero the
dependence of the fields on the Killing isometry direction (say $y$)
in adapted coordinates, can be restored and the usual massless
11-dimensional supergravity theory is recovered. The purpose of this
Section is to present the massive 11-dimensional supergravity theory,
as a preliminary to the construction of the worldvolume sigma models
in Section~\ref{sec-mbraneactions}.  In the Conclusion we will compare
with the work of \cite{kn:HLW} and discuss a possible connection with
M-theory.

The {\it massive} 11-dimensional theory has the same field content as
the massless one\footnote{Hats on spacetime fields and indices
  indicate they are 11-dimensional. Absence of hats indicates they are
  10-dimensional.}:

\begin{equation}
\left\{\hat{g}_{\hat{\mu}\hat{\nu}},
\hat{C}_{\hat{\mu}\hat{\nu}\hat{\rho}} \right\}\, .  
\end{equation}

The action for these fields is manifestly 11-dimensional Lorentz
covariant but it does not correspond to a proper 11-dimensional theory
because, in order to write down the action, we need to introduce an
auxiliary non-dynamic vector field $\hat{k}^{\hat{\mu}}$ such that the
Lie derivatives of the metric and 3-form potential with respect to it
are zero\footnote{We will heavily rely on this property. In
  combination with the identity

\begin{equation}
i_{\hat{k}}\partial \hat{S}^{(r)}= 
{\textstyle\frac{(-1)^{r}}{r+1}}\pounds_{\hat{k}}    
\hat{S}^{(r)} +{\textstyle\frac{r}{r+1}}
\partial\left(i_{\hat{k}}\hat{S}^{(r)} \right)\, 
\end{equation}

\noindent for $r$-forms, it will allow us to pull $\hat{k}$ through partial
derivatives. $i_{\hat{k}}\hat{T}$ indicates the contraction of
$\hat{k}$ with the last (by convention) index of the tensor
$\hat{T}$.}:

\begin{equation}
\pounds_{\hat{k}} \hat{g}_{\hat{\mu}\hat{\nu}}
=\pounds_{\hat{k}} \hat{C}_{\hat{\mu}\hat{\nu}\hat{\rho}}
=0.
\end{equation}

To construct the action we start by defining the infinitesimal massive
gauge transformations on any rank $r$ 11-dimensional tensor with lower
indices $\hat{L}_{\hat{\mu}_{1}\ldots\hat{\mu}_{r}}$ (not necessarily
an $r$-form) with vanishing Lie derivative with respect to $\hat{k}$
(except for $\hat{C}_{\hat{\mu}\hat{\nu}\hat{\rho}}$, whose
transformation law we will define later) as follows\footnote{All
  tensors in this theory transform according to this rule, except for
  $\hat{C}$ and $\hat{\tilde{C}}$.  However, tensors of higher rank
  also transform under {\it dual massive transformations}. These have
  as infinitesimal parameter the 6-form
  $\hat{\tilde{\lambda}}_{\hat{\mu}_{1}\ldots\hat{\mu}_{6}}$ (for more
  details, see Section~\ref{sec-dual10}).}:

\begin{equation}
\label{eq:generalmasstrans}
\delta_{\hat{\chi}} \hat{L}_{\hat{\mu}_{1}\ldots\hat{\mu}_{r}}
= m\hat{\lambda}_{\hat{\mu}_{1}}\hat{k}^{\hat{\nu}}
\hat{L}_{\hat{\nu}\hat{\mu}_{2}\ldots\hat{\mu}_{r}}
+\ldots +m\hat{\lambda}_{\hat{\mu}_{r}}\hat{k}^{\hat{\nu}}
\hat{L}_{\hat{\mu}_{1}\ldots\hat{\mu}_{r-1}\hat{\nu}}\, .
\end{equation}

\noindent Here the 1-form parameter  $\hat{\lambda}_{\hat{\mu}}$ is 
related to the infinitesimal 2-form parameter
$\hat{\chi}_{\hat{\mu}\hat{\nu}}$, which generates the gauge
transformations of the 3-form $\hat C$ and must also have vanishing
Lie derivative $\pounds_{\hat{k}} \hat{\chi}_{\hat{\mu}\hat{\nu}}=0$,
by

\begin{equation}
\hat{\lambda}_{\hat{\mu}}\equiv
-{\textstyle\frac{1}{2}}\left(i_{\hat{k}}\hat{\chi}\right)_{\hat{\mu}} 
=-{\textstyle\frac{1}{2}}
\hat{k}^{\hat{\nu}} \hat{\chi}_{\hat{\mu}\hat{\nu}}\, .
\end{equation}

In particular, for the metric $\hat{g}_{\hat{\mu}\hat{\nu}}$ and for
any $r$-form $\hat{S}_{\hat{\mu}_{1}\ldots\hat{\mu}_{r}}
=\hat{S}_{[\hat{\mu}_{1}\ldots\hat{\mu}_{r}]}$ (different from the
3-form $\hat{C}$) we have, according to the rule
(\ref{eq:generalmasstrans})

\begin{equation}
\label{eq:massg}
\left\{
\begin{array}{rcl}
\delta_{\hat{\chi}}\hat{g}_{\hat{\mu}\hat{\nu}} & = & 
2m  \hat{\lambda}_{(\hat{\mu}}
\left(i_{\hat{k}}\hat{g}\right)_{\hat{\nu})}\, ,\\
& & \\
\delta_{\hat{\chi}}\hat{S}_{\hat{\mu}_{1}\ldots\hat{\mu}_{r}} & = & 
(-1)^{r-1} rm  \hat{\lambda}_{[\hat{\mu}_{1}}
\left(i_{\hat{k}}\hat{S}\right)_{\hat{\mu}_{2}\ldots\hat{\mu}_{r}]}
\, .\\
\end{array}
\right.
\end{equation}

\noindent Observe that the above transformation laws imply

\begin{equation}
\left\{
\begin{array}{rcl}
\delta_{\hat{\chi}}\sqrt{|\hat{g}|} & = & 0\, ,\\
& & \\
\delta_{\hat{\chi}}\hat{S}^{2} & = & 0\, .\\
\end{array}
\right.
\end{equation}

The 3-form is going to play the role of a connection-field with
respect to the massive gauge transformations and therefore it does not
transform covariantly:

\begin{equation}
\label{eq:massC}
\delta_{\hat{\chi}} \hat{C}_{\hat{\mu}\hat{\nu}\hat{\rho}}
 =  3\partial_{[\hat{\mu}}\hat{\chi}_{\hat{\nu}\hat{\rho}]}
+3m \hat{\lambda}_{[\hat{\mu}}
\left(i_{\hat{k}}\hat{C} \right)_{\hat{\nu}\hat{\rho}]}\, .
\end{equation}

For further use we also quote the transformation laws

\begin{equation}
\left\{
\begin{array}{rcl}
\delta_{\hat{\chi}}\left(i_{\hat{k}}\hat{C}\right)_{\hat{\mu}\hat{\nu}}
& = & -4 \partial_{[\hat{\mu}}\hat{\lambda}_{\hat{\nu}]}\, , \\
& & \\
\delta_{\hat{\chi}} \left(i_{\hat{k}}\hat{g}\right)_{\hat{\mu}}
& = & m\hat{\lambda}_{\hat{\mu}}\hat{k}^{2}\, .
\end{array} 
\right.
\end{equation}

The next step in our construction is to build a connection for the
massive gauge transformations. The new total connection is

\begin{equation}
\label{connection}
\hat{\Omega}_{\hat{a}}{}^{\hat{b}\hat{c}} 
= \hat{\omega}_{\hat{a}}{}^{\hat{b}\hat{c}} (\hat{e})
+\hat{K}_{\hat{a}}{}^{\hat{b}\hat{c}}\, ,
\end{equation}

\noindent where the piece that we have added is defined by

\begin{equation}
\hat{K}_{\hat{a}\hat{b}\hat{c}}= 
{\textstyle\frac{m}{4}} \left[\hat{k}_{\hat{a}}
\left(i_{\hat{k}}\hat{C}\right)_{\hat{b}\hat{c}}
+\hat{k}_{\hat{b}}
\left(i_{\hat{k}}\hat{C}\right)_{\hat{a}\hat{c}}
-\hat{k}_{\hat{c}}
\left(i_{\hat{k}}\hat{C}\right)_{\hat{a}\hat{b}}\right]\, .
\end{equation}

\noindent The tensor $\hat{K}$ coincides with the contorsion tensor while the
torsion tensor is

\begin{equation}
\hat{T}_{\hat{\mu}\hat{\nu}}{}^{\hat{\rho}} = 
-2 \hat{K}_{[\hat{\mu}\hat{\nu}]}{}^{\hat{\rho}} 
={\textstyle\frac{m}{2}}
\left(i_{\hat{k}}\hat{C}\right)_{\hat{\mu}\hat{\nu}} 
\hat{k}^{\hat{\rho}}\, .
\end{equation}

Observe that in the limit $m\rightarrow 0$ one recovers the standard
11-dimensional supergravity connections and gauge transformation laws
(assuming that the dependence of the fields on the isometry direction
has also been restored).

The exterior covariant derivative on the 3-form
$\hat{C}$ is defined by

\begin{equation}
D_{[\hat{\mu}} \hat{C}_{\hat{\nu}\hat{\rho}\hat{\sigma}]} 
=\partial_{[\hat{\mu}}\hat{C}_{\hat{\nu}\hat{\rho}\hat{\sigma}]}
-{\textstyle\frac{3}{2}}
\hat{\Gamma}_{[\hat{\mu}\hat{\nu}}{}^{\hat{\alpha}}
\hat{C}_{\hat{\rho}\hat{\sigma}]\hat{\alpha}}
=\partial_{[\hat{\mu}}\hat{C}_{\hat{\nu}\hat{\rho}\hat{\sigma}]}
+{\textstyle\frac{3}{8}}m 
\left(i_{\hat{k}}\hat{C}\right)_{[\hat{\mu}\hat{\nu}}
\left(i_{\hat{k}}\hat{C}\right)_{\hat{\rho}\hat{\sigma}]}\, .
\end{equation}

\noindent Using the vanishing of the Lie derivative with respect to 
$\hat{k}$ of $\hat C$ it is easy to prove that the covariant
derivative does indeed transform covariantly under massive gauge
transformations.  The 4-form field strength $\hat{G}$ is defined in
terms of this covariant derivative

\begin{equation}
\hat{G} =4D\hat{C}\, ,
\Rightarrow \delta_{\hat{\chi}} \hat{G} = -4m\hat{\lambda}\left(
i_{\hat{k}}\hat{G} \right)\, ,
\end{equation}

\noindent and thus the obvious kinetic term for $\hat{C}$ in the
action, $\hat{G}^{2}$, is invariant.

We are now ready to write the 11-dimensional action:

\begin{equation}
\begin{array}{rcl}
\hat{S} & = & \frac{1}{16\pi G_{N}^{(11)}}\int d^{11}x\
\sqrt{|\hat{g}|}\ \left\{\hat{R}(\hat{\Omega})
-{\textstyle\frac{1}{2\cdot 4!}}\hat{G}^{2}
-{\textstyle\frac{1}{8}}m^{2}|\hat{k}^{2}|^{2}
\right. \\
& & \\
& &
\hspace{-1.5cm}
\left.
+{\textstyle\frac{1}{(144)^{2}}} \frac{1}{\sqrt{|\hat{g}|}}
\hat{\epsilon} \left[2^{4} \partial\hat{C} \partial\hat{C} \hat{C}
+3^{2}m\partial\hat{C}\hat{C}\left(i_{\hat{k}}\hat{C}\right)^{2}
+\frac{3^{3}}{20}m^{2}\hat{C}\left(i_{\hat{k}}\hat{C}\right)^{4}   
\right]
\right\}\, .
\end{array}
\end{equation}

\noindent When $m=0$ and upon restoring the dependence on $y$
this action becomes that of the usual 11-dimensional
supergravity theory.

Now we are going to perform the dimensional reduction of this action
in the direction associated to the Killing vector $\hat{k}$,
parametrized by the coordinate $y$. From now on we will use
coordinates adapted to it, so $\hat{k}^{\hat{\mu}}
=\delta^{\hat{\mu}y}$.  First we define the 10-dimensional fields. We
use the notation and conventions for the gauge transformations
explained in Appendix~\ref{sec-fields}. The bosonic fields of the
10-dimensional type~IIA theory are

\begin{equation}
\left\{g_{\mu\nu},B_{\mu\nu},\phi,C^{(1)},C^{(3)} \right\}\, .  
\end{equation}

\noindent The R-R 1-form $C^{(1)}$ plays the role of a Stueckelberg 
field in the massive theory and could in principle be eliminated by an
appropriate gauge choice, but it is important for us to keep it.
Later we will also introduce the fields dual to $B,C^{(1)},C^{(3)}$,
namely $\tilde{B}_{\rm IIA}, C^{(7)},C^{(5)}$. These fields occur in
the effective worldvolume actions we will deal with in
Section~\ref{sec-mbraneactions} and it will be important to introduce
them properly.

The relation between the 11- and the above 10-dimensional fields is
the same as in the massless case (see, e.g.~\cite{kn:BHO}).  In
particular, we can use the same elfbein basis:

\begin{equation}
\begin{array}{rcl}
\left( \hat{e}_{\hat{\mu}}{}^{\hat{a}} \right) & = &
\left(
\begin{array}{cc}
e^{-\frac{1}{3}\phi} e_{\mu}{}^{a}
&
e^{\frac{2}{3}\phi} C^{(1)}{}_{\mu}
\\
&
\\
0
&
e^{\frac{2}{3}\phi}
\\
\end{array}
\right)
\, , \\
& & \\
\left(\hat{e}_{\hat{a}}{}^{\hat{\mu}} \right) & = &
\left(
\begin{array}{cc}
e^{\frac{1}{3}\phi} e_{a}{}^{\mu}
&
-e^{\frac{1}{3}\phi} C^{(1)}{}_{a}
\\
&
\\
0
&
e^{-\frac{2}{3}\phi}
\\
\end{array}
\right)\, . \\
\end{array}
\label{eq:basis}
\end{equation}

The 11-dimensional fields are expressed in terms of the
10-dimensional ones by

\begin{equation}
\label{eq:11versus10}
\left\{
\begin{array}{rcl}
\hat{g}_{yy} & = & -e^{\frac{4}{3}\phi}\, ,\\
& & \\
\hat{g}_{\mu y} & = & \left(i_{\hat{k}}\hat{g}\right)_{\mu} = 
-e^{\frac{4}{3}\phi}C^{(1)}{}_{\mu}\, , \\
& & \\
\hat{g}_{\mu\nu} & = & e^{-\frac{2}{3}\phi} g_{\mu\nu} 
-e^{\frac{4}{3}\phi}C^{(1)}{}_{\mu}C^{(1)}{}_{\nu}\, , \\
\end{array}
\right.
\hspace{.5cm}
\left\{
\begin{array}{rcl}
\hat{C}_{\mu\nu\rho} & = & C^{(3)}{}_{\mu\nu\rho}\, , \\
& & \\
\hat{C}_{\mu\nu y} & = & \left(i_{\hat{k}}\hat{C} \right)_{\mu\nu}=
B_{\mu\nu}\, .\\
\end{array}
\right.
\end{equation}

Now we are going to perform the reduction of the action.  We first
consider the Ricci scalar term.  To reduce this term we use a slight
generalization of Palatini's identity which in $d$ dimensions takes
the form

\begin{eqnarray}
\lefteqn{
\int d^{d}x\ \sqrt{|g|}\, e^{-2\varphi}\ [R]=}
\nonumber \\
& &
\nonumber \\
& &
\int d^{d}x \ \sqrt{|g|}\ \left\{ -e^{-2\varphi}\, \left[
\omega_{b}{}^{ba}\omega_{c}{}^{c}{}_{a} +\omega_{a}{}^{bc}
\omega_{bc}{}^{a} +4\omega_{b}{}^{ba}(\partial_a\varphi)
\right]\right\}\, .
\label{eq:Pal}
\end{eqnarray}

With the above ansatz for the elfbeins the non-vanishing components of
the 11-dimensional connection are

\begin{equation}
\begin{array}{rclrcl}
\hat{\Omega}_{yay}
&
=
&
-\frac{2}{3} e^{\frac{1}{3}\phi} \partial_{a}\phi\, ,
\hspace{1cm}
&
\hat{\Omega}_{yab}
&
=
&
-\frac{1}{2} e^{\frac{4}{3} \phi} G^{(2)}{}_{ab}\, ,
\\
& & & & &
\\
\hat{\Omega}_{aby}
&
=
&
\frac{1}{2}  e^{\frac{4}{3}\phi} G^{(2)}{}_{ab}\, ,
&
\hat{\Omega}_{abc}
&
=
&
e^{\frac{1}{3}\phi} \left( \omega_{abc}(e)
+\frac{2}{3} \delta_{a[b} \partial_{c]}
\phi \right)\, ,
\\
\end{array}
\end{equation}

\noindent where\footnote{When indices are not shown explicitly 
  and partial derivatives are used, we assume that all indices are
  completely antisymmetrized in the obvious order.  For instance the
  equation below means

\begin{equation}
G^{(2)}{}_{\mu\nu}
=2\partial_{[\mu}C^{(1)}{}_{\nu]} 
+{\textstyle\frac{m}{2}} B_{\mu\nu}\, .
\end{equation}
}
\begin{equation}
G^{(2)} =2\partial C^{(1)}+{\textstyle\frac{m}{2}}B\, ,
\end{equation}

\noindent is the field strength of the 10-dimensional vector
field $C^{(1)}{}_{\mu}$ and the term linear in $m$ is a Chern-Simons
term\footnote{All the definitions for the field strengths and gauge
  transformations can be found in Appendix~\ref{sec-fields} in
  differential form language.}.  Using

\begin{equation}
\sqrt{|\hat{g}|}\ =\sqrt{|g|}\ e^{-\frac{8}{3}\phi}\, ,
\end{equation}

\noindent plus Palatini's identity Eq.~(\ref{eq:Pal}) for $d=11$ and
$\varphi=0$, and the fact that the coordinate $y$ conventionally lives
in a circle of radius equal to the string length
$\ell_{s}=\sqrt{\alpha^{\prime}}$ we find

\begin{eqnarray}
\int d^{11}x\
\sqrt{|\hat{g}|}\ [\hat{R}(\hat{\Omega})]=
\hspace{-3cm}
& &  \nonumber \\
& & \nonumber \\
& &
2\pi \ell_{s}
\int d^{10}x\ \sqrt{|g|} \left\{ -e^{-2\phi}
\left[ \left( \omega_{b}{}^{ba}
+2\partial^{a}\phi \right)^{2}
+\omega_{a}{}^{bc}\omega_{bc} {}^{a}\right]
\right.
\nonumber \\
& & \nonumber \\
& &
\left.
-{\textstyle\frac{1}{4}} \left( G^{(2)} \right)^{2}\right\}\, .
\end{eqnarray}

\noindent Finally, using Palatini's identity Eq.~(\ref{eq:Pal}) again,
but now for $d=10$ and $\varphi = \phi$, we get for the Ricci scalar
term:

\begin{eqnarray}
\lefteqn{\int d^{11}x\
\sqrt{|\hat{g}|}\ \left[ \hat{R}(\hat{\Omega}) \right]=}
\nonumber \\
& &
\nonumber \\
& &
2\pi \ell_{s}\int d^{10}x\
\sqrt{|g|}\ \left\{ e^{-2\phi}
\left[ R -4\left( \partial\phi \right)^{2}\right]
-{\textstyle\frac{1}{4}} \left( G^{(2)} \right)^{2}\right\}\, .
\end{eqnarray}

Now we have to reduce the $\hat{G}^{2}$-term in the action.  We
identify field strengths in eleven and ten dimensions with flat
indices (this automatically ensures gauge invariance) taking into
account the scaling of the 10-dimensional metric

\begin{equation}
G^{(4)}{}_{abcd}=e^{-\frac{4}{3}\phi}\ \hat{G}_{abcd}\, ,
\end{equation}

\noindent which leads to

\begin{equation}
G^{(4)} = 4\left(\partial C^{(3)} -3\partial B C^{(1)} 
+{\textstyle\frac{3}{8}}mB^{2}\right)\, .
\end{equation}

\noindent The remaining components of ${\hat G}$ are given by

\begin{equation}
\hat{G}_{abcy} =  e^{\frac{1}{3}\phi} H_{abc}\, ,
\end{equation}

\noindent where $H$ is the field strength of the two-form $B$

\begin{equation}
H = 3\partial B\, ,
\end{equation}

\noindent and the contribution of the $\hat{G}$-term to the
10-dimensional action becomes

\begin{eqnarray}
\int d^{11} x \ \sqrt{|\hat{g}|}\
\left[ -{\textstyle\frac{1}{2\cdot 4!}} \left(
\hat{G} \right)^{2} \right] & = & \nonumber \\
& & \\
& &
\hspace{-4cm}
2\pi \ell_{s} \int d^{10} x\ \sqrt{|g|}\
\left[ {\textstyle\frac{1}{2\cdot 3!}}
e^{-2\phi} H^{2} 
-{\textstyle\frac{1}{2\cdot 4!}} \left(G^{(4)}\right)^{2} \right]\, .
\nonumber
\end{eqnarray}

\noindent Applying the identity $\hat{k}^{2}=-e^{\frac{4}{3}\phi}$ the
cosmological constant term contribution is

\begin{equation}
\int d^{11} x \ \sqrt{|\hat{g}|}\
\left[ -{\textstyle\frac{1}{8}} m^{2} |\hat{k}^{2}|^{2} \right]
=2\pi \ell_{s} \int d^{10} x\ \sqrt{|g|}\
\left[ -{\textstyle\frac{1}{8}} m^{2} \right]\, .
\end{equation}

Next we consider the first term in the WZ term. Taking into account
that 

\begin{equation}
\hat{\epsilon}^{\ \mu_{0}\ldots\mu_{9} y} =
\epsilon^{\ \mu_{0}\ldots\mu_{9}}\, ,
\end{equation}

\noindent and using curved indices we find that:

\begin{equation}
2^{4}\hat{\epsilon}\partial\hat{C}\partial\hat{C}\hat{C}=
2^{4}\cdot 3 \epsilon\partial C^{(3)}\partial C^{(3)} B
-2^{5}\cdot 3\epsilon\partial C^{(3)} \partial B C^{(3)}\, .
\end{equation}

\noindent Integrating by parts we get

\begin{equation}
\int d^{11} x\
\left[{\textstyle\frac{1}{(144)^{2}}}\
\hat{\epsilon}\ 2^{4}\partial\hat{C}\partial\hat{C}\hat{C}
\right] 
=2\pi \ell_{s}\int d^{10} x\
\left[{\textstyle\frac{1}{144}} \epsilon
\partial C^{(3)} \partial C^{(3)} B\right]\, .
\end{equation}
Next we have

\begin{equation}
3^{2}m\hat{\epsilon}\partial\hat{C}\hat{C}
\left(i_{\hat{k}}\hat{C}\right)^{2}= 
3^{3}m\epsilon\left[\partial C^{(3)} B^{3} 
-\partial B C^{(3)} B^{2} \right]\, ,
\end{equation}

\noindent and integrating by parts the second term we find

\begin{equation}
\int d^{11} x\
\left[{\textstyle\frac{1}{(144)^{2}}}\
\hat{\epsilon}\ 3^{2}m\partial\hat{C}\hat{C}
\left(i_{\hat{k}}\hat{C}\right)^{2} \right]= 
2\pi \ell_{s}\int d^{10} x\
\left[{\textstyle\frac{1}{144}}\ \epsilon\
{\textstyle\frac{1}{4}} m \partial C^{(3)} B^{3} \right]\, .
\end{equation}

\noindent Finally,  

\begin{equation}
{\textstyle\frac{3^{3}}{20}}m^{2}\hat{\epsilon}\hat{C}
\left(i_{\hat{k}}\hat{C}\right)^{4} 
={\textstyle\frac{3^{4}}{20}} m^{2}\epsilon B^{5}\, ,
\end{equation}

\noindent and 

\begin{equation}
\int d^{11} x\
\left[{\textstyle\frac{1}{(144)^{2}}}
\hat{\epsilon} {\textstyle\frac{3^{3}}{20}}m^{2}\hat{C}
\left(i_{\hat{k}}\hat{C}\right)^{4} \right]= 
2\pi \ell_{s}\int d^{10} x\
\left[{\textstyle\frac{1}{144}} \epsilon
{\textstyle\frac{3^{2}}{320}}m^{2}B^{5}
\right]\, .
\end{equation}

Putting all these results together we find the bosonic part of the
massive $N=2A,d=10$ supergravity action \cite{kn:R} in the string
frame, as given in \cite{kn:BRGPT}

\begin{equation}
\begin{array}{rcl}
S & = &
\frac{1}{16\pi G_{N}^{(10)}} \int d^{10}x\
\sqrt{|g|} \left\{ e^{-2\phi}
\left[ R(\omega) -4\left( \partial\phi\right)^{2}
+{\textstyle\frac{1}{2\cdot 3!}} H^{2}\right]\right.\\
& & \\
& & 
- \left[
{\textstyle\frac{1}{4}} \left( G^{(2)} \right)^{2}
+{\textstyle\frac{1}{2\cdot 4!}} \left( G^{(4)}\right)^{2}
+{\textstyle\frac{1}{8}}m^{2} \right]\\
& & \\
& & 
+{\textstyle\frac{1}{144}} \frac{1}{\sqrt{|g|}}\
\epsilon \left[\partial C^{(3)}\partial C^{(3)} B
+{\textstyle\frac{1}{4}}m\partial C^{(3)} B^{3}
+{\textstyle\frac{9}{320}}m^{2} B^{5}
\right]
\biggr \}\, ,
\end{array}
\end{equation}

\noindent where 

\begin{equation}
G^{(10)}_{N}=(2\pi \ell_{s})^{-1} G^{(11)}_{N}\, .  
\end{equation}

This action is invariant under the gauge transformations of the R-R fields

\begin{equation}
\left\{
\begin{array}{rcl}
\delta_{\Lambda^{(0)}} C^{(1)} & = & 
\partial\Lambda^{(0)}\, ,\\
& & \\
\delta_{\Lambda^{(0)}} C^{(3)} & = & 
3\partial\Lambda^{(0)} B\, ,\\
& & \\
\delta_{\Lambda^{(2)}} C^{(3)} & = & 3\partial\Lambda^{(2)}\, .\\
\end{array}
\right.
\end{equation}

The $\delta_{\Lambda^{(0)}}$ transformations are associated to
reparametrizations of the compact coordinate of the form

\begin{equation}
\delta_{\Lambda^{(0)}}\hat{X}^{\hat{\mu}} = -\Lambda^{(0)}(x)
\delta^{\hat{\mu}y}\, ,
\end{equation}

\noindent while the gauge transformation of the 3-form potential
is inherited from the massless gauge transformation of the
11-dimensional 3-form potential.

The above action is also invariant under the massive gauge
transformations

\begin{equation}
\left\{
\begin{array}{rcl}
\delta_{\lambda} C^{(1)} & = & m \lambda\, ,\\
& & \\
\delta_{\lambda} C^{(3)} & = & 3m \lambda B\, ,\\
& & \\
\delta_{\lambda} B & = & -4\partial\lambda\, ,\\
\end{array}
\right.
\end{equation}

\noindent which are directly inherited from the 11-dimensional massive 
gauge transformations. Observe that $C^{(1)}$ transforms by shifts of
the parameter of the gauge transformations of $B$. Thus, $C^{(1)}$ can
be completely eliminated by a massive gauge transformation, leaving in
the action a mass term for $B$: $m^{2}B^{2}$. It is usually said that
$C^{(1)}$ is the Stueckelberg field which is ``eaten up'' by $B$,
which then gets a mass.

This concludes our description of the massive 11-dimensional
supergravity theory.


\subsection{Dual Massless 11-Dimensional Supergravity}
\label{sec-dual11}

It is known that in the worldvolume effective actions of
supersymmetric solitons all the potentials of the theory may appear:
not only those that occur in the usual formulation but also their
(electro-magnetic) duals.  For this reason we explore here the
dualization of 11-dimensional supergravity. Since additional
subtleties occur in the massive case we will first discuss in this
Subsection the massless case. Next we will discuss in the next
Subsection dual massive IIA supergravity and, finally, discuss in
Section~\ref{sec-dualmassive11} dual massive 11-dimensional
supergravity.

The usual procedure for formulating a theory in terms of the dual of
some potential consists in finding an intermediate first-order action
with an extra field.  Integration of the extra field yields the
Bianchi identity for the original potential field strength and
integration of the field strength leads to the dual action with the
auxiliary field playing now the role of dual potential.

There is a systematic way of constructing the first-order intermediate
action: consider the action as a function of the field strength and
add a Lagrange multiplier term to enforce its Bianchi identity.  This
recipe fails when the action cannot be written in terms of the field
strength only and the potential itself (i.e.~not its derivative)
appears explicitly in it. This is what happens in the action of
massless d=11 supergravity and this is the reason why a formulation of
11-dimensional supergravity in terms of the dual 6-form
$\hat{\tilde{C}}$ only has not been found.

Strictly speaking, for the sake of constructing worldvolume actions,
it is enough to introduce dual potentials only on-shell \cite{kn:Ah}
and one can also allow for the introduction of potentials and dual
potentials at the same time.  However, inspired by the work of
\cite{kn:L}, we would like to point out that there is a way of finding
the dual theory, even if in the original action the potential does not
only occur as a derivative.  The idea consists in finding an
intermediate action with an auxiliary field such that the 3-form
potential $\hat{C}$ only appears through its derivatives. Elimination
of the auxiliary field leads to the usual 11-dimensional supergravity
action. To this intermediate action one can now apply the usual
procedure for dualization, eliminating the auxiliary field in the last
stage.

An intermediate action with the required properties is 

\begin{equation}
\begin{array}{rcl}
S[\hat{g},\hat{C},\hat{L}] & = & 
\frac{1}{16\pi G_{N}^{(11)}} \int d^{11}x\ \sqrt{|\hat{g}|}\
\left\{ R -\frac{1}{2\cdot 4!}\hat{G}^{2}
\right. \\
& & \\
& & 
\left.
-\frac{1}{(144)^{2}} \frac{\hat{\epsilon}}{\sqrt{|\hat{g}|}}
\left[
\hat{G}\hat{G}\hat{L}
+\frac{4}{3}\hat{G}\partial\hat{L}\hat{L}
+\frac{16}{27}\partial\hat{L}\partial\hat{L}\hat{L}
\right]
\right\}\, .\\
\end{array}
\end{equation}

The equation of motion for the auxiliary field 
$\hat{L}_{\hat{\mu}_{1}\hat{\mu}_{2}\hat{\mu}_{3}}$ is

\begin{equation}
\hat{\epsilon}^{\hat{\alpha}_{1}\hat{\alpha}_{2}\hat{\alpha}_{3}
\hat{\mu}_{1}\ldots \hat{\mu}_{4} 
\hat{\nu}_{1}\ldots \hat{\nu}_{4}}  
\left(\partial\hat{C} +{\textstyle\frac{1}{3}} \partial\hat{L}
\right)_{\hat{\mu}_{1}\ldots \hat{\mu}_{4}}
\left(\partial\hat{C} +{\textstyle\frac{1}{3}} \partial\hat{L}
\right)_{\hat{\nu}_{1}\ldots \hat{\nu}_{4}}=0\, .
\end{equation}

\noindent Substituting the solution

\begin{equation}
\hat{L}= -3 \hat{C}
\end{equation}

\noindent into the above auxiliary action one recovers the usual action
of 11-dimensional supergravity.

We can now consider this action as an action for $\hat{G}$ and add to
it the Lagrange multiplier term 

\begin{equation}
{\textstyle\frac{1}{16\pi G_{N}^{(11)}}}
\int  d^{11}x {\textstyle\frac{1}{4!\cdot6!}}\hat{\epsilon} 
\partial\hat{\tilde{C}}\hat{G}\, ,
\end{equation}

\noindent where we have already integrated it by parts, 
and try to eliminate $\hat{G}$ by using its equation of motion.
However, in this equation of motion both $\hat{G}$ and
${}^{\star}\hat{G}$ appear:

\begin{equation}
\label{eq:dualrel}
\hat{G}= {}^{\star}\left\{7\left[\partial\hat{\tilde{C}}
-{\textstyle\frac{5}{3}}\left(\hat{G}
+{\textstyle\frac{2}{3}}\partial \hat{L}\right)\hat{L} \right] 
\right\}\, .  
\end{equation}

After some tensor gymnastics we find the following equation
for $\hat{G}$ as a function of $\hat{\tilde{C}}$ and $\hat{L}$:

\begin{equation}
\hat{G}^{\hat{\alpha}_{1}\ldots\hat{\alpha}_{4}}
= \left(f^{-1}\right)^{\hat{\alpha}_{1}\ldots\hat{\alpha}_{4}}
{}_{\hat{\beta}_{1}\ldots\hat{\beta}_{4}} 
\left(\hat{B}^{\hat{\beta}_{1}\ldots\hat{\beta}_{4}}  
-{}^{\star}\hat{B}^{\hat{\beta}_{1}\ldots\hat{\beta}_{7}}
\hat{A}_{\hat{\beta}_{5}\hat{\beta}_{6}\hat{\beta}_{7}}\right)\, ,
\end{equation}

\noindent where

\begin{equation}
\begin{array}{rcl}
f_{\hat{\alpha}_{1}\ldots\hat{\alpha}_{4}}
{}^{\hat{\beta}_{1}\ldots\hat{\beta}_{4}} 
& = & 
\delta_{\hat{\alpha}_{1}\ldots\hat{\alpha}_{4}}
{}^{\hat{\beta}_{1}\ldots\hat{\beta}_{4}} 
-\frac{7!}{4!}\delta_{\hat{\alpha}_{1}\ldots\hat{\alpha}_{7}}
{}^{\hat{\beta}_{1}\ldots\hat{\beta}_{7}} 
\hat{A}_{\hat{\beta}_{5}\hat{\beta}_{6}\hat{\beta}_{7}}
\hat{A}^{\hat{\alpha}_{5}\hat{\alpha}_{6}\hat{\alpha}_{7}}\, .\\
& & \\
\hat{A}_{\hat{\alpha}_{1}\hat{\alpha}_{2}\hat{\alpha}_{3}}
& = &
\frac{1}{3^{3}\cdot 2^{4}} 
\hat{L}_{\hat{\alpha}_{1}\hat{\alpha}_{2}\hat{\alpha}_{3}}\, ,\\
& & \\
\hat{B}^{\hat{\alpha}_{1}\ldots\hat{\alpha}_{4}} & = & 
\frac{1}{6!\sqrt{|g|}}\hat{\epsilon}^{\hat{\alpha}_{1}
\ldots\hat{\alpha}_{10}} \left(\partial\hat{\tilde{C}}
-\frac{10}{9}\partial\hat{L}\hat{L} 
\right)_{\hat{\alpha}_{5}\ldots\hat{\alpha}_{10}}\, .\\
\end{array}
\end{equation}

The next step in the dualization procedure, namely the elimination of
the auxiliary field, can now be done. The result (too complicated to
be meaningful) would be $\hat{L}$ as a complicated non-linear and
perhaps non-local function of $\hat{\tilde{C}}$ and
$\partial\hat{\tilde{C}}$. This relation, together with the relation
between $\hat{L}$ and $\hat{C}$ would give us the relation between
$\hat{C}$ and its dual field $\hat{\tilde{C}}$.  The lesson to be
learned here is that an 11-dimensional supergravity theory formulated
in terms of only the 6-form $\hat{\tilde{C}}$ does exist although it
is highly non-linear.

What we know about $\hat{\tilde{C}}$ is, however, enough to determine
its gauge transformation laws. We can use Eq.~(\ref{eq:dualrel}) plus
the relation between $\hat{L}$ and $\hat{C}$ and the fact that it has
to be gauge-invariant to find that \cite{kn:Ah}

\begin{equation}
\label{eq:gauge6}
\left\{
\begin{array}{rcl}
\delta_{\hat{\tilde{\chi}}}\hat{\tilde{C}} & = & 
6\partial\hat{\tilde{\chi}}\, ,\\
& & \\
\delta_{\hat{\chi}}\hat{\tilde{C}} & = & 30\partial\hat{\chi}\hat{C}\, ,\\
\end{array}
\right.
\end{equation}

\noindent with the understanding that $\hat{C}$ is 
a complicated function of $\hat{\tilde{C}}$. We can write the field
strength of the latter as

\begin{equation}
\hat{\tilde{G}}=7\left(\partial\hat{\tilde{C}} 
+10\hat{C}\partial\hat{C} \right)\, ,  
\end{equation}

\noindent so Eq.~(\ref{eq:dualrel}) is nothing but the usual
 duality relation

\begin{equation}
\hat{G}={}^{\star}\hat{\tilde{G}}\, .
\end{equation}


\subsection{Dual Massive IIA Supergravity}
\label{sec-dual10}

In this Section we are going to discuss several aspects concerning the
duals of the potentials appearing in massive IIA supergravity.

First we consider the massless theory as a special case.  It is easy
to see that in dualizing the NS/NS 2-form $B$ or the R-R 3-form
$C^{(3)}$ one runs into the non-linearity problem we found in eleven
dimensions. The R-R 1-form $C^{(1)}$ cannot be immediately dualized
off-shell since it appears explicitly in $G^{(4)}$. However, it can be
reabsorbed into a redefinition of $C^{(3)}$ (after an integration by
parts in the action). In the end, though, the same problem occurs.
One can only define field strengths for the dual potentials if one
uses also the original potentials (understood as functions of the dual
ones) and the gauge transformation laws can then be determined. We
will do so after discussing the massive case.

In the massive theory on top of the above-mentioned problems we have
the problem of having to dualize the {\sl massive} NS/NS 2-form $B$.
In principle it is not clear what the dual of a massive $k$-form
potential is, since a $(d-k-2)$-form potential (which is the dual in
absence of mass) always has a different number of degrees of freedom.
However, a massive $(d-k-1)$-form does describe the same number of
degrees of freedom as a massive $k$-form. Dualization of the
corresponding action can be achieved through the use of an
intermediate action (see Ref.~\cite{kn:Q} and references therein). The
construction of this intermediate action with an auxiliary field is
not systematic and has to be made on a case by case basis, which
obscures the meaning of the dualization.

In order to set up a more systematic procedure, it is convenient to
rewrite the action for the massive $k$-form using an auxiliary
$(k-1)$-form which plays the role of a Stueckelberg field. The
resulting action is invariant under massive gauge transformations.
That permits the elimination of the Stueckelberg field so that the
usual action is recovered. One now dualizes first the Stueckelberg
$(k-1)$-form field and then the $k$-form potential.  The latter
dualization is possible because after the first dualization the action
only depends on the $k$-form potential via its field strength.  This
procedure is described in full detail in Appendix~\ref{sec-massdual}.
Here we only need to know the qualitative result: after the two
dualizations we find that the $(d-k-2)$ dual (in the massless sense)
of the massive $k$-form field becomes the Stueckelberg field of the
$(d-k-1)$ dual (in the massless sense) of the original Stueckelberg
$(k-1)$-form. The mass parameter is the same after dualization.

In our case, the R-R 1-form $C^{(1)}$ dualizes into the R-R 7-form
$C^{(7)}$ which now is massive, while the NS/NS 2-form $B$, which is
massive, dualizes into the NS/NS 6-form $\tilde{B}_{\rm IIA}$ which is
nothing but the Stueckelberg field for $C^{(7)}$. There is a dual
massive transformation but the mass parameter is the same. This result
has to be taken into account when trying to find the field strength of
the dual fields, specially of $\tilde{B}_{\rm IIA}$. From now on,
since confusion with the Heterotic and IIB cases is not possible, we
will denote ${\tilde B}_{\rm IIA}$ simply as $\tilde B$.

In all other respects the dualization of the massive type~IIA theory
proceeds as in the massless case. To find the field strengths we have
to allow for the presence of the original potentials. Now we are going
to construct those field strengths starting with the massless case.
For $\tilde{B}$ and $C^{(5)}$ we can use the dimensional reduction of
the field strength of $\hat{\tilde C}$.  This 11-dimensional field
splits as follows in terms of 10-dimensional fields:

\begin{equation}
\left\{
\begin{array}{rcl}
\hat{\tilde{C}}_{\mu_{1}\ldots\mu_{5}y} & = & 
\left(i_{\hat{k}}\hat{\tilde{C}} \right)_{\mu_{1}\ldots\mu_{5}} \\
& & \\
& = &
C^{(5)}{}_{\mu_{1}\ldots\mu_{5}} -5 C^{(3)}{}_{[\mu_{1}\mu_{2}\mu_{3}}
B_{\mu_{4}\mu_{5}]}\, , \\
& & \\
& & \\
\hat{\tilde{C}}_{\mu_{1}\ldots\mu_{6}} & = &
-\tilde{B}_{\mu_{1}\ldots\mu_{6}} \, .\\
\end{array}
\right.
\end{equation}

\noindent It is now immediate to find the massless field 
strengths\footnote{The subscript $(0)$ indicates that they 
correspond to the massless case.}:

\begin{equation}
\left\{
\begin{array}{rcl}
G^{(6)}_{(0)} & = & 
6\left[\partial C^{(5)} -10\partial BC^{(3)}\right]\, ,\\
& & \\
\tilde{H}_{(0)} & = & 7\left[\partial\tilde{B} 
+G^{(6)}_{(0)}C^{(1)} -10C^{(3)}\partial C^{(3)}\right]\, .\\
\end{array}
\right.
\end{equation}

To find the massless field strength of the R-R 7-form $C^{(7)}$ (dual
to $C^{(1)}$) we rewrite the equation of motion of $C^{(1)}$ as
follows:

\begin{equation}
\partial\left\{{}^{\star}G^{(2)}_{(0)} 
-28\left[{}^{\star}G^{(4)}_{(0)}B  -30\partial C^{(3)}B^{2}\right] 
\right\} =0  \, .
\end{equation}

This equation should be equivalent to the Bianchi identity of the dual
$C^{(7)}$. We can identify the expression in curly brackets with
$8\partial C^{(7)}$ up to a total derivative. (The factor of 8 ensures
canonical normalization of $C^{(7)}$.) This total derivative is
arbitrary and it amounts to a different definition of $C^{(7)}$ with
different gauge transformations. The choice that gives a $C^{(7)}$ in
the form proposed in Refs.~\cite{kn:GHT,kn:BCT} is

\begin{equation}
\begin{array}{rcl}
8 \partial \left[C^{(7)}+3\cdot 5\cdot 7 C^{(3)}B^{2} 
-3\cdot 7 C^{(5)}B  \right] =\hspace{-4cm}&  & \\
& & \\
& & 
= {}^{\star}G^{(2)}_{(0)} 
-28\left[{}^{\star}G^{(4)}_{(0)}B -30\partial C^{(3)} B^{2}\right]\, ,
\end{array}
\end{equation}

\noindent and so, using that ${}^{\star}G^{(2)}_{(0)}= G^{(8)}_{(0)}$ 
we find 

\begin{equation}
G^{(8)}_{(0)} = 8 \left[\partial C^{(7)}-3\cdot 7 
\partial B C^{(5)}\right]\, .
\end{equation}

The gauge transformation laws of $C^{(5)},\tilde{B},C^{(7)}$ can be
found by using the gauge invariance (by construction) of the above
field strengths. We find:

\begin{equation}
\left\{
\begin{array}{rcl}
\delta C^{(5)} & = & 15\partial \Lambda^{(0)} B^{2} 
+30\partial\Lambda^{(2)} B +5\partial\Lambda^{(4)}\, ,\\
& & \\
\delta \tilde{B} & = & 6\partial\Lambda^{(0)} 
\left(C^{(5)} -5C^{(3)} B \right)
-30\partial \Lambda^{(2)} C^{(3)} 
+6\partial\tilde{\Lambda}\, ,\\
& & \\
\delta C^{(7)} & = & 3\cdot 5\cdot 7\partial \Lambda^{(0)} B^{3} 
+3^{2}\cdot 5\cdot 7 \partial\Lambda^{(2)} B^{2}\\
& & \\
& &
+3\cdot 5\cdot 7\partial\Lambda^{(4)} B + 7 \partial\Lambda^{(6)}\, .\\
\end{array}
\right.
\end{equation}

The massive field strengths are found by adding extra terms
proportional to $m$, which are uniquely determined by requiring
invariance under massive gauge transformations. The gauge
transformations in the massive theory can be found from those of the
massless theory (above) by the replacements

\begin{equation}
\left\{
\begin{array}{rcl}
\Lambda & \rightarrow & -2\lambda\, ,\\
& & \\
\partial\Lambda^{(0)} & \rightarrow & \partial\Lambda^{(0)}+m\lambda\, ,\\
\end{array}
\right.
\hspace{1cm}
\left\{
\begin{array}{rcl}
\Lambda^{(6)} & \rightarrow & 2\tilde{\lambda}\, ,\\
& & \\
\partial\tilde{\Lambda} & \rightarrow & \partial\tilde{\Lambda}
+\frac{m}{6}\tilde{\lambda}\, ,\\
\end{array}
\right.
\end{equation}

\noindent where we have taken into account that now 
$\tilde{B}$ is the Stueckelberg field for $C^{(7)}$ (for this reason
we expect a term $mC^{(7)}$ in $\tilde{H}$). The result
is\footnote{All these formulae are collected in
  Appendix~\ref{sec-fields}.}

\begin{equation}
\left\{
\begin{array}{rcl}
G^{(6)} & = & 6\left[\partial C^{(5)} -10\partial BC^{(3)}
+\frac{5}{4}mB^{3}\right]\, ,\\
& & \\
\tilde{H} & = & 7\left[\partial\tilde{B} +G^{(6)}C^{(1)} 
-10C^{(3)}\partial C^{(3)}
\right.\\
& & \\
& & 
\left.
-\frac{1}{14}m \left(  C^{(7)} -3\cdot 7 C^{(5)}B
+3\cdot 5\cdot 7 C^{(3)}B^{2} \right)\right]\, ,\\
& & \\
G^{(8)} & = & 8 \left[\partial C^{(7)}-3\cdot 7 \partial B C^{(5)}
+ \frac{3\cdot 5 \cdot 7}{16}m B^{4}\right]\, .\\
\end{array}
\right.
\end{equation}

It is also useful to have the relation between the infinitesimal gauge
transformation parameters of the 11- and 10-dimensional theories:

\begin{equation}
\left\{
\begin{array}{rcl}
\Lambda^{(2)}{}_{\mu\nu} & = & \hat{\chi}_{\mu\nu}\, ,\\
& & \\
\lambda_{\mu} & = & \hat{\lambda}_{\mu} = \\
& & \\
& = &
-\frac{1}{2}\left(i_{\hat{k}}\hat{\chi}\right)_{\mu} = 
-\frac{1}{2}\hat{\chi}_{\mu y}\, ,\\
\end{array}
\right.
\hspace{.5cm}
\left\{
\begin{array}{rcl}
\Lambda^{(4)}{}_{\mu_{1}\ldots\mu_{4}} & = & 
\left(i_{\hat{k}}\hat{\tilde{\chi}} \right)_{\mu_{1}\ldots\mu_{4}} \\
& & \\
& = &
\hat{\tilde{\chi}}_{\mu_{1}\ldots\mu_{4}y}\, ,\\
& & \\
\tilde{\Lambda}_{\mu_{1}\ldots\mu_{5}} & = & 
-\hat{\tilde{\chi}}_{\mu_{1}\ldots\mu_{5}}\, .\\
\end{array}
\right.
\label{eq:useful}
\end{equation}

The gauge parameter $\Lambda^{(0)}$ is related to general coordinate
transformations of the compact eleventh coordinate $y$. The parameters
$\Lambda^{(6)}$ and $\tilde{\lambda}$ will be discussed in the next
Subsection in which we will try to understand the higher-dimensional
origin of $C^{(7)}$.


\subsection{Dual Massive 11-Dimensional Supergravity}
\label{sec-dualmassive11}

Dual massive 11-dimensional supergravity is by definition the theory
that upon dimensional reduction leads to dual massive IIA
supergravity.  Using the relations between 10- and 11-dimensional
fields and gauge-transformation parameters it is immediate to find the
massive gauge-transformation rules for $\hat{\tilde{C}}$, which we
write together with those of $\hat{C}$ for convenience:

\begin{equation}
\left\{
\begin{array}{rcl}
\delta \hat{C}_{\hat{\mu}\hat{\nu}\hat{\rho}}
 & =  & 3\partial_{[\hat{\mu}}\hat{\chi}_{\hat{\nu}\hat{\rho}]}
+3m \hat{\lambda}_{[\hat{\mu}}
\left(i_{\hat{k}}\hat{C} \right)_{\hat{\nu}\hat{\rho}]}\, , \\
& & \\
\delta \hat{\tilde{C}}_{\hat{\mu}_{1}\ldots\hat{\mu}_{6}} & = & 
6\partial_{[\hat{\mu}_{1}}
\hat{\tilde{\chi}}_{\hat{\mu}_{2}\ldots\hat{\mu}_{6}]}
+30\partial_{[\hat{\mu}_{1}}\hat{\chi}_{\hat{\mu}_{2}\hat{\mu}_{3}}
\hat{C}_{\mu_{4}\hat{\mu}_{5}\hat{\mu}_{6}]} \\
& & \\
& &    
-6m\hat{\lambda}_{[\hat{\mu}_{1}}
\left(i_{\hat{k}}\hat{\tilde{C}}
\right)_{\hat{\mu}_{2}\ldots\hat{\mu}_{6}]}
-m \hat{\tilde{\lambda}}_{\hat{\mu}_{1}\ldots\hat{\mu}_{6}}
\, .\\
\end{array}
\right.
\end{equation}

The first and second terms are also present in the massless case.  The
third term is just the massive gauge transformation of a 6-form
according to the general rule
Eq.~(\ref{eq:generalmasstrans})\footnote{Observe, though, that just as
  $\hat{C}$, $\hat{\tilde{C}}$ has additional terms and does not
  transform covariantly under massive gauge transformations. Extra
  terms are needed to construct a covariant gauge field strength.}.
The fourth term corresponds to a {\it dual massive gauge
  transformation}.  In order to understand its origin, we first
observe that the above transformation law only reduces to those of
$\tilde{B},C^{(5)}$ that we found in the previous Subsection if

\begin{equation}
\left(i_{\hat{k}}\hat{\tilde{\lambda}}\right)_{\mu_{1}\ldots\mu_{5}}=
\hat{\tilde{\lambda}}_{\mu_{1}\ldots\mu_{5}y}=0\, ,
\end{equation}

\noindent because otherwise $C^{(5)}$ would have an extra symmetry under
massive shifts by a 5-form.  The same property is also satisfied by
$\hat{\lambda}$ precisely because it is equal to
$(i_{\hat{k}}\hat{\chi})$. This analogy then leads us to identify

\begin{equation}
\hat{\tilde{\lambda}}\equiv a\left(i_{\hat{k}}\hat{\Sigma}\right)\, ,
\end{equation}

\noindent where $\hat{\Sigma}$ is a 7-form and $a$ a constant to
be determined. A 7-form gauge parameter is associated to an 8-form
gauge potential. The only such potential available is the dual of the
Killing vector, that we can consider as a 1-form potential
$\hat{k}_{\hat{\mu}}$.

The same conclusion is reached if one writes the 11-dimensional
massive field strength of $\hat{\tilde{C}}$:

\begin{equation}
\hat{\tilde{G}}= 7\left\{D\hat{\tilde{C}} +10\hat{C}D\hat{C}
-{\textstyle\frac{5}{4}}m \left(i_{\hat{k}}\hat{C} \right)
\left(i_{\hat{k}}\hat{C} \right)\hat{C} +{\textstyle\frac{1}{14}}m 
\left(i_{\hat{k}}\hat{\tilde{N}} \right) \right\}\, ,
\end{equation}

\noindent where $\hat{\tilde{N}}$ is an 8-form potential such that

\begin{equation}
\left(i_{\hat{k}}\hat{\tilde{N}} \right)_{\mu_{1}\ldots\mu_{7}}
= C^{(7)}{}_{\mu_{1}\ldots\mu_{7}} -5\cdot 7 
C^{(3)}{}_{[\mu_{1}\mu_{2}\mu_{3}} B_{\mu_{4}\mu_{5}} B_{\mu_{6}\mu_{7}]}\, .
\end{equation}

A 7-form potential was needed to absorb the dual massive gauge
transformations of $\hat{\tilde{C}}$ but the contraction of the 7-form
potential with the Killing vector would again be zero. Thus, the
7-form potential has to be the contraction of the Killing vector with
an 8-form potential that we have denoted by $\hat{\tilde{N}}$.

The field strength of $\hat{\tilde{C}}$ does transform covariantly
under massive gauge transformations:

\begin{equation}
\delta_{\hat{\chi}}\hat{\tilde G}= 7m \hat{\lambda} 
\left(i_{\hat{k}} \hat{\tilde{G}} \right)\, .
\end{equation}

To complete the picture we have to determine the gauge transformation
laws and field strength of $\hat{\tilde{N}}$ and then check that it
leads to the field strength of $C^{(7)}$.

In a first step we find that

\begin{equation}
\begin{array}{rcl}
\delta  \hat{\tilde{N}}_{\mu_{1}\ldots\mu_{7}y} & = &
16 a \partial_{[\mu_{1}}\hat{\Sigma}_{\mu_{2}\ldots\mu_{7}y]}
+ \frac{8!}{3\cdot 4!} \partial_{[\mu_{1}}\hat{\chi}_{\mu_{2}\mu_{3}}
\hat{C}_{\mu_{4}\mu_{5}\mu_{6}} \left(i_{\hat{k}} \hat{C} \right)_{\mu_{7}y]}
\\
& & \\
& & 
+3 \frac{8!}{6!}  \partial_{[\mu_{1}}\hat{\tilde{\chi}}_{\mu_{2}\ldots\mu_{6}}
\left(i_{\hat{k}} \hat{C} \right)_{\mu_{7}y]}\, ,\\
\end{array}
\end{equation}

\noindent which looks somewhat strange because one would expect terms
of the form $\left(i_{\hat{k}}\hat{C} \right)$ to appear only at order
$m$.  However, since $\hat{\tilde{N}}$ is the dual of the Killing
vector it only exists when there is an isometry and, thus, those terms
are allowed (since they do not appear in the massless case).

This suggests the following 11-dimensional transformation law

\begin{equation}
\delta  \hat{\tilde{N}} =
16 a \partial\hat{\Sigma} + {\textstyle\frac{8!}{3\cdot 4!}} 
\partial\hat{\chi} \hat{C} \left(i_{\hat{k}} \hat{C} \right)
+3 {\textstyle\frac{8!}{6!}}  \partial\hat{\tilde{\chi}}
\left(i_{\hat{k}} \hat{C} \right)
-8m\hat{\lambda} \left(i_{\hat{k}}\hat{\tilde{N}}\right)\, ,
\end{equation}

\noindent where we have added a term corresponding to massive gauge 
transformations of an 8-form. The contraction of this term with the
Killing vector vanishes and one recovers the previous formula.

If we accept the existence of an 8-form in 11 dimensions then this
leads naturally to a 7-form and an 8-form in 10 dimensions.  The
7-form is $C^{(7)}$, the dual of $C^{(1)}$. Similarly, it seems that
the 8-form naturally corresponds to the dual of the dilaton. If true,
this suggests the existence of a related NS/NS 7-brane to which this
8-form couples. We will not pursue these ideas further here.


\section{Worldvolume Fields}
\label{sec-worldvolume}

To construct the worldvolume effective actions of massive branes in
the next Section a knowledge of the background fields and worldvolume
fields will be necessary. In the previous Section we have studied the
former and in this Section we are going to study the gauge
transformation laws and field strengths of the different worldvolume
fields. The results of this Section are valid both for type~IIA and
type~IIB worldvolume theories.

All D-p-brane worldvolume theories contain the BI vector field, that
we call here $b$ because it transforms by shifts of $\Lambda$, the
gauge transformation parameter of $B$.

\begin{table}
\begin{center}
\begin{tabular}{||c|c|c|c|c||}
\hline\hline
Target space & Field     & Gauge     & World Volume &  Field \\
Field        & Strength  & Parameter & Field        & Strength \\
\hline\hline
$B$        & $H$       & $\Lambda$  ($\lambda$) & $b$       
& ${\cal F}$ \\
\hline
$C^{(1)}$  & $G^{(2)}$ & $\Lambda^{(0)}$                     & $c^{(0)}$ & 
${\cal G}^{(1)}$ \\
\hline
$C^{(3)}$  & $G^{(4)}$ & $\Lambda^{(2)}$                     & $c^{(2)}$ & 
${\cal G}^{(3)}$ \\   
\hline
$C^{(5)}$  & $G^{(6)}$ & $\Lambda^{(4)}$                     & $c^{(4)}$ & 
${\cal G}^{(5)}$ \\   
\hline
$C^{(7)}$  & $G^{(8)}$ & $\Lambda^{(6)}$ ($\tilde{\lambda}$) & $c^{(6)}$ & 
${\cal G}^{(7)}$ \\   
\hline \hline
\end{tabular}
\end{center}  
\caption{\label{eso}This table shows the correspondence between type~IIA 
target-space potentials and worldvolume potentials, together with 
their field strengths and the gauge parameters. The worldvolume fields 
transform by shifts of the gauge parameter of the associated 
target-space potential.}
\end{table}

We are going to see that other worldvolume fields may occur in p- and
D-p-brane actions. In the case of D-p-branes, these are the duals of
the BI field $b$.  They are forms of different rank depending on the
dimension of the worldvolume at hand. These worldvolume fields
transform by shifts of the gauge parameter of a R-R form. We will
denote them by $c^{(n)}$, since they are associated to $C^{(n+1)}$
(see Table~\ref{eso} for some type~IIA examples).

For instance, the D-4-brane can be described using the BI 1-form $b$
(in the ``1-form formalism'') but it may also be described by a 2-form
$c^{(2)}$ together with $b$ (``1-2-form formalism'') \cite{kn:BRO} or
by a 2-form field $c^{(2)}$ alone (``2-form formalism'')
\cite{kn:APPS}. The worldvolume 2-form $c^{(2)}$ transforms by shifts
of $\Lambda^{(2)}$, the gauge parameter associated to $C^{(3)}$.

Our goal in this Section is to find an homogeneous description ({\it
  independent of the worldvolume dimension}) of the gauge
transformation laws and field strengths of all worldvolume p-form
fields $c^{(p)}$ that will be needed in the next Section.

Our work will be greatly simplified by the following observation: the
WZ terms of D-p-branes $WZ^{(p+1)}$ are $(p+1)$-forms that are
invariant up to a total derivative\footnote{In this Section and in
  Appendix~\ref{sec-fields} we will use, as opposed to the rest of the
  paper, a differential form notation. The relation between both
  notations is explained in Appendix~\ref{sec-fields}. Some of the
  formulae in this Section can be found in Appendix~\ref{sec-wvfields}
  in component notation.}

\begin{equation}
\delta (WZ)^{(p+1)} \equiv d A^{(p)}\, .
\end{equation}

\noindent In particular, $A^{(p)}$ contains $\Lambda^{(p)}$.
Furthermore, $A^{(p)}$ contains only the pullbacks of the
$\Lambda^{(p)}$'s and worldvolume fields but no background fields
whatsoever.  Thus, it is natural to identify $A^{(p)}$ with minus the
variation of $c^{(p)}$ up to a total derivative which we denote by
$d\kappa^{(p-1)}$.  Then, by construction we have\footnote{Note that,
  strictly speaking, we first introduce each $c^{(p)}$ for a fixed
  value (p+1) of the worldvolume. Next, we may consider the same
  $c^{(p)}$ for other worldvolume dimensions.}

\begin{equation}
\left\{
\begin{array}{rcl}
{\cal G}^{(p+1)} & = & dc^{(p)} 
+\frac{1}{2 \pi\alpha^{\prime}} WZ^{(p+1)}\, ,\\
& & \\
\delta c^{(p)} & = & d\kappa^{(p-1)}
-\frac{1}{2\pi\alpha^{\prime}} A^{(p)}\, .\\
\end{array}
\right.
\end{equation}

In order to give explicit expressions in the simplest possible way we
use the formalism introduced in Refs.~\cite{kn:Douglas,kn:GHT} in
which k-forms of different degrees are formally combined into a single
entity:

\begin{equation}
\left\{
\begin{array}{rcl}
C & = & C^{(0)} + C^{(1)} + C^{(2)} +\ldots\, , \\
& & \\
G & = & G^{(0)} + G^{(1)} + G^{(2)} +\ldots\, , \\
& & \\
\Lambda^{(\cdot)}  & = & \Lambda^{(0)} +\Lambda^{(1)} + \ldots\, , \\
\end{array}
\right.
\end{equation}

\noindent and we extend it to the worldvolume fields\footnote{This
formal sum has also been introduced in \cite{kn:BeTo}.}:

\begin{equation}
\left\{
\begin{array}{rcl}
c & = & c^{(0)} + c^{(1)} + c^{(2)} +\ldots\, , \\
& & \\
{\cal G} & = & {\cal G}^{(0)} +{\cal G}^{(1)} +{\cal G}^{(2)} +\ldots\, , \\
& & \\
\kappa  & = & \kappa^{(0)} +\kappa^{(1)} + \ldots\, . \\
\end{array}
\right.
\end{equation}

In this language the gauge transformations and field strengths of the
R-R fields and the $B$ field can be written in the more compact
form\footnote{Observe that our $\Lambda^{(\cdot)}$ differs form that
  of Ref.~\cite{kn:GHT} by a factor of $e^{B}$. All products of forms
  here are exterior products.}

\begin{equation}
\left\{
\begin{array}{rcl}
\delta C & = & \left(d\Lambda^{(\cdot)} +m\lambda\right)e^{B}\, ,\\
& & \\
\delta B & = & -2d\lambda\, ,\\
& & \\
G & = & d C - dB C + \frac{m}{2}e^{B}\, .\\
& & \\
H & = & dB\, .\\
\end{array}
\right.
\end{equation}

\noindent
The BI field gauge transformation rule and field strength are\footnote{
It is worth remarking that $\kappa^{(0)}\neq \rho^{(0)}$.}

\begin{equation}
\left\{
\begin{array}{rcl}
\delta b & = &  \frac{1}{2\pi\alpha^{\prime}}2\lambda +d\rho^{(0)}\, ,\\
& & \\
{\cal F} & = & db +\frac{1}{2\pi\alpha^{\prime}}B\, .\\
\end{array}
\right.
\end{equation}

\noindent
We also define 

\begin{equation}
\omega(b) = \sum_{r=0}{\textstyle\frac{(-1)^{r+1}}{(r+1)!}}
 (2\pi\alpha^{\prime})^{r+1} b(db)^{r}\, ,
\end{equation}

\noindent which has the property 

\begin{equation}
d\omega =e^{-(2\pi\alpha^{\prime})db} -1= 
e^{-(2\pi\alpha^{\prime}){\cal F}}e^{B} -1\, .  
\end{equation}

With these elements we find that the gauge transformation law and field 
strength of $c$ are given by

\begin{equation}
\label{transc}
\left\{
\begin{array}{rcl}
\delta c & = &  d\kappa -\frac{1}{2\pi\alpha^{\prime}}
\Lambda^{(\cdot)}e^{-(2\pi\alpha^{\prime})db}  
-\frac{m}{2}\rho^{(0)}\sum_{r=0}\frac{(-1)^{r+1}}{(r+1)!} 
(2\pi\alpha^{\prime})^{r} (db)^{r} \\
& & \\
& &
-\frac{m}{2}2\lambda \sum_{r=1}\frac{(-1)^{r+1}r}{(r+1)!} (2\pi\alpha^{\prime})^{r-1}
b(db)^{r-1}\, ,\\
& & \\
{\cal G} & = & dc +\frac{1}{2\pi\alpha^{\prime}}
\left\{Ce^{-(2\pi\alpha^{\prime}){\cal F}} +\frac{m}{2}\omega \right\}\, .\\
\end{array}
\right.
\end{equation}

\noindent
Observe that the above definition of ${\cal G}$ has the following 
property

\begin{equation}
G e^{-(2\pi\alpha^{\prime}){\cal F}} =(2\pi\alpha^{\prime})d{\cal G}\, ,
\end{equation}

\noindent where $G$ is the pullback of the R-R field strengths.

Knowing the existence of the p-form worldvolume fields it is natural to
redefine the WZ terms of the D-p-brane effective actions as

\begin{equation}
\label{WZprime}
WZ^{(p+1)\ \prime} =(2\pi\alpha^{\prime}){\cal G}^{(p+1)}\, .  
\end{equation}

\noindent The new WZ term is {\it exactly gauge-invariant}, 
which permits the extension of previous results to non-trivial
worldvolume topologies.  Observe that the new worldvolume field is
non-dynamical and no new degrees of freedom are added. Furthermore,
the equations of motion remain the old ones.

An alternative way to motivate the introduction of the worldvolume
p-form fields $\{c^{(p)}\}$ is to consider Poincar\'e-duality
transformations of D-p-brane actions.

To perform the Poincar\'e-duality transformation with respect to $b$
one first has to eliminate any explicit dependence on $b$ so that it
only appears through its derivative $db$. It turns out that $b$ occurs
explicitly in the WZ term for non zero mass.  Now, it is
straightforward to obtain a form of the WZ term in which there is no
explicit dependence on $b$ by introducing an auxiliary vector field,
that we call $e$, such that its integration gives back the original WZ
term. The resulting WZ term is

\begin{equation}
\label{WZmass}
\widetilde{WZ}^{(p+1)}=(2\pi\alpha^\prime)\left\{dc+
{\textstyle\frac{1}{2\pi\alpha^\prime}}
\left[C+{\textstyle\frac{m}{2}}\omega(e)
e^{2\pi\alpha^\prime{\cal F}(e)}\right]
e^{-2\pi\alpha^\prime {\cal F}(b)}\right\}\, .
\end{equation}

\noindent Here we have specified the definitions of $\omega$ and 
$\cal F$ in terms of the BI field (the auxiliary field $e$) as
$\omega(b)$, ${\cal F}(b)$ ($\omega(e)$, ${\cal F}(e)$).  Integration
over $e$ implies $e=b$ and the usual WZ term (\ref{WZprime}) is
obtained. 

Although the full Poincar\'e-dualization of a general D-brane action
is complicated, given that this transformation is consistent with
gauge-invariance it can in principle be performed in the quadratic
approximation, and the field strength of the field dual to $b$ (in a
$(p+1)$-dimensional worldvolume, a $(p-2)$-form) can be immediately
read off and turns out to be precisely ${\cal G}^{(p-1)}$.  Thus, the
$c^{(p-2)}$ worldvolume fields are nothing but the duals of the BI
field in the quadratic approximation and one can also say that the
field strength ${\cal G}^{(p+1)}$ is obtained via Poincar\'e duality
from the BI field strength ${\cal F}$ in the $(p+3)$-dimensional
worldvolume effective action of the D-$({\rm p}+2)$-brane.

The set of worldvolume fields $\{c^{(p)}\}$ that we have just
introduced is, of course, not unique. In order to build duals of
D-brane actions keeping track of all total derivatives it is more
appropriate to use a basis of worldvolume fields $\{a^{(p)}\}$ that do
not transform under Hodge duality of the BI field\footnote{The basis
  that we have introduced so far does not satisfy this requirement
  because there is an explicit dependence on $b$ in the gauge
  variations of the fields. Note that $b$ appears explicitly in the
  massive contribution to the gauge transformation law of the new
  fields, however this dependence disappears when in the actual
  dualization one introduces the auxiliary field $e$ as indicated
  above.}.  The gauge transformation rules and field strengths (${\cal
  H}^{(p+1)}$) of the $a^{(p)}$'s are given by

\begin{equation}
\left\{  
\begin{array}{rcl}
\delta a & = & d\mu -\frac{1}{2\pi\alpha^{\prime}}\Lambda^{(\cdot)}
-2da\lambda \\
& & \\
& & 
-\frac{m}{2}e^{(2\pi\alpha^{\prime})db}
\left[\rho^{(0)}\sum_{r=0}\frac{(-1)^{r+1}}{(r+1)!} 
(2\pi\alpha^{\prime})^{r} (db)^{r} \right.\\
& & \\
& &
\left.
+2\lambda \sum_{r=1}\frac{(-1)^{r+1}r}{(r+1)!} (2\pi\alpha^{\prime})^{r-1}
b(db)^{r-1}\right]\, ,\\
& & \\
{\cal H} & = & d a\ e^{B} +\frac{1}{2\pi\alpha^{\prime}}C 
+\frac{1}{2\pi\alpha^{\prime}}\frac{m}{2}
e^{2\pi\alpha^{\prime}{\cal F}(b)}
\omega(b)\, .
\end{array}
\right.
\end{equation}

\noindent ${\cal H}$ satisfies

\begin{equation}
G = (2\pi\alpha^{\prime}) d{\cal H} +(2\pi\alpha^{\prime}){\cal H}H\, .  
\end{equation}

\noindent The relation between the two sets of worldvolume fields is:

\begin{equation}
\label{relaciones}
\left\{  
\begin{array}{rcl}
a(c,b) & = & c\  e^{(2\pi\alpha^{\prime})db}\, ,\\
& & \\
{\cal H}(a) & = & e^{(2\pi\alpha^{\prime}){\cal F}}
{\cal G}\left[c(a,b)\right]\, ,\\
& & \\
\mu & = & \kappa e^{(2\pi\alpha^{\prime})db}\, .\\
\end{array}
\right.
\end{equation}

\noindent From the above relations the gauge-invariance of 
${\cal H}$ is evident. Observe also that $a^{(0)}=c^{(0)}$.

The main difference between the $\{c^{(p)}\}$ and $\{a^{(p)}\}$ basis
is that, while in the field strengths of the $c^{(n)}$'s many R-R
potentials $C^{(r)}$ appear but only one $c^{(n)}$, in the field
strength of each $a^{(n)}$ many other worldvolume fields $a^{(r)}$
appear but only one R-R potential $C^{(n+1)}$.

This second basis of worldvolume fields can be used as follows: if, in
the WZ term of the {\it massless} D-p-brane action one substitutes
everywhere the R-R field $C^{(n)}$ by the field strength ${\cal
  H}^{(n)}$

\begin{equation}
\label{modifiedc}
C\rightarrow (2\pi\alpha^{\prime}) {\cal H}\, ,
\end{equation}

\noindent one gets a new WZ term which is invariant under all gauge
transformations, including the massive ones.  Conversely, given a
massive D-brane action we can absorb all the mass dependence into
modified RR fields as above, and actually eliminate the explicit
dependence on $b$ by the introduction of the auxiliary field $e$ as:

\begin{equation}
{\cal H}  =  d a\ e^{B} +{\textstyle\frac{1}{2\pi\alpha^{\prime}}}C 
+{\textstyle\frac{1}{2\pi\alpha^{\prime}}}{\textstyle\frac{m}{2}}
e^{2\pi\alpha^{\prime}{\cal F}(e)}
\omega(e).
\end{equation}

\noindent With this trick we can now apply the usual duality
procedure to build the dual action, since the resulting action is a
massless D-brane with modified RR fields. In the end we just
substitute in the dual action these RR fields and read the final
action in terms of the dual variable and the auxiliary fields. One can
then check that the resulting action is gauge invariant, {\sl
  including} total derivative terms.

We can extend our reasoning to the p-brane case. In the case of the
fundamental string, the WZ term can be corrected by the inclusion of a
total derivative $db$. The new WZ term is now

\begin{equation}
WZ_{string}^{\prime}= (2\pi\alpha^{\prime}){\cal F}\, ,
\end{equation}

\noindent and satisfies

\begin{equation}
H=(2\pi\alpha^{\prime})d{\cal F}\, .  
\end{equation}

The case of the (solitonic) p-5-brane is different for each string
theory for the reasons explained in the Introduction and here we will
only consider the p-5A-brane. One can introduce again a worldvolume
5-form field $\tilde{b}$ whose role is in a sense dual to that of the
BI vector field $b$. In the massless case, its field strength and
gauge transformation law are:

\begin{equation}
\left\{
\begin{array}{rcl}
\tilde{\cal F}_{(0)} & = & d\tilde{b} 
+{\textstyle\frac{1}{2\pi\alpha^{\prime}}}\tilde{B}
-\left(C^{(5)}-{\textstyle\frac{1}{2}}C^{(3)}B \right) dc^{(0)}\\
& & \\
& & 
-{\textstyle\frac{1}{2}}da^{(2)} \left(C^{(3)} 
+(2\pi\alpha^{\prime})Bdc^{(0)}\right)\, ,\\
& & \\
\delta \tilde{b} & = & -\frac{1}{2\pi\alpha^{\prime}}\tilde{\Lambda}
 +\Lambda^{(4)}dc^{(0)}\\
& & \\
& & 
+\frac12 (2\pi\alpha^{\prime})da^{(2)}\left( \delta_{(0)} a^{(2)} 
-d\mu^{(1)}\right)\, ,\\
\end{array}
\right.
\end{equation}

\noindent and the gauge field strength satisfies

\begin{equation}
(2\pi\alpha^{\prime})d\tilde{\cal F}_{(0)}=
\tilde{H}_{(0)} -(2\pi\alpha^{\prime}) G^{(6)}_{(0)}{\cal G}^{(1)}_{(0)}
+3(2\pi\alpha^{\prime})^{2}{\cal H}_{(0)}^{(3)}d{\cal H}_{(0)}^{(3)}\, .
\end{equation}

In the massive case, $\tilde{B}$ transforms under shifts of the 6-form
$\tilde{\lambda}$. There is only one way to make the field strength of
$\tilde{b}$ invariant: to include $c^{(6)}$ in it. This, in turn
forces $\tilde{b}$ to transform under shifts of the 5-form
$\rho^{(5)}$, the gauge parameter for $c^{(6)}$. Thus, $c^{(6)}$ may
become massive by ``eating'' $\tilde{b}$, which would be completely
eliminated by a gauge transformation. In more detail, we find

\begin{equation}
\left\{
\begin{array}{rcl}
\tilde{\cal F} & = & d\tilde{b} 
+{\textstyle\frac{1}{2\pi\alpha^{\prime}}}\tilde{B}
-\left(C^{(5)}-{\textstyle\frac{1}{2}}C^{(3)}B \right) 
\left(dc^{(0)}-\frac{m}{2}b\right)\\
& & \\
& & 
-{\textstyle\frac{1}{2}}\left[da^{(2)}
-\frac{m}{2}(2\pi\alpha^{\prime})bdb\right]\times\\
& & \\
& &  \left[C^{(3)} 
+(2\pi\alpha^{\prime})B\left(dc^{(0)}
-\frac{m}{2}b\right)
-\frac{m}{4}(2\pi\alpha^{\prime})^{2}bdb\right]\\
& & \\
& & -\frac{1}{6}\frac{m}{2}(2\pi\alpha^{\prime})^3dc^{(0)}bdbdb
+\frac{m}{2}c^{(6)}\, ,\\
& & \\
\delta \tilde{b} & = & -\frac{1}{2\pi\alpha^{\prime}}\tilde{\Lambda}
+\frac{m}{2}\rho^{(5)} +\Lambda^{(4)}dc^{(0)}\\
& & \\
& & 
+\frac12 (2\pi\alpha^{\prime})da^{(2)}
\left( \delta a^{(2)} -d\mu^{(1)}\right)\, ,\\
\end{array}
\right.
\end{equation}

\noindent and the gauge field strength satisfies now

\begin{equation}
\tilde{H}=
(2\pi\alpha^{\prime})d\tilde{\cal F}
-(2\pi\alpha^{\prime}) G^{(6)}{\cal G}^{(1)}
+3(2\pi\alpha^{\prime})^{2}{\cal H}^{(3)}d{\cal H}^{(3)}
+{\textstyle\frac{m}{2}}{\cal G}^{(7)}\, .
\end{equation}

This ends our description of the worldvolume fields of 10-dimensional
extended objects. Explicit formulae for the worldvolumes fields we
will be considering in the next Section can be found in
Appendix~\ref{sec-wvfields}.

We will see in the next sections that additional world-volume fields
need also be introduced for eleven dimensional branes.  Their
corresponding gauge transformations and field strengths will be
introduced case by case.


\section{Effective Actions of Massive Branes}
\label{sec-mbraneactions}

The purpose of this Section is to construct effective worldvolume
actions describing the dynamics of branes whose target space is one of
the massive supergravity theories described in
Section~\ref{sec-masssugras}.  Our strategy will be to first construct
the action for a massive M-brane. These actions turn out to have as a
common characteristic that they are gauged sigma models. The gauged
isometry coincides with the isometry needed to define the massive
11-dimensional supergravity theory. The original (ungauged) sigma
model describes an object of the same dimension in the massless
theory, i.e.~the corresponding massless brane. The worldvolume actions
are invariant under 11-dimensional massive gauge transformations.

\begin{figure}[!ht]
\begin{center}
\leavevmode
\epsfxsize= 12cm
\epsffile{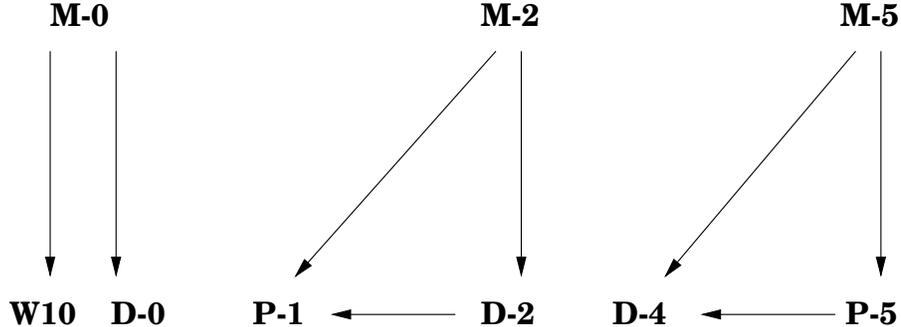}
\caption{\scriptsize Each M-brane gives rise to a pair of 
  type~IIA branes both in the massive and massless cases. Vertical
  arrows indicate direct dimensional reduction (i.e.~target-space
  dimensional reduction), slanted arrows indicate double dimensional
  reduction and horizontal arrows indicate worldvolume dimensional
  reduction together with gauge-fixing of some symmetries. As the
  figure indicates, double dimensional reduction is equivalent to
  direct dimensional reduction followed by worldvolume dimensional
  reduction and gauge-fixing.  Upon direct dimensional reduction, the
  M-0-brane gives the D-0-brane or the ten-dimensional Brinkmann wave
  W10 depending on whether the M-0-brane has momentum along the 11th
  dimension or not, respectively.}
\label{fig:reduction}
\end{center}
\end{figure}

In a second stage we consider different kinds of dimensional
reductions of the massive M-brane actions. In particular, we will show
that the so-called direct and double dimensional reductions in the
isometry direction lead to a pair of massive branes of IIA superstring
theory as represented in Figure~\ref{fig:reduction}. This shows that a
single massive M-brane unifies a pair of massive branes of string
theory.

Below, we will consider in successive order the massive M-0-brane,
M-2-brane and M-5-brane. The case of the d=11 KK monopole (or
M-6-brane) and a conjectured M-9-brane will be discussed in the
Conclusions.


\subsection{The Massive M-0-Brane}
\label{sec-M0}

Consider the action of the massless 11-dimensional particle
(M-0-brane):

\begin{equation}
\label{eq:usual0}
\hat{S}\ [\hat{X}^{\hat{\mu}},\gamma] =
-{\textstyle\frac{p}{2}} \int d\xi \sqrt{|\gamma|}\ \gamma^{-1}\
\partial_{\xi}\hat{X}^{\hat{\mu}}
\partial_{\xi}\hat{X}^{\hat{\nu}} 
\hat{g}_{\hat{\mu}\hat{\nu}}\, ,
\end{equation}

\noindent where $p$ is a constant with dimensions of mass.  This
action is known to give upon direct dimensional reduction the action
of the D-0-brane of Type~IIA superstring theory (see
e.g.~\cite{kn:BeTo2}).  Our goal is to obtain an effective action with
11-dimensional target space from which one can derive the effective
action of the massive D-0-brane. This object moves in 11-dimensional
spacetime and will henceforth be referred to as the {\it massive}
M-0-brane\footnote{We reserve the name massive M-branes for those
  branes that move in an 11-dimensional spacetime and have the
  property that they give upon dimensional reduction the massive
  branes of Type~IIA superstring theory.}.  We proceed by analogy with
the massive D-2-brane case, whose effective worldvolume action
\cite{kn:BR,kn:GHT} can be obtained from the M-2-brane effective
action by gauging an isometry and introducing the massive gauge
transformations in their 11-dimensional form \cite{kn:L,kn:O}.

It is reasonable to expect that the same procedure will give us the
action we are looking for.  Thus, we assume that the metric has an
isometry generated by a Killing vector $\hat{k}^{\hat{\mu}}$, so the
above action is invariant under the infinitesimal transformations

\begin{equation}
\label{eq:trans}
\left\{
\begin{array}{rcl}
\delta_{\eta}\hat{X}^{\hat{\mu}} & = & 
\eta\ \hat{k}^{\hat{\mu}}(\hat{X})\, ,\\
& & \\
\delta_{\eta}\hat{g}_{\hat{\mu}\hat{\nu}} & = & 
\eta\ \hat{k}^{\hat{\lambda}}\partial_{\hat{\lambda}}
\hat{g}_{\hat{\mu}\hat{\nu}}\, , \\
\end{array}
\right.
\end{equation}

\noindent with constant $\eta$. To make the action invariant under the
same transformations but now with $\eta(\xi)$ being an arbitrary
worldline function, we simply substitute the derivative with respect
to $\xi$ by the covariant derivative

\begin{equation}
D_{\xi}\hat{X}^{\hat{\mu}} =\partial_{\xi}\hat{X}^{\hat{\mu}} 
+A_{\xi}(\xi) \hat{k}^{\hat{\mu}}\, ,
\end{equation}

\noindent where $A_{\xi}$ is an auxiliary worldline variable
transforming as follows

\begin{equation}
\delta_{\eta}A_{\xi}=-\partial_{\xi}\eta\, 
\Rightarrow  \delta_{\eta} D_{\xi}\hat{X}^{\hat{\mu}}
=\eta\ D_{\xi}\hat{X}^{\hat{\nu}}
\partial_{\hat{\nu}}\hat{k}^{\hat{\mu}}\, .
\end{equation}

We obtain the following action:

\begin{equation}
\label{eq:gauged}
\hat{S}_{\rm gauged}\ [\hat{X}^{\hat{\mu}},A_{\xi},\gamma] =
-{\textstyle\frac{p}{2}} \int d\xi \sqrt{|\gamma|}\ \gamma^{-1}\
D_{\xi}\hat{X}^{\hat{\mu}}
D_{\xi}\hat{X}^{\hat{\nu}} 
\hat{g}_{\hat{\mu}\hat{\nu}}\, .
\end{equation}

\noindent On top of being invariant under local transformations 
of the form (\ref{eq:trans}), this action turns out to be invariant
under the massive gauge transformations of the metric (\ref{eq:massg})
and the auxiliary variable

\begin{equation}
\delta_{\hat{\chi}}A_{\xi} = -m\hat{\lambda}_{\xi}\, . 
\end{equation}

In order to make more uniform our notation and write all actions in
such a way that the massless $m\rightarrow 0$ limit coincides with the
ungauged limit, it is convenient to make the following redefinitions:

\begin{equation}
\begin{array}{rcl}
A_{\xi} & = & -{\textstyle\frac{m}{2}}(2\pi \alpha^{\prime})\ 
b_{\xi}\, ,\\
& & \\
\eta & = & {\textstyle\frac{m}{2}}\ (2\pi\alpha^{\prime})\ 
\rho^{(0)}\, .\\
\end{array}
\end{equation}

\noindent After these redefinitions the action is

\begin{equation}
\label{eq:gauged2}
\hat{S}_{\rm gauged}\ [\hat{X}^{\hat{\mu}},b_{\xi},\gamma] =
-{\textstyle\frac{p}{2}} \int d\xi \sqrt{|\gamma|}\ \gamma^{-1}\
D_{\xi}\hat{X}^{\hat{\mu}}
D_{\xi}\hat{X}^{\hat{\nu}} 
\hat{g}_{\hat{\mu}\hat{\nu}}\, ,
\end{equation}

\noindent where

\begin{equation}
D_{\xi}\hat{X}^{\hat{\mu}} = 
\partial_{\xi}\hat{X}^{\hat{\mu}} -{\textstyle\frac{m}{2}}\
(2\pi\alpha^{\prime})\ b_{\xi}\ \hat{k}^{\hat{\mu}}\, ,
\end{equation}

\noindent and the fields transform as follows:

\begin{equation}
\label{eq:trans2}
\left\{
\begin{array}{rcl}
\delta \hat{X}^{\hat{\mu}} & = &  {\textstyle\frac{m}{2}} \
(2\pi\alpha^{\prime})\ \rho^{(0)}\left(\xi \right)\  
\hat{k}^{\hat{\mu}}(\hat{X})\, ,\\
& & \\
\delta \hat{g}_{\hat{\mu}\hat{\nu}} & = & 
{\textstyle\frac{m}{2}}\ (2\pi\alpha^{\prime})\ \rho^{(0)}\
\hat{k}^{\hat{\lambda}}\partial_{\hat{\lambda}}
\hat{g}_{\hat{\mu}\hat{\nu}}\, , \\
& & \\
\delta b_{\xi} & = & \frac{1}{2\pi\alpha^{\prime}}2\hat{\lambda}_{\xi} 
+\partial_{\xi} \rho^{(0)}\, .\\
\end{array}
\right.
\end{equation}

Now we want to perform the dimensional reduction\footnote{Strictly
  speaking it is not ``direct'' dimensional reduction because we are
  going to eliminate one coordinate. Since this is not a field theory,
  but a quantum-mechanical theory, there is no counting of degrees of
  freedom and we cannot say how many of them we are eliminating.} of
the gauged action in the direction associated to the isometry. As a
first step, using a coordinate system adapted to the isometry
($\hat{k}^{\hat{\mu}}=\delta^{\hat{\mu} y}$) and using
Eqs.~(\ref{eq:11versus10}) we rewrite the background fields in
10-dimensional form, obtaining

\begin{equation}
\label{eq:usual0red}
\hat{S}\ [X^{\mu},c^{(0)},b_{\xi},\gamma] =
-{\textstyle\frac{p}{2}} \int d\xi\ \sqrt{|\gamma|}\ \gamma^{-1}
\left[ e^{-\frac{2}{3}\phi} g_{\xi\xi} -(2\pi\alpha^{\prime})^{2}
e^{\frac{4}{3}\phi} \left({\cal G}^{(1)}{}_{\xi}\right)^{2}
\right]\, ,
\end{equation}

\noindent with 

\begin{equation}
\left\{
\begin{array}{rcl}
g_{\xi\xi} & = & \partial_{\xi}X^{\mu}\partial_{\xi}X^{\nu}g_{\mu\nu}\, ,\\
& & \\
C^{(1)}{}_{\xi} & = & \partial_{\xi}X^{\mu}\ C^{(1)}{}_{\mu}\, ,\\
& & \\
{\cal G}^{(1)}{}_{\xi} & = & \partial_{\xi}c^{(0)} 
+\frac{1}{2\pi\alpha^{\prime}} C^{(1)}{}_{\xi} -\frac{m}{2}b_{\xi}\, .\\
\end{array}
\right.
\end{equation}

\noindent ${\cal G}^{(1)}_{\xi}$ is  the gauge-invariant ``field''
strength of $c^{(0)}$.  The worldvolume 0-form $c^{(0)}$ is related to
the original 11-dimensional coordinate $Y$ by

\begin{equation}
Y= (2\pi\alpha^{\prime}) c^{(0)}\, ,
\end{equation}

\noindent and transforms as given by (\ref{transc}):

\begin{equation}
\delta c^{(0)} = 
-{\textstyle\frac{1}{2\pi\alpha^{\prime}}} \Lambda^{(0)} 
+{\textstyle\frac{m}{2}}\rho^{(0)}\, .
\end{equation}

Now we want to eliminate $c^{(0)}$ (or, equivalently, $Y$) by using
its equation of motion, which essentially says that the momentum of
the particle in the direction $Y$, $P_{y}$, is constant.  The right
way of doing this \cite{kn:BJO3} is to first perform the Legendre
transformation of the action with respect to $Y$ (or $c^{(0)}$)

\begin{equation}
\tilde{S}^{\prime}\ [X^{\mu},P_{y},b_{\xi},\gamma]
=\int d\xi \left[ -P_{y}\partial_{\xi} Y
+{\cal L} \right]\, ,
\end{equation}

\noindent with 

\begin{equation}
P_{y}=\frac{\partial {\cal L} }{\partial (\partial_{\xi} Y)}\, .
\end{equation}

\noindent We obtain

\begin{equation}
\label{eq:usual0redprime}
\begin{array}{rcl}
\tilde{S}^{\prime}\ [X^{\mu},P_{y},b_{\xi},\gamma] & = &
-{\textstyle\frac{p}{2}} \int d\xi\ \sqrt{|\gamma|}\ \gamma^{-1}
\left[ e^{-\frac{2}{3}\phi} g_{\xi\xi} 
+\gamma e^{-\frac{4}{3}\phi} \left(\frac{P_{y}}{p}\right)^{2}
\right] \\
& & \\
& & 
+\int d\xi P_{y} \left[C^{(1)}{}_{\xi} 
-\frac{m}{2}\ (2\pi\alpha^{\prime})\ b_{\xi}\right]\, . \\
\end{array}
\end{equation}

\noindent We can now set $P_{y}$ to a constant value and eliminate the 
auxiliary metric. For $P_{y}\neq 0$ (that is, the M-0-brane has some
momentum in the compact dimension) we get\footnote{Note that the field
  equation for $b_\xi$ seems to lead to $m=0$.  However, one should
  keep in mind that in the supersymmetric case $b_\xi$ couples to
  further fermionic terms. We thank M.B.~Green for a discussion on
  this point.}:

\begin{equation}
\label{eq:usualD0}
\tilde{S}^{\prime}\ [X^{\mu},b_{\xi}] =
-|P_{y}| \int d\xi\ e^{-\phi} \sqrt{|g_{\xi\xi}|} 
+P_{y}\int d\xi \left(C^{(1)}{}_{\xi} 
-{\textstyle\frac{m}{2}}\ (2\pi\alpha^{\prime})\ b_{\xi}\right) \, ,
\end{equation}

\noindent which is the standard effective action of the massive D-0-brane 
\cite{kn:BR}, $b_{\xi}$ being the BI vector ``field''. The mass of the
D-0-brane is $|P_{y}|$ and its R-R charge is $-P_{y}$. Note that the
constant $p$ has disappeared from the action. As we said in
Section~\ref{sec-worldvolume} this action is gauge-invariant only up
to total derivatives. The gauge-invariance of the Lagrangian has been
lost in the Legendre transformation. It is convenient to introduce an
auxiliary scalar field $c^{(0)}$ to make the above action exactly
gauge invariant. Doing this, we have

\begin{equation}
\label{eq:usualD0withauxiliary}
\tilde{S}^{\prime}\ [X^{\mu},b_{\xi}] =
-|P_{y}| \int d\xi\ e^{-\phi} \sqrt{|g_{\xi\xi}|} 
+(2\pi\alpha^{\prime})P_{y}\int d\xi\ {\cal G}^{(1)}{}_{\xi} \, .
\end{equation}



When $P_{y}=0$ (the M-0-brane only moves in directions orthogonal to
$Y$) we obtain the action for a null D-0-brane \cite{kn:LvonU,kn:BeTo}:

\begin{equation}
\tilde{S}\ [X^{\mu},\gamma] = 
-{\textstyle\frac{p}{2}} \int d\xi\ \sqrt{|\gamma|}\ \gamma^{-1}
e^{-\frac{2}{3}\phi} g_{\xi\xi}\, .
\end{equation}

\noindent Observe that the factor  $e^{-\frac{2}{3}\phi}$
cannot be reabsorbed into $\gamma$ at least when the above effective
action is acting as a source for supergravity. The null D-0-brane does
not qualify as fundamental, solitonic or D-type.

%
%
%
%
%
%
%

\subsection{The Massive M-2-Brane}
\label{sec-massM2}

Starting from the massless M-2-brane action we perform a similar
gauging as in the particle case. At the same time however we want the
resulting gauged action to remain invariant under the gauge
transformations of the 3-form $\hat{C}$.  A straightforward Noether
procedure leads to an action \cite{kn:O} that we rewrite as
follows\footnote{In this Section we will consider a double dimensional
  reduction which involves not only a target space reduction (from
  d=11 to d=10) but also a worldvolume reduction (from 3 dimensions to
  2 dimensions). To distinguish fields before and after the
  worldvolume reduction we use a notation where worldvolume hatted
  indices (Latin) and fields are 3-dimensional, while unhatted
  worldvolume indices and fields are 2-dimensional. The split is
  $\hat{\imath}=(i,2)$, with $i=0,1$. In the next Section we will
  discuss the massive M-5-brane and use a similar notation.}
\footnote{The three-dimensional gauged sigma model given below also
  occurs in \cite{kn:HullSpence}. The authors of \cite{kn:HullSpence}
  also note that the Killing isometry condition can be slightly
  weakened so that only the curvatures are Lie-invariant.}

\begin{equation}
\label{eq:M2gauged}
\begin{array}{rcl}
\hat{S}_{\rm gauged}\ [\hat{X}^{\hat{\mu}},
\hat{b}_{\hat{\imath}}] & = & 
-T_{M2}\int d^{3}\hat{\xi}\ 
\sqrt{|D_{\hat{\imath}}\hat{X}^{\hat{\mu}}
D_{\hat{\jmath}}\hat{X}^{\hat{\nu}} \hat{g}_{\hat{\mu}\hat{\nu}}|} \\
& & \\
& &
-(2\pi\alpha^{\prime})\frac{T_{M2}}{3!}\int d^{3}\hat{\xi}\ 
\hat{\epsilon}^{\hat{\imath}\hat{\jmath}\hat{k}} 
\hat{\cal K}^{(3)}{}_{\hat{\imath}\hat{\jmath}\hat{k}}\, ,
\end{array}
\end{equation}

\noindent where

\begin{equation}
\label{eq:Ddef}
D_{\hat{\imath}}\hat{X}^{\hat{\mu}} 
=  \partial_{\hat{\imath}} \hat{X}^{\hat{\mu}} 
-{\textstyle\frac{m}{2}}\ (2\pi\alpha^{\prime})\
\hat{b}_{\hat{\imath}}\ \hat{k}^{\hat{\mu}}\, ,
\end{equation}

\noindent $\hat{k}^{\hat{\mu}}$ is a fixed (i.e.~non-dynamical) 
spacetime vector and

\begin{equation}
\hat{\cal K}^{(3)}  =  3\left[\partial\hat{\omega}^{(2)}
+{\textstyle\frac{1}{3(2\pi\alpha^{\prime})}}
D\hat{X}^{\hat{\mu}}
D\hat{X}^{\hat{\nu}}
D\hat{X}^{\hat{\rho}}
\hat{C}_{\hat{\mu}\hat{\nu}\hat{\rho}} 
-{\textstyle\frac{m}{2}} (2\pi\alpha^{\prime}) 
\hat{b}\partial\hat{b} \right]\, ,
\end{equation}

\noindent is the gauge-invariant field-strength of the auxiliary
worldvolume field $\hat{\omega}^{(2)}$ that transforms as follows:

\begin{equation}
\delta \hat{\omega}{}^{(2)}_{\hat{\imath}\hat{\jmath}} = 
-{\textstyle\frac{1}{2\pi\alpha^{\prime}}}
\hat{\chi}_{\hat{\imath}\hat{\jmath}}
-{\textstyle\frac{m}{2}} (2\pi\alpha^{\prime}) 
\left( {\textstyle\frac{1}{2\pi\alpha^{\prime}}}
2\hat{\lambda} +\partial\hat{\rho}^{(0)}\right)_{[\hat{\imath}} 
\hat{b}_{\hat{\jmath}]} 
+2\partial_{[\hat{\imath}}\hat{\rho}^{(1)}{}_{\hat{\jmath}]}\, .
\end{equation}

\noindent Finally, $|M_{\hat{\imath}\hat{\jmath}}|$ stands
for the absolute value of the determinant of the matrix
$M_{\hat{\imath}\hat{\jmath}}$.  Like in the case of the particle we
use conventions such that the massless and ungauged limits coincide.

Without the auxiliary worldvolume field $\hat{\omega}^{(2)}$, the
above massive M-2-brane action would be invariant up to total
derivatives under the infinitesimal massive gauge transformations with
parameter $\hat{\chi}$ of $\hat{g}_{\hat{\mu}\hat{\nu}}$ and
$\hat{C}_{\hat{\mu}\hat{\nu}\hat{\rho}}$ given in
Eqs.~(\ref{eq:massg},\ref{eq:massC}) together with the massive
transformation of the auxiliary field $\hat{b}_{\hat{\imath}}$
 
\begin{equation}
\label{eq:massA}
\delta_{\hat{\chi}}\hat{b}_{\hat{\imath}} =
{\textstyle\frac{1}{2\pi\alpha^{\prime}}}
2\hat{\lambda}_{\hat{\imath}}\, , 
\end{equation}

\noindent and the $\delta_{{\hat{\rho}}^{(0)}}$ transformations (for
which $\hat{b}_{\hat{\imath}}$ plays the role of gauge field)

\begin{equation}
\label{etatrans}
\left\{
\begin{array}{rcl}
\delta_{\hat{\rho}^{(0)}}\hat{X}^{\hat{\mu}} & = & 
{\textstyle\frac{m}{2}}\ (2\pi\alpha^{\prime})\ 
\hat{\rho}^{(0)}(\hat{\xi})\ \hat{k}^{\hat{\mu}}(\hat{X})\, ,\\
& & \\
\delta_{\hat{\rho}^{(0)}}\hat{b}_{\hat{\imath}} & = & 
\partial\hat{\rho}^{(0)}\, ,\\
& & \\
\delta_{\hat{\rho}^{(0)}}\hat{g}_{\hat{\mu}\hat{\nu}} & = & 
{\textstyle\frac{m}{2}}\ (2\pi\alpha^{\prime})\ 
\hat{\rho}^{(0)}
\hat{k}^{\hat{\lambda}}\partial_{\hat{\lambda}}
\hat{g}_{\hat{\mu}\hat{\nu}}\, , \\
& & \\
\delta_{\hat{\rho}^{(0)}} \hat{C}_{\hat{\mu}\hat{\nu}\hat{\rho}}
& = & 
{\textstyle\frac{m}{2}}\ (2\pi\alpha^{\prime})\ 
\hat{\rho}^{(0)} \hat{k}^{\hat{\lambda}}\partial_{\hat{\lambda}}
\hat{C}_{\hat{\mu}\hat{\nu}\hat{\rho}}\, , \\
\end{array}
\right.
\end{equation}

\noindent assuming the conditions

\begin{equation}
\pounds_{\hat{k}} \hat{g}_{\hat{\mu}\hat{\nu}} 
=\pounds_{\hat{k}} \hat{C}_{\hat{\mu}\hat{\nu}\hat{\rho}}
=\pounds_{\hat{k}} \hat{\chi}_{\hat{\mu}\hat{\nu}}=0
\end{equation}

\noindent hold. We see that, in particular, $\hat{k}^{\hat{\mu}}$ 
must be a Killing vector. These conditions coincide with those
necessary to build the massive M-0-brane and are exactly those
satisfied by the massive 11-dimensional supergravity theory described
in Section~\ref{sec-massiveM}. This was our main motivation for
constructing the massive 11-dimensional supergravity theory in
Section~\ref{sec-massiveM}.

Observe that the gauge transformation laws of the worldvolume field
$\hat{b}_{\hat{\imath}}$ are identical to those of the auxiliary field
$b_{\xi}$ that we introduced in the M-0-brane case (taking into
account the different dimensionality of the respective worldvolumes).

The auxiliary field $\hat{\omega}^{(2)}$ has been introduced to
compensate for the total derivatives so the action is exactly
gauge-invariant (the WZ term is precisely its field-strength). A field that
transforms precisely in the same way and with the same field strength
occurs in the M-5-brane effective action. There it is a dynamical
field whose equation of motion is the anti-self-duality condition.
Following our philosophy, we denote both worldvolume 2-forms by
$\hat{\omega}^{(2)}$.


\subsubsection{Direct Dimensional Reduction: the Massive D-2-Brane}

By assumption the eleven dimensional background has an isometry and we
can perform a direct dimensional reduction of the action
(\ref{eq:M2gauged}) in the direction associated to the isometry that
we have gauged. This amounts to a simple rewriting of the background
fields from 11-dimensional to 10-dimensional form. Using a coordinate
system adapted to the isometry ($\hat{k}^{\hat{\mu}}=\delta^{\hat{\mu}
  y}$) and the relations between 11- and 10-dimensional fields
Eqs.~(\ref{eq:11versus10}) we obtain the dual action for the D-2-brane
of massive type~IIA theory found in \cite{kn:L} (plus the auxiliary
field). Taking into account that $\hat{\omega}^{(2)}$ transforms as
the $\partial \hat{a}^{(2)}$ defined in Section~\ref{sec-worldvolume}
and substituting accordingly, we can write that action as
follows:

\begin{equation}
\label{eq:dual}
\begin{array}{rcl}
\tilde{S}\ [X^{\mu},\hat{c}^{(0)},\hat{b}_{\hat{\imath}},
{\hat a}^{(2)}_{\hat{\imath}\hat{\jmath}}]
& = & 
-T_{M2} \int d^{3}\hat{\xi}\ e^{-\phi}\ \sqrt{|g_{\hat{\imath}\hat{\jmath}} 
-(2\pi\alpha^{\prime})^{2}e^{2\phi} \hat{\cal G}^{(1)}{}_{\hat{\imath}}
{\cal G}^{(1)}{}_{\hat{\jmath}}|} \\
& & \\
& & 
-(2\pi\alpha^{\prime})\frac{T_{M2}}{3!}\int d^{3}\hat{\xi}\ 
\hat{\epsilon}^{\hat{\imath}\hat{\jmath}\hat{k}}
\hat{\cal H}^{(3)}{}_{\hat{\imath}\hat{\jmath}\hat{k}}\, , \\
\end{array}
\end{equation}

\noindent where

\begin{equation}
g_{\hat{\imath}\hat{\jmath}} =  \partial_{\hat{\imath}}  X^{\mu}
\partial_{\hat{\jmath}} X^{\nu} g_{\mu\nu}\, , 
\end{equation}

\noindent and $\hat{\cal G}^{(1)}{}_{\hat{\imath}}$ and 
$\hat{\cal H}^{(3)}{}_{\hat{\imath}\hat{\jmath}\hat{k}}$ are defined
in Section~\ref{sec-worldvolume} and given explicitly in
Appendix~\ref{sec-wvfields}.


As for the D-particle the worldvolume 0-form $\hat{c}^{(0)}$, whose
transformation properties are also defined in
Section~\ref{sec-worldvolume}, is related to the original 11-dimensional
coordinate $Y$ by

\begin{equation}
Y= (2\pi\alpha^{\prime}) \hat{c}^{(0)}\, .
\end{equation}



The equivalence between this action and the usual effective action for
the massive D-2-brane which contains the BI vector field can be proven
via the intermediate action of Ref.~\cite{kn:L} or by eliminating
$\hat{c}^{(0)}$ in the above action. This can be done in two
alternative but equivalent ways. First, one can try to use the
equation of motion of $\hat{b}_{\hat{\imath}}$. To do this
consistently one has to Legendre-transform the above action with
respect to $\hat{c}^{(0)}$\footnote{This is more easily done using an
  auxiliary worldvolume metric
  $\hat{\gamma}_{\hat{\imath}\hat{\jmath}}$. The corresponding
  worldvolume action can be found in Ref.~\cite{kn:O}.}:

\begin{equation}
\tilde{S}^{\prime}\ 
[X^{\mu},\hat{P}_{(0)}{}^{\hat{\imath}},\hat{b}_{\hat{\imath}},
\hat{\gamma}_{\hat{\imath}\hat{\jmath}}]  
=\int d^{3}\hat{\xi} \left[ -{\hat P}_{(0)}{}^{\hat{\imath}}
\partial_{\hat{\imath}}
\hat{c}^{(0)} +\hat{\cal L} \right]\, ,
\end{equation}

\noindent with 

\begin{equation}
\hat{P}_{(0)}{}^{\hat{\imath}} 
=\frac{\partial \hat{\cal L} }{\partial (\partial_{\hat{\imath}} 
\hat{c}^{(0)})}\, .
\end{equation}

\noindent The result is

\begin{equation}
\label{eq:dualprime}
\begin{array}{rcl}
\tilde{S}^{\prime}\ 
[X^{\mu},{\hat {\cal P}}_{(0)}{}^{\hat{\imath}},
\hat{b}_{\hat{\imath}},
\hat{\gamma}_{\hat{\imath}\hat{\jmath}}] = \hspace{-4cm} &  & \\
& & \\
& & 
-\frac{T_{M2}}{2} \int d^{3}\hat{\xi}\ 
\sqrt{|\hat{\gamma}|}\
\hat{\gamma}^{\hat{\imath}\hat{\jmath}}
\left\{ e^{-\frac{2}{3}\phi} g_{\hat{\imath}\hat{\jmath}} 
+e^{-\frac{4}{3}\phi} 
\left({\hat {\cal P}}_{(0)}+{}^{\star}B\right)_{\hat{\imath}}
\left({\hat {\cal P}}_{(0)}+{}^{\star}B\right)_{\hat{\jmath}} 
\right. \\
& & \\
& & 
\left.
\hspace{2cm}-2\left({\hat {\cal P}}_{(0)}+{}^{\star}
B\right)_{\hat{\imath}}
\left(C^{(1)} -\frac{m}{2}\ (2\pi\alpha^{\prime})\ 
\hat{b}\right)_{\hat{\jmath}}
-1 \right\} \\
& & \\
& & 
-\frac{T_{M2}}{3!}\int d^{3}\hat{\xi}\ 
\hat{\epsilon}^{\hat{\imath}\hat{\jmath}\hat{k}}
\left\{
C^{(3)}{}_{\hat{\imath}\hat{\jmath}\hat{k}}
-3\ \frac{m}{2}\ (2\pi\alpha^{\prime})^{2}\hat{b}_{\hat{\imath}}
{\hat{\cal F}}_{\hat{\jmath}\hat{k}}
+\frac{3}{2}\ \frac{m}{2}\ (2\pi\alpha^{\prime})\ 
{\hat b}_{\hat{\imath}}B_{\hat{\jmath}\hat{k}}
\right. \\
& & \\
& &
\left.
+3(2\pi\alpha^\prime)\partial_{\hat{\imath}}
\hat{c}^{(2)}{}_{\hat{\jmath}\hat{k}} \right\}\, ,\\
\end{array}
\end{equation}

\noindent where $\hat{\cal F}$ is defined in 
Section~\ref{sec-worldvolume}, and we are using for simplicity

\begin{equation}
{\hat {\cal P}}_{(0)}{}^{\hat{\imath}} = 
{\textstyle\frac{1}{2\pi\alpha^{\prime} T_{M2} \sqrt{|\hat{\gamma}|}}}
{\hat P}_{(0)}{}^{\hat{\imath}}\, .
\end{equation}

\noindent We have also substituted the auxiliary field $\hat{a}^{(2)}$ by
the auxiliary field $\hat{c}^{(2)}$ to compensate for the lack of
exact gauge-invariance introduced by the Legendre transformation.

\noindent Now, the equation of motion for $\hat{b}_{\hat{\imath}}$ 
is a purely algebraic constraint for $\hat{\cal
  P}_{(0)}{}^{\hat{\imath}}$:

\begin{equation}
\left({\hat {\cal P}}_{(0)} +{}^{\star}B\right)^{\hat{\imath}}
=(2\pi\alpha^{\prime})\ {}^{\star}\hat{\cal F}^{\hat{\imath}}\, .
\end{equation}

Substituting this constraint into the action Eq.~(\ref{eq:dualprime})
and eliminating again the auxiliary metric we readily obtain the
action of the massive D-2-brane \cite{kn:BR,kn:GHT}:

\begin{equation}
\label{eq:usual}
\begin{array}{rcl}
S\ [X^{\mu},\hat{b}_{\hat{\imath}}] & = & 
-T_{M2} \int d^{3}\hat{\xi}\ e^{-\phi} 
\sqrt{|g_{\hat{\imath}\hat{\jmath}} 
+(2\pi \alpha^{\prime}) \hat{\cal F}_{\hat{\imath}\hat{\jmath}}|} \\
& & \\
& & 
\hspace{-1cm}
-(2\pi\alpha^{\prime})\frac{T_{M2}}{3!}\int d^{3}\hat{\xi}\ 
\epsilon^{\hat{\imath}\hat{\jmath}\hat{k}}
\hat{\cal G}^{(3)}{}_{\hat{\imath}\hat{\jmath}\hat{k}}\, ,\\
\end{array}
\end{equation}

\noindent where $\hat{\cal G}^{(3)}{}_{\hat{\imath}\hat{\jmath}\hat{k}}$
is defined in Section~\ref{sec-worldvolume}.


In this action the worldvolume 1-form $\hat{b}_{\hat{\imath}}$ plays
the role of the Born-Infeld vector field and the auxiliary field
$\hat{c}^{(2)}$ makes the action exactly gauge-invariant.

A more elegant way to proceed, keeping track as well of the total
derivative induced in the dualization, is to perform the worldvolume
canonical transformation \cite{kn:L2}:

\begin{equation}
\left\{
\begin{array}{rcl}
{\hat P}_{(0)} & = & -(2\pi \alpha^{\prime})^2 T_{M2}
\hat{\epsilon}^{0\hat{\alpha}\hat{\beta}}
\partial_{\hat{\alpha}} \hat{V}_{\hat{\beta}}\, ,\\
& & \\
\hat{\Pi}^{\hat{\alpha}} & = & (2\pi \alpha^{\prime})^2 T_{M2}
\hat{\epsilon}^{0\hat{\alpha}\hat{\beta}}
\partial_{\hat{\beta}}{\hat c}^{(0)}\, ,
\end{array}
\right.
\end{equation}

\noindent from $\{{\hat c}^{(0)},{\hat P}_{(0)}\}$ to
$\{{\hat V}_{\hat \alpha},{\hat \Pi}^{\hat \alpha}\}$, where we have
split the three dimensional index $\hat{\imath}=(0,\hat{\alpha})$
and ${\hat P}_{(0)}$ ($\equiv {\hat P}_{(0)}^{0}$),
$\hat{\Pi}^{\hat{\alpha}}$ are the canonically conjugate momenta of
${\hat c}_{(0)}$ and $\hat{V}_{{\hat \alpha}}$ respectively.  This
transformation is generated by the functional

\begin{equation}
\Psi=(2\pi \alpha^{\prime})^2 T_{M2}\int_{{\rm t\, fixed}}
\hat{\epsilon}^{0\hat{\alpha}\hat{\beta}}
\hat{V}_{\hat{\alpha}}\partial_{\hat{\beta}}{\hat c}^{(0)}\, .
\end{equation}

\noindent The canonically transformed action is given by:

\begin{equation}
\label{eq:unusual}
\begin{array}{rcl}
S[X^\mu,{\hat b}_{\hat \imath},{\hat V}_{\hat \imath}] & = &
-T_{M_2} \int d^3{\hat \xi}e^{-\phi}
\sqrt{|g_{\hat{\imath}\hat{\jmath}}
+(2\pi\alpha^\prime) {\hat{\cal F}}_{\hat{\imath}
\hat{\jmath}}|} \\
& & \\
& &
-\frac{T_{M_2}}{3!}\int d^3\hat{\xi}\epsilon^{\hat{\imath}
\hat{\jmath}\hat{k}}
\left[
C^{(3)}_{\hat{\imath}\hat{\jmath}\hat{k}}
-3(2\pi\alpha^\prime)C^{(1)}_{\hat{\imath}}
\hat{\cal F}_{\hat{\jmath}\hat{k}}
\right.\\
& & \\
& &
\left.
+6m\pi\alpha^\prime(2\pi\alpha^\prime){\hat b}_{\hat{\imath}}
\partial_{\hat{\jmath}}{\hat V}_{\hat{k}}-3\frac{m}{2}
(2\pi\alpha^\prime)^2{\hat b}_{\hat{\imath}}
\partial_{\hat{\jmath}} {\hat b}_{\hat{k}}
\right.\\
& & \\
& &
\left.
+3(2\pi\alpha^\prime)\partial_{\hat{\imath}}
\hat{a}^{(2)}{}_{\hat{\jmath}\hat{k}}-6(2\pi\alpha^\prime)^2
\partial_{\hat{\imath}}{\hat c}^{(0)}\partial_{\hat{\jmath}}
V_{\hat{k}}
\right]\, ,
\end{array}
\end{equation}

\noindent where $\hat{\cal F}_{\hat{\imath}\hat{\jmath}}=
2\partial_{[\hat{\imath}}{\hat V}_{\hat{\jmath}]}+
\frac{1}{2\pi\alpha^\prime} B_{\hat{\imath}\hat{\jmath}}$ and
we have included the total derivative generated in the
dualization procedure, namely

\begin{equation}
-(2\pi\alpha^\prime)^2T_{M_2}\int d^3\hat{\xi}
\epsilon \partial {\hat c}^{(0)}\partial {\hat V}\, .
\end{equation}

\noindent Elimination of $\hat{b}_{\hat{\imath}}$ through its
equation of motion gives ${\hat b}_{\hat{\imath}}={\hat
  V}_{\hat{\imath}}$ and the usual action for the D-2-brane is
recovered where now ${\hat V}_{\hat{\imath}}$ is the BI field and
there are additional world-volume fields compensating for all the
total derivatives coming from the gauge transformations.  
Note that ${\hat c}^{(0)}$ is now playing the role of an auxiliary
field needed to compensate for the total derivatives, and not of the
eleventh coordinate as in (\ref{eq:dual}).
In
(\ref{eq:unusual}) ${\hat b}$ is an auxiliary field whose role is to
eliminate the explicit dependence on the BI field (other than through
its derivatives), such that the standard duality transformation can be
applied even in the massive case. Once it is integrated out the usual
action is recovered.


\subsubsection{Double Dimensional Reduction: the Type~IIA String}

In this Section we consider the double dimensional reduction of the
massive M-2-brane worldvolume effective action. We will see that this
reduction leads to the usual (``massless'') type~IIA string action.

It is convenient to perform the double dimensional reduction of
Eq.~(\ref{eq:M2gauged}) in three steps. First, we perform a direct
dimensional reduction, obtaining Eq.~(\ref{eq:dual}). The second step
consists in a partial gauge-fixing of the action Eq.~(\ref{eq:dual}).
We set

\begin{equation}
\label{eq:gaugefix}
\hat{c}^{(0)}={\textstyle\frac{1}{2\pi\alpha^{\prime}}}\hat{\xi}^{2}\, ,
\end{equation}

\noindent while all other worldvolume fields (and gauge parameters)
 are independent of $\hat{\xi}^{2}$:

\begin{equation}
\partial_{2} X^{\mu}=\partial_{2} \hat{b}_{\hat{\imath}}
=\partial_{2} {\hat a}^{(2)}{}_{\hat{\imath}\hat{\jmath}}=0\, ,
\end{equation}

\noindent etc. We want to keep track of all gauge transformations.
Thus, we introduce a compensating gauge transformation to keep our
gauge choice Eq.~(\ref{eq:gaugefix}). This transformation is the
g.c.t.

\begin{equation}
\delta\hat{\xi}^{\hat{\imath}}=\delta^{\hat{\imath}\ 2}
\left[-\Lambda^{(0)} +{\textstyle\frac{m}{2}}(2\pi\alpha^{\prime})
\hat{\rho}^{(0)} \right]\, .
\end{equation}

\noindent Including this compensating gauge transformation, we find
the modified gauge transformations

\begin{equation}
\left\{
\begin{array}{rcl}
\delta_{M} \hat{b}_{2} & = & 0\, ,\\
& & \\
\delta_{M} \hat{b}_{i} & = & 
\frac{1}{2\pi\alpha^{\prime}} 2\lambda_{i}
+\partial_{i}\hat{\rho}^{(0)} 
+\left[\partial_{i}\Lambda^{(0)} 
-\frac{m}{2}(2\pi\alpha^{\prime})\partial_{i}\hat{\rho}^{(0)} 
\right] \hat{b}_{2}\, , \\
& & \\
\delta_{M} \hat{a}^{(2)}{}_{i2} & = & 
\partial_{i}\hat{\mu}^{(1)}{}_{2}
+\frac{1}{2\pi\alpha^{\prime}} 2\lambda_{i}
+\frac{m}{4}(2\pi\alpha^{\prime})\hat{\rho}^{(0)}\partial_{i} \hat{b}_{2}
-\frac{m}{2}\lambda_{i}\hat{b}_{2}\, ,\\
& & \\
\delta_{M} \hat{a}^{(2)}{}_{ij} & = & 
2\partial_{[i}\hat{\mu}^{(1)}{}_{j]} 
-\frac{1}{2\pi\alpha^{\prime}}\Lambda^{(2)}{}_{ij}
+\frac{m}{2}(2\pi\alpha^{\prime})\hat{\rho}^{(0)}
\partial_{[i} \hat{b}_{j]} -m\lambda_{[i}\hat{b}_{j]} \\
& & \\
& & 
-2 \left[\partial\Lambda^{(0)} 
-\frac{m}{2}(2\pi\alpha^{\prime})\partial\hat{\rho}^{(0)} 
\right]_{[i} \hat{a}^{(2)}{}_{j]2}\, .\\
\end{array}
\right.
\end{equation}

Now, we perform a field redefinition consisting of a $\Lambda^{(0)}$
gauge transformation  where the gauge parameter $\Lambda^{(0)}$ is a new
scalar field $c^{(0)}$. Since we are dealing with a gauge-invariant
action, the final result is guaranteed not to depend on $c^{(0)}$, no
matter how it transforms. It is convenient to asign to $c^{(0)}$ the
gauge transformation rules introduced in Section~\ref{sec-worldvolume} since then
the redefined fields also happen to transform as combinations of the standard
worldvolume fields introduced in Section~\ref{sec-worldvolume}. More
explicitly, we find the following decompositions:

\begin{equation}
\label{nuevasredi}
\left\{
\begin{array}{rcl}
\hat{b}_{2} & = & v^{(0)}\, ,\\
& & \\
\hat{b}_{i} & = & 
b_{i} -(2\pi\alpha^{\prime})\partial_{i}c^{(0)}v^{(0)}\, ,\\
& & \\
\hat{a}^{(2)}{}_{i2} & = & 
\left[1-\frac{m}{4}(2\pi\alpha^{\prime})v^{(0)}\right]b_{i}
+v^{(1)}{}_{i}\, ,\\
& & \\
\hat{a}^{(2)}{}_{ij} & = & 
a^{(2)}{}_{ij} 
+2(2\pi\alpha^{\prime}) \partial_{[i} c^{(0)}
\left\{\left[1-\frac{m}{4}(2\pi\alpha^{\prime})v^{(0)}\right]b
+v^{(1)}\right\}_{j]}\, ,\\
\end{array}
\right.
\end{equation}

\noindent where $v^{(1)}{}_{i}$ is a vector field transforming according
to

\begin{equation}
\left\{
\begin{array}{rcl}
\delta v^{(1)} & = & \partial \Delta^{(0)}\, ,\\
& & \\
\Delta^{(0)} & = & \hat{\mu}^{(1)}{}_{2} 
-\left[1-\frac{m}{4}(2\pi\alpha^{\prime})v^{(0)}\right]
\hat{\rho}^{(0)}\, ,\\
\end{array}
\right.
\end{equation}

\noindent and we have identified $\hat{\rho}^{(0)}=\rho^{(0)}$ and 
$\hat{\mu}^{(1)}{}_{i}=\mu^{(1)}{}_{i}$.

Given the above identifications of the potentials, we find, for the
field strength

\begin{equation}
\hat{\cal H}^{(3)}{}_{ij2}
=\left[1 -{\textstyle\frac{m}{2}} (2\pi \alpha^{\prime}) v^{(0)} \right]
{\cal F}_{ij} +{\cal L}^{(2)}{}_{ij}\, ,
\end{equation}

\noindent where ${\cal F}$ is the BI field strength defined in
 Section~\ref{sec-worldvolume} and 

\begin{equation}
{\cal L}^{(2)}=2\partial v^{(1)}\, .   
\end{equation}

Substituting these results and integrating over $\hat{\xi}^{2}$ the
massive M-2-brane action one arrives at the action

\begin{equation}
\label{eq:dualred1}
\begin{array}{rcl}
\tilde{S}\ [X^{\mu},b,v^{(1)}]
& = & 
-T_{M2}l \int d^{2}\xi\ 
\left[1- \frac{m}{2} (2\pi \alpha^{\prime}) v^{(0)} \right] 
\sqrt{|g_{ij}|} \\
& & \\
& & 
\hspace{-1.5cm}
-\frac{T_{M2}l}{2}\int d^{2}\xi\ \epsilon
\left\{
\left[1 -{\textstyle\frac{m}{2}} (2\pi \alpha^{\prime}) v^{(0)} \right]
{\cal F}_{ij} +{\cal L}^{(2)}{}_{ij} 
\right\}\, .\\
\end{array}
\end{equation}

Now, the equation of motion for $b$ implies that the scalar $v^{(0)}$
is constant. Substituting this result into the action and redefining

\begin{equation}
b+\left[1-{\textstyle\frac{m}{2}}
(2\pi\alpha^{\prime})v^{(0)}\right]^{-1} v^{(1)}
\rightarrow b\, ,
\end{equation}

\noindent (plus an analogous redefinition of the gauge parameters) we
get the action of the type~IIA string in the Nambu-Goto form with an
extra term which does not change the equations of motion but makes the
WZ term exactly gauge-invariant\footnote{Substituting the equation of
  motion of $v^{(0)}$ into the action one gets an action from which
  one cannot derive any equation of motion. As we have stressed
  before, the elimination of a field through its equation of motion
  can not always be done directly in the action and the equations of
  motion always have to be checked.}:

\begin{equation}
\label{eq:IIAstring}
S\ [X^{\mu}, b_{i}]  =  -T \int d^{2}\xi\ \sqrt{|g_{ij}|}
-(2\pi\alpha^{\prime}){\textstyle\frac{T}{2}}
\int d^{2}\xi\ \epsilon {\cal F}\, ,
\end{equation}

\noindent where the tension $T$ is now given by:

\begin{equation}
T = \left[ 1-{\textstyle{\frac{m}{2}}}
(2\pi\alpha^{\prime})v^{(0)}\right] T_{M_2}l\, .
\end{equation}

Observe that the new tension depends on the constant value of the gauge 
vector field $\hat{b}$ in the compact direction $v^{(0)}=\hat{b}_{2}$ (a
Wilson line). The Wilson line has a critical value for which the tension
vanishes. It would be interesting to investigate  the physical mechanism
behind this phenomenon.


\subsubsection{Reduction in a Non-Isotropic Direction}

Let us now assume that there exists an additional isometry realized by
translations of a coordinate other than the isotropic coordinate $Y$,
and perform the double dimensional reduction along this direction.  
In this subsection we will ignore total derivatives and only analyze
in detail the $\lambda$, $\rho^{(0)}$ gauge transformations for 
the sake of simplicity.  
We split the eleven coordinates into $(Z,X^{\mu})$ and use the ansatz:

\begin{equation}
Z={\hat \xi}^{2}\, ,
\hspace{1cm}  \partial_2 X^{\mu}=
\partial_2 \hat{b}_{\hat \imath}=0\, . 
\end{equation}

\noindent We also split the vector field $\hat{b}$ 

\begin{equation}
\hat{b}_{i} = b_{i}\, ,
\hspace{1cm} 
\hat{b}_{2} = v^{(0)}\, .  
\end{equation}

Notice that now the isotropic coordinate $Y$ is among the $X^{\mu}$
coordinates. As before, the ansatz above is compatible with the
equations of motion of the action (\ref{eq:dual}) and the resulting
action.  The latter is given by:

\begin{equation}
\label{nonisoac}
\begin{array}{rcl}
{\tilde S}[X^\mu,b_i,v^{(0)}] & = & 
-T_{M2}l\int d^2\xi \sqrt{|D_{i} X^{\mu}
D_{j} X^{\nu} g_{\mu\nu}|}\\
& & \\
& &
\hspace{-2cm}
-\frac{T_{M2}l}{2}\int d^2 \xi \epsilon^{ij}
\left[
D_{i} X^{\mu} D_{j}X^{\nu} B_{\mu\nu}
-m\pi\alpha^{\prime} v^{(0)} \left(i_k C^{(3)}\right)_{\mu\nu}
\partial_{i} X^{\mu} \partial_{j} X^{\nu}
\right.
\\
& & \\
& &
\hspace{-2cm}
\left.
-m (2\pi\alpha^{\prime})^{2}v^{(0)}
\partial_{i} {b}_{j}\right]\, . \\
\end{array}
\end{equation} 

\noindent Integrating out $v^{(0)}$ we obtain the constraint:

\begin{equation}
\epsilon^{ij}\partial_i b_j=-
{\textstyle\frac{1}{2(2\pi\alpha^\prime)}}
\epsilon^{ij}(i_k C^{(3)})_{ij}\, .
\end{equation}

\noindent Substituting this constraint back into (\ref{nonisoac})
the action of a gauged superstring in ten dimensions is obtained:

\begin{equation}
\begin{array}{rcl}
{\tilde S}[X^{\mu},b_i] & =  &
-T_{M2}l\int d^2\xi \sqrt{|D_{i} X^{\mu} D_{j} X^{\nu} g_{\mu\nu}|}\\
& & \\
& & 
-{\textstyle\frac{T_{M2}l}{2}}\int d^2\xi 
\epsilon^{ij}D_{i} X^{\mu} D_{j}X^{\nu} B_{\mu\nu}\, . \\
\end{array}
\end{equation}

\noindent This action is invariant under the transformations:

\begin{equation}
\left\{
\begin{array}{rcl}
\delta g_{\mu\nu} & = & 2m\lambda_{(\mu}
\left(i_k g\right)_{\nu)}+\frac{m}{2}(2\pi\alpha^\prime)\rho^{(0)}
k^\gamma \partial_\gamma g_{\mu\nu}\, , \\
& & \\
\delta B_{\mu\nu} & =& -2m\lambda_{[\mu}
\left(i_k B\right)_{\nu]}+2\partial_{[\mu}\lambda_{\nu]}
+\frac{m}{2}(2\pi\alpha^\prime)\rho^{(0)}
k^\gamma \partial_\gamma B_{\mu\nu}\, ,\\
& & \\
\delta b & = & \frac{1}{2\pi\alpha^\prime}2\lambda+\partial
\rho^{(0)}\, ,\\
\end{array}
\right.
\end{equation}

\noindent assuming the conditions:

\begin{equation}
\pounds_{k}g_{\mu\nu}=\pounds_{k} B_{\mu\nu}
=\pounds_{k}\lambda_\mu=0
\end{equation}

\noindent hold.

The above transformations are to be interpreted within the nine
dimensional theory obtained by dimensional reduction along the
isotropic coordinate. The ansatz for the decomposition of the ten
dimensional NS-NS fields is\footnote{We split the ten dimensional
  coordinates as $X^{\mu}=({\tilde X}^{{\tilde \mu}},Y)$. In this
  section tilded fields are nine dimensional.} \cite{kn:BHO}:

\begin{equation}
\label{three}
\left\{
\begin{array}{rcl}
g_{{\tilde \mu}{\tilde \nu}} & = & {\tilde g}_{{\tilde \mu}{\tilde \nu}}
-{\tilde k}^2 {\tilde A}^{(1)}{}_{\tilde \mu}
{\tilde A}^{(1)}{}_{\tilde \nu}\, , \\
& & \\
g_{y{\tilde \mu}} & = & -{\tilde k}^2 
{\tilde A}^{(1)}{}_{{\tilde \mu}}\, ,\\
& & \\
g_{yy} & = & -{\tilde k}^2\, , \\
\end{array}
\right.
\hspace{1cm}
\left\{
\begin{array}{rcl}
B_{{\tilde \mu}{\tilde \nu}} & = &
{\tilde B}^{(2)}{}_{{\tilde \mu}{\tilde \nu}}
+2{\tilde A}^{(1)}{}_{[{\tilde \mu}}
{\tilde B}^{(1)}{}_{{\tilde \nu}]}\, ,\\
& & \\
B_{y{\tilde \mu}} & = & {\tilde B}^{(1)}{}_{{\tilde \mu}}\, ,\\
& & \\
\phi & = & {\tilde \phi}+\frac{1}{2} \log{{\tilde k}}\, ,\\
\end{array}
\right.
\end{equation}

\noindent and the dimensionally reduced action reads:

\begin{equation}
\label{four}
\begin{array}{rcl}
S & = & -T_{M_2}l\int d^2\xi \sqrt{|{\tilde g}_{ij}
-\tilde{k}^2 (2\pi\alpha^\prime)^2 \tilde{\cal G}^{(1)}{}_{i}
\tilde{\cal G}^{(1)}{}_{j}|}\\
& & \\
& & 
-\frac{T_{M_2}l}{2}\int d^2\xi
\epsilon^{ij} \left[\tilde{B}^{(2)}{}_{ij}+
2(2\pi\alpha^\prime)\tilde{\cal G}^{(1)}{}_{i}
\tilde{B}^{(1)}{}_{j}\right]\, ,\\
\end{array}
\end{equation}

\noindent where now the field strength $\tilde{\cal G}^{(1)}$
is defined by (recall $Y=(2\pi\alpha^\prime) c^{(0)}$):

\begin{equation}
\tilde{\cal G}^{(1)}=\partial c^{(0)}+
{\textstyle\frac{1}{2\pi\alpha^\prime}}\tilde{A}^{(1)}
-{\textstyle\frac{m}{2}} b\, .
\end{equation}

\noindent It is easy to see that this action is invariant under
the nine dimensional massive transformations:

\begin{equation}
\left\{
\begin{array}{rcl}
\delta_\lambda {\tilde A}^{(1)} & = &m\lambda\, ,\\
& & \\
\delta_\lambda {\tilde B}^{(1)} & =& 0\, ,\\
\end{array}
\right.
\hspace{1.5cm}
\left\{
\begin{array}{rcl}
\delta_\lambda {\tilde B}^{(2)} & = & 2\partial\lambda\, ,\\
& & \\
\delta_\lambda b & = & \frac{1}{2\pi\alpha^\prime}2\lambda\, .\\
\end{array}
\right.
\end{equation}

The existence of a string solution invariant under nine dimensional
massive transformations seems to imply that a similar type of solution
will exist in the type~IIB theory after performing a T-duality
transformation.  However we will now show that after the T-duality
transformation the auxiliary vector field $b$ decouples from the other
fields and its integration fixes the mass parameter to zero, in such a
way that the resulting action corresponds to the dimensional reduction
of the massless type~IIB superstring.

In order to construct the T-dual of the action (\ref{four}), which is
nothing but a Poincar\'e duality transformation of the redundant
coordinate, we add to it the following Lagrange multiplier term
enforcing the Bianchi identity for $c^{(0)}$ (after an integration by
parts):

\begin{equation}
T_{M_2}l\int d^2\xi (2\pi\alpha^\prime)\epsilon^{ij}
\partial_i \varrho\ \left({\tilde {\cal G}}^{(1)}_j-
{\textstyle\frac{1}{2\pi\alpha^\prime}}{\tilde A}^{(1)}{}_{j}
+{\textstyle\frac{m}{2}} b_{j}\right)\, .
\end{equation} 

\noindent Integration over $\tilde{\cal G}^{(1)}$
gives the following dual action:

\begin{equation}
\begin{array}{rcl}
S_{{\rm IIB}} & = & -T_{M_2}l\int d^2\xi\sqrt{|{\tilde g}_{ij}
-\frac{1}{{\tilde k}^2}
(\partial_i \varrho+{\tilde B}^{(1)}_i)
(\partial_j \varrho+{\tilde B}^{(1)}_j)|}\\
& & \\
& &
-\frac{T_{M_2}l}{2}\int d^2\xi
\epsilon^{ij}\left[\tilde{B}^{(2)}{}_{ij}+
2\partial_{i}\varrho\ \left({\tilde A}^{(1)}{}_{i}+
m\pi\alpha^{\prime} b_{j}\right)\right]\, .\\
\end{array}
\end{equation}

\noindent Here the auxiliary field $b$ can be integrated out, 
giving the constraint that $m$ must be equal to zero\footnote{Or
  $\varrho={\rm constant}$, which is a particular case in the final
  action.}.  The resulting action is the dimensional reduction of the
ten dimensional type~IIB superstring along the $\varrho$ coordinate
(see \cite{kn:BHO} for more details).


\subsection{The Massive M-5-Brane}
\label{sec-M5brane}

We start by reviewing the massless case.  The worldvolume fields of the
massless M-5-brane are the eleven embedding scalars
$\hat{X}^{\hat{\mu}}\ (\hat{\mu} = 0,1,\ldots, 10)$ and an
(anti-self-dual\footnote{With our conventions the 3-form field
  strength has to be anti-self-dual, otherwise the kinetic term would
  have the wrong sign. In the absence of the WZ term, imposing self-
  or anti-self-duality is a matter of convention and either choice is
  possible. In the presence of the WZ term, with our conventions, only
  anti-self-duality can be consistently imposed.})  worldvolume
two-form $\hat{\omega}^{(2)}{}_{\hat{\imath}\hat{\jmath}}\ (i=0,
\ldots, 5)$.  The curvature $\hat{\cal K}^{(3)}$ of
$\hat{\omega}^{(2)}$ is given by\footnote{In this Section, worldvolume
  hatted indices (Latin) and fields are 6-dimensional, while unhatted
  worldvolume indices and fields are 5-dimensional. The split is
  $\hat{\imath}=(i,5)$, with $i=0,1,\ldots,4$. Note also that we have
  adapted our conventions from those in \cite{kn:BRO}.}
 
\begin{equation}
\hat{\cal K}^{(3)} = 3\left[\partial \hat{\omega}^{(2)} 
+{\textstyle\frac{1}{3(2\pi\alpha^{\prime})}}\hat{C}\right]\, ,
\end{equation}

\noindent where $\hat{C}$ is the 3-form potential of 11-dimensional
supergravity, with gauge transformation

\begin{equation}
\label{mgtrC}
\delta \hat{C} = 3\partial \hat{\chi}\, 
\end{equation}

\noindent under a 2-form gauge parameter $\hat{\chi}$.
The gauge invariance of $\hat{\cal K}^{(3)}$ implies the following gauge
transformation law of $\hat{\omega}^{(2)}$:

\begin{equation}
\label{mgtrW}
\delta \hat{\omega}^{(2)} = 
-{\textstyle\frac{1}{2\pi\alpha^{\prime}}}\hat{\chi} 
+2\partial\hat{\rho}^{(1)}\, .
\end{equation}

The anti-self-duality condition of the field strength can be written
in the linear approximation (which we will use throughout this
section) as follows:

\begin{equation}
\hat{\cal K}^{(3)} = -{}^{*}\hat{\cal K}^{(3)}\, ,
\end{equation}

\noindent and therefore it is not possible to write a covariant 
action for $\hat{\omega}^{(2)}$\footnote{A covariant action can be
  constructed by introducing an auxiliary scalar field \cite{Kh}.  The
  nonlinear form of the equations of motion of the massless M-5-brane
  has been given in \cite{Kh,Ca,Se}.  We expect that the results given
  in this Section on the massive M-5-brane automatically carry over to
  the nonlinear case simply by replacing the massless curvatures in
  \cite{Kh,Ca,Se} by their massive extensions.}.  We can write however
an action in which this constraint is only imposed at the level of the
equations of motion derived from it.  This can be done consistently if
the action is Poincar\'e anti-self-dual.  To second order in
$\hat{\cal K}^{(3)}$ the unique action with the required properties is
\cite{kn:BRO}

\begin{equation}
\label{M5-action}
\begin{array}{rcl}
\hat{S}[\hat{X}^{\hat{\mu}},\hat{\omega}^{(2)}{}_{\hat{\imath}\hat{\jmath}}] 
& = & 
-T_{M5}\int d^{6}\hat{\xi}\ \sqrt {|\hat{g}_{\hat{\imath}\hat{\jmath}}|}\
\left\{1-{\textstyle\frac{1}{4\cdot 3!}} (2\pi\alpha^{\prime})^{2}\
\left(\hat{\cal K}^{(3)}\right)^{2}+\cdots \right\}\\
& & \\
& & 
+(2\pi\alpha^{\prime})\frac{T_{M5}}{6!}\int d^{6}\hat{\xi}\
\hat{\epsilon}^{\hat{\imath}_{1}\cdots 
\hat{\imath}_{6}}\hat{\cal K}^{(6)}{}_{\hat{\imath}_{1}\cdots 
\hat{\imath}_{6}}\, ,\\
\end{array}
\end{equation}

\noindent where $\hat{\cal K}^{(6)}$ is the gauge-invariant 
6-form field strength of the worldvolume 5-form field
$\hat{\omega}^{(5)}$:

\begin{equation}
\hat{\cal K}^{(6)} = 6\left[\partial\hat{\omega}^{(5)} 
+{\textstyle\frac{1}{6(2\pi\alpha^{\prime})}}\hat{\tilde{C}} 
+{\textstyle\frac{5}{3}} \hat{\cal K}^{(3)}\hat{C}\right]\, ,
\end{equation}

\noindent whose mission is to make the above action exactly 
gauge invariant. Thus

\begin{equation}
\delta \hat{\omega}^{(5)} =
-{\textstyle\frac{1}{2\pi\alpha^{\prime}}}\hat{\tilde{\chi}}
+15 \partial\hat{\omega}^{(2)} \hat{\chi}
+5\partial\hat{\rho}^{(4)}\, .
\end{equation}

\noindent This worldvolume 5-form has also been recently considered 
in Ref.~\cite{kn:CeNi}. Observe that the M-5-brane couples naturally
to the dual 6-form potential of d=11 supergravity \cite{kn:Ah}.  The
action (\ref{M5-action}) is invariant (including total derivatives)
under the massless gauge transformations (\ref{mgtrC}), (\ref{mgtrW})
together with the gauge transformations of the 6-form
(\ref{eq:gauge6}), and it gives the massless D-4-brane action upon
double dimensional reduction and the massless p-5A-brane action upon
direct dimensional reduction \cite{kn:BRO}.

Our goal is to construct a massive M-5-brane action which is invariant
under the 11-dimensional massive gauge transformations. The double
dimensional reduction of this action should give the massive D-4-brane
action and the direct dimensional reduction the massive p-5A-brane
action. These will be derived as by-products of our construction in a
second step.

To construct the massive M-5-brane action we proceed as we did with
the M-0- and M-2-brane actions: we assume that the background admits
an isometry and fulfills all the requirements to be considered a
``massive 11-dimensional background''. Then we gauge the isometry in
the usual M-5-brane action, substituting all partial derivatives of
the embedding scalars $\partial_{\hat{\imath}} \hat{X}^{\hat{\mu}}$ by
covariant derivatives $D_{\hat{\imath}} \hat{X}^{\hat{\mu}}$ with
gauge field $\hat{b}_{\hat{\imath}}$, trying to keep gauge invariance
as well as Poincar\'e anti-self-duality (which are both broken by
naive gauging). As we saw in the M-2-brane case this implies the
introduction of new terms (in the present case not only the auxiliary
gauge fields $\hat{b}_{\hat{\imath}},\hat{\omega}^{(5)}$ but also an
additional worldvolume 6-form $\hat{\omega}^{(6)}$), as well as a
modification of the gauge transformations of both the background
fields and the worldvolume fields.  The new transformations of the
background fields turn out to be precisely those of massive (dual)
11-dimensional supergravity.

The action that we get to second order is:

\begin{equation}
\label{massiveM5-action}
\begin{array}{rcl}
\hat{S}[\hat{X}^{\hat{\mu}},\hat{\omega}^{(2)}{}_{\hat{\imath}\hat{\jmath}},
\hat{b}_{\hat{\imath}}] 
& = & \\
& & \\
& & 
\hspace{-3cm}
-T_{M5}\int d^{6}\hat{\xi}\ \sqrt {|D_{\hat{\imath}}\hat{X}^{\hat{\mu}}
D_{\hat{\jmath}}\hat{X}^{\hat{\nu}}
\hat{g}_{\hat{\mu}\hat{\nu}}|}\
\left\{1-{\textstyle\frac{1}{4\cdot 3!}} (2\pi\alpha^{\prime})^{2}\
\left(\hat{\cal K}^{(3)}\right)^{2}+\cdots \right\}\\
& & \\
& & 
\hspace{-3cm}
+(2\pi\alpha^{\prime})\frac{T_{M5}}{6!}\int d^{6}\hat{\xi}\
\hat{\epsilon}^{\hat{\imath}_{1}\ldots\hat{\imath}_{6}}
\hat{\cal K}^{(6)}{}_{\hat{\imath}_{1}\ldots\hat{\imath}_{6}}\, ,
\end{array}
\end{equation}

\noindent where $\hat{\cal K}^{(3)}$ and $\hat{\cal K}^{(6)}$ 
are now the massive field strengths of $\hat{\omega}^{(2)}$
and $\hat{\omega}^{(5)}$:

\begin{equation}
\left\{
\begin{array}{rcl}
\hat{\cal K}^{(3)} & = & 3\left[\partial\hat{\omega}^{(2)} 
+{\textstyle\frac{1}{3(2\pi\alpha^{\prime})}}
D\hat{X}^{\hat{\mu}}
D\hat{X}^{\hat{\nu}}
D\hat{X}^{\hat{\rho}}
\hat{C}_{\hat{\mu}\hat{\nu}\hat{\rho}} 
-{\textstyle\frac{m}{2}} (2\pi\alpha^{\prime}) 
\hat{b}\partial\hat{b} \right]\, ,\\
& & \\
\hat{\cal K}^{(6)}
& = & 
6\partial\hat{\omega}^{(5)} +\frac{m}{2}\hat{\omega}^{(6)}
+\frac{1}{2\pi\alpha^{\prime}}
D\hat{X}^{\hat{\mu}_{1}}\cdots D\hat{X}^{\hat{\mu}_{6}} 
\hat{\tilde{C}}_{\hat{\mu}_{1}\cdots\hat{\mu}_{6}}\\
& & \\
& &       
+10\hat{\cal K}^{(3)}
\left[
D\hat{X}^{\hat{\mu}}
D\hat{X}^{\hat{\nu}}
D\hat{X}^{\hat{\rho}}
\hat{C}_{\hat{\mu}\hat{\nu}\hat{\rho}} 
-3\frac{m}{2} (2\pi\alpha^{\prime})^{2}\hat{b}\partial \hat{b}
\right]\\
& & \\
& & 
+30\frac{m}{2} (2\pi\alpha^{\prime})
D\hat{X}^{\hat{\mu}}
D\hat{X}^{\hat{\nu}}
D\hat{X}^{\hat{\rho}}
\hat{C}_{\hat{\mu}\hat{\nu}\hat{\rho}} 
\hat{b}\partial\hat{b}\, ,\\
\end{array}
\right.
\end{equation}

\noindent and the indices of $\hat{\cal K}^{(3)}$ in the kinetic 
term are contracted using the metric

\begin{equation}
D_{\hat{\imath}}\hat{X}^{\hat{\mu}}
D_{\hat{\jmath}}\hat{X}^{\hat{\nu}}
\hat{g}_{\hat{\mu}\hat{\nu}}\, .
\end{equation}

\noindent The covariant derivative is defined by

\begin{equation}
D_{\hat{\imath}}\hat{X}^{\hat{\mu}} =  
\partial_{\hat{\imath}}\hat{X}^{\hat{\mu}} 
-{\textstyle\frac{m}{2}}(2\pi\alpha^{\prime}) \hat{b}_{\hat{\imath}}
\hat{k}^{\hat{\mu}}\, .
\end{equation}

\noindent Note that the metric above is also used to raise 
the indices of $\hat{\cal K}^{(3)}$ in the anti-self-duality constraint,
which now takes the form\footnote{In our conventions the antisymmetric
  tensor $\hat{\epsilon}$ is defined to be independent of the metric
  with upper indices $\hat{\epsilon}^{012345}=+1$.}

\begin{equation}
\hat{\cal K}^{(3)\ \hat{\imath}_{1}\hat{\imath}_{3}\hat{\imath}_{3}}=
-\frac{\hat{\epsilon}^{\hat{\imath}_{1}\ldots \hat{\imath}_{6}}}{3!
\sqrt {|D_{\hat{\imath}}\hat{X}^{\hat{\mu}}
D_{\hat{\jmath}}\hat{X}^{\hat{\nu}}\hat{g}_{\hat{\mu}\hat{\nu}}|} }
\hat{\cal K}^{(3)}{}_{\hat{\imath}_{4}\hat{\imath}_{5}\hat{\imath}_{6}}\, .
\end{equation}

The massive field strengths $\hat{\cal K}^{(3)}$ and $\hat{\cal
K}^{(6)}$, and therefore the full action, are invariant  under the
following transformations of spacetime fields

\begin{equation}
\left\{
\begin{array}{rcl}
\delta \hat{g}_{\hat{\mu}\hat{\nu}} & = & 
-\left[\Lambda^{(0)} -\frac{m}{2} (2\pi\alpha^{\prime}) 
\hat{\rho}^{(0)}\right] \hat{k}^{\hat{\rho}}
\partial_{\hat{\rho}}\hat{g}_{\hat{\mu}\hat{\nu}}
\\
& & \\
& &
+2\left(\partial\Lambda^{(0)} +m\hat{\lambda} \right)_{(\hat{\mu}}
\left(i_{\hat{k}}\hat{g}\right)_{\hat{\nu})}\, ,\\
& & \\
\delta \hat{C}_{\hat{\mu}\hat{\nu}\hat{\rho}} & = & 
3\partial_{[\hat{\mu}}\hat{\chi}_{\hat{\nu}\hat{\rho}]} 
-\left[\Lambda^{(0)} -\frac{m}{2} (2\pi\alpha^{\prime}) 
\hat{\rho}^{(0)}\right] \hat{k}^{\hat{\lambda}}
\partial_{\hat{\lambda}}\hat{C}_{\hat{\mu}\hat{\nu}\hat{\rho}}
\\
& & \\
& & 
+3\left(\partial\Lambda^{(0)} +m\hat{\lambda} \right)_{[\hat{\mu}}
\left(i_{\hat{k}}\hat{C}\right)_{\hat{\nu}\hat{\rho}]}\, ,\\
& & \\
\delta \hat{\tilde{C}}_{\hat{\mu}_{1}\ldots\hat{\mu}_{6}} & = & 
6\partial_{[\hat{\mu}_{1}}
\hat{\tilde{\chi}}_{\hat{\mu}_{2}\ldots\hat{\mu}_{6}]} 
+30\partial_{[\hat{\mu}_{1}}\hat{\chi}_{\hat{\mu}_{2}\hat{\mu}_{3}}
\hat{C}_{\hat{\mu}_{4}\hat{\mu}_{5}\hat{\mu}_{6}]}\\
& & \\
& & 
-\left[\Lambda^{(0)} -\frac{m}{2} (2\pi\alpha^{\prime}) 
\hat{\rho}^{(0)}\right] \hat{k}^{\hat{\lambda}}
\partial_{\hat{\lambda}}\hat{\tilde{C}}_{\hat{\mu}_{1}\ldots\hat{\mu}_{6}}
\\
& & \\
& &
-6\left(\partial\Lambda^{(0)} +m\hat{\lambda} \right)_{[\hat{\mu}_{1}}
\left(i_{\hat{k}}\hat{\tilde{C}}
\right)_{\hat{\mu}_{2}\ldots\hat{\mu}_{6}]}
-m\hat{\tilde{\lambda}}_{\hat{\mu}_{1}\ldots\hat{\mu}_{6}}\, ,\\
\end{array}
\right.  
\end{equation}

\noindent and worldvolume fields

\begin{equation}
\left\{
\begin{array}{rcl}
\delta \hat{X}^{\hat{\mu}} & = & 
-\left[\Lambda^{(0)} -\frac{m}{2} (2\pi\alpha^{\prime}) 
\hat{\rho}^{(0)}\right] \hat{k}^{\hat{\mu}}\, , \\
& & \\
\delta \hat{b}_{\hat{\imath}} & = & \frac{1}{2\pi\alpha^{\prime}}
2\hat{\lambda}_{\hat{\imath}} +\partial_{\hat{\imath}}\hat{\rho}^{(0)}\, ,\\
& & \\
\delta \hat{\omega}^{(2)}{}_{\hat{\imath}\hat{\jmath}} & = & 
-\frac{1}{2\pi\alpha^{\prime}}\hat{\chi}_{\hat{\imath}\hat{\jmath}}
-\frac{m}{2} (2\pi\alpha^{\prime}) 
\left( \frac{1}{2\pi\alpha^{\prime}}
2\hat{\lambda} +\partial\hat{\rho}^{(0)}\right)_{[\hat{\imath}} 
\hat{b}_{\hat{\jmath}]} 
+2\partial_{[\hat{\imath}}\hat{\rho}^{(1)}{}_{\hat{\jmath}]}\, ,\\
& & \\
\delta {\hat \omega}^{(5)}{}_{\hat{\imath}_{1}\ldots\hat{\imath}_{5}}
& = & 
-{\textstyle\frac{1}{2\pi\alpha^{\prime}}}
\hat{\tilde{\chi}}_{\hat{\imath}_{1}\ldots\hat{\imath}_{5}}
-\frac{m}{2} \hat{\rho}^{(5)}
+5\partial_{[\hat{\imath}_{1}}
\hat{\rho}^{(4)}{}_{\hat{\imath}_{2}\ldots\hat{\imath}_{5}]}\\
& & \\
& & 
+15 \partial\hat{\omega}^{(2)}{}_{[\hat{\imath}_{1}\hat{\imath}_{2}} 
\left[\hat{\chi} +\frac{m}{2} (2\pi\alpha^{\prime}) 
\left( \frac{1}{2\pi\alpha^{\prime}} 2\hat{\lambda}
+\partial\hat{\rho}^{(0)} \right)\hat{b}
\right]_{\hat{\imath}_{3}\hat{\imath}_{4}\hat{\imath}_{5}]}\, ,\\
& & \\
\delta {\hat \omega}^{(6)}{}_{\hat{\imath}_{1}\ldots\hat{\imath}_{6}} 
& = & \frac{1}{2\pi\alpha^{\prime}}
2\hat{\tilde{\lambda}}_{\hat{\imath}_{1}\ldots\hat{\imath}_{6}} \\
& & \\
& & 
+90 \frac{m}{2} (2\pi\alpha^{\prime})^{3}
\left( \frac{1}{2\pi\alpha^{\prime}}
2\hat{\lambda} +\partial\hat{\rho}^{(0)}\right)_{[\hat{\imath}_{1}}
\hat{b}_{\hat{\imath}_{2}}
\partial_{\hat{\imath}_{3}} 
\hat{b}_{\hat{\imath}_{4}}
\partial_{\hat{\imath}_{5}} 
\hat{b}_{\hat{\imath}_{6}]} \\
& & \\
& &
-30 \hat{b}_{[\hat{\imath}_{1}}
\partial_{\hat{\imath}_{2}}
\left(i_{\hat{k}}\hat{\tilde{\chi}} 
\right)_{\hat{\imath}_{3}\ldots\hat{\imath}_{6}]} 
-180 (2\pi\alpha^{\prime}) \partial_{[\hat{\imath}_{1}}
\hat{\chi}_{\hat{\imath}_{2}\hat{\imath}_{3}}
\hat{b}_{\hat{\imath}_{4}}
\partial_{\hat{\imath}_{5}} 
\hat{b}_{\hat{\imath}_{6}]} \\
& & \\
& &
+6\partial_{[ \hat{\imath}_{1}}
\hat{\rho}^{(5)}{}_{\hat{\imath}_{2}\ldots\hat{\imath}_{6}]}\, .\\
\end{array}
\right.
\end{equation}

\noindent We have included in these transformations
infinitesimal reparametrizations with parameter $\Lambda^{(0)}$ in the
direction of the Killing vector $\hat{k}^{\hat{\mu}}$ since they will
become gauge transformations after dimensional reduction. Observe that
$\hat{\omega}^{(5)}$ transforms as a Stueckelberg field for
$\hat{\omega}^{(6)}$.

We want to comment on a subtlety in the construction of the above
action and transformation rules which will be relevant when discussing
the double dimensional reduction of the massive M-5-brane. The point
is the following. In the present construction we have used that the
target-space 6-form $\hat {\tilde C}$ transforms with a shift under a
dual massive gauge transformation with parameter ${\hat
  {\tilde\lambda}}_{\hat \mu_1\cdots \hat \mu_6}$. This shift was
canceled, in the variation of the WZ term, by the introduction of the
6-form worldvolume field ${\hat \omega}^{(6)}$. However, we know that
the parameter ${\hat {\tilde\lambda}}_{\hat \mu_1\cdots \hat \mu_6}$
is a constrained parameter satisfying

\begin{equation}
{\hat k}^{\hat \mu_6}
{\hat {\tilde\lambda}}_{\hat \mu_1\cdots \hat\mu_5\hat \mu_6} = 0\, .
\end{equation}

\noindent This means that, after a worldvolume reduction in the isometry
direction, there is no shift variation to cancel in the WZ term and,
consequently, there is no need to introduce a new worldvolume 5-form
to cancel such a variation. Nevertheless, as we will see, our
reduction procedure leads to the appearance of another 5-form,
essentially decoupled from the rest of the action. This is consistent
with the fact that the known standard massive D-4-brane action, which
is the double dimensional reduction of the massive M-5-brane action,
does not contain a worldvolume 5-form in its WZ term.  Therefore, in
order to obtain the standard D-4-brane action one must perform by hand
a truncation of the theory.  It is only with this understanding that
the massive M-5-brane action given above leads to the massive
D-4-brane action.  At present it is not clear to us how to impose this
truncation {\sl before} the double dimensional reduction, at the level
of the massive M-5-brane action itself.

We now take the massive M-5-brane action as our starting point and
construct in the following two subsections two massive brane actions
of type~IIA superstring theory. First, we perform a direct dimensional
reduction to obtain the massive p-5A-brane effective action and next
we perform a double dimensional reduction to get the massive D-4-brane
effective action.  This last action is obtained in the so-called
``1-2-form formalism'', which uses a 1- and a 2-form worldvolume fields
that are related to each other by a duality constraint (inherited from
the anti-self-duality of $\hat{\cal K}^{(3)}$).  Using this constraint
we then obtain the usual formulation of the  massive D-4-brane
action  which only uses a 1-form BI field.


\subsubsection{\bf Direct Dimensional Reduction: The Massive p-5A-Brane}

We assume that the background fields do not depend on the coordinate
$Y$ and rewrite them in 10-dimensional form. At the same time the
worldvolume scalar $Y$ can be treated differently to the other
embedding coordinate scalars. We rename it

\begin{equation}
Y \equiv (2\pi\alpha^{\prime}) \hat{c}^{(0)}\, .  
\end{equation}

\noindent This $\hat{c}^{(0)}$ transforms as the worldvolume scalar 
defined in Section~\ref{sec-worldvolume}.

Furthermore, it is convenient to redefine
$\hat{\omega}^{(2)},\hat{\omega}^{(5)}$ and $\hat{\omega}^{(6)}$ so
that their relation with the worldvolume fields defined in
Section~\ref{sec-worldvolume} is manifest.  Using 10-dimensional
notation, the transformation rules of these fields read\footnote{We
  use that ${\hat \lambda}_\mu = \lambda_\mu$ and ${\hat \lambda}_y =
  0$. Similarly, ${\hat {\tilde \lambda}}_{\mu_1\cdots \mu_6} =
  {\tilde \lambda}_{\mu_1\cdots \mu_6} $ and ${\hat {\tilde
      \lambda}}_{\mu_1\cdots \mu_5y} = 0$.  The relations between the
  other gauge parameters of the 11- and 10-dimensional theories are
  given in (\ref{eq:useful}).}

\begin{equation}
\left\{
\begin{array}{rcl}
\delta \hat{\omega}^{(2)} & = & 
-{1\over 2\pi\alpha^\prime}\Lambda^{(2)} + 4\lambda
\partial {\hat c}^{(0)}\\
& & \\
& & -\frac{m}{2}(2\pi\alpha^{\prime})
\left[\frac{1}{2\pi\alpha^{\prime}}2\lambda +\partial{\hat \rho}^{(0)}
\right]{\hat b} + 2\partial {\hat \rho}^{(1)}\, ,\\
& & \\
\delta \hat{\omega}^{(5)} & = & 
\frac{1}{2\pi\alpha^{\prime}}\tilde{\Lambda}
-\frac{m}{2}\hat{\rho}^{(5)}
-5\Lambda^{(4)}\partial \hat{c}^{(0)} \\
& & \\
& & 
-15(2\pi\alpha^\prime) \partial\hat{a}^{(2)}
\left(\delta\hat{a}^{(2)}
-2\partial\hat{\mu}^{(1)} \right) 
+5\partial\hat{\rho}^{(4)}\, ,\\
& & \\
\delta \hat{\omega}^{(6)} & = & 
\frac{1}{2\pi\alpha^{\prime}}2\tilde{\lambda}
+90 \frac{m}{2}(2\pi\alpha^{\prime})^{3} 
\left[\frac{1}{2\pi\alpha^{\prime}}2\lambda 
+\partial\hat{\rho}^{(0)}\right]\hat{b}
\partial\hat{b}\partial\hat{b}\\
& & \\
& &
-180 (2\pi\alpha^{\prime}) \partial\Lambda^{(2)}
\hat{b}\partial\hat{b}  
+720 (2\pi\alpha^{\prime})^{2} \partial\lambda \partial\hat{c}^{(0)}
\hat{b}\partial\hat{b}\\  
& & \\
& &
-30 \hat{b} \partial\Lambda^{(4)}+6\partial\hat{\rho}^{(5)}\, .\\
\end{array}
\right.
\end{equation} 

It is easy to see that $\hat{\omega}^{(2)}$ transforms as
$\hat{a}^{(2)}$ (just as it happened in the M-2-brane case), and
$\hat{\omega}^{(5)}$ transforms as $-\hat{\tilde{b}}$.  Similarly, one
may verify that the transformation rule of $\hat{\omega}^{(6)}$ is
related to that of $\hat{c}^{(6)}$ via the relation

\begin{equation}
\label{uceseis}
\hat{\omega}^{(6)} \equiv -{\hat c}^{(6)}+120(2\pi\alpha^\prime)^3
\partial {\hat c}^{(0)}{\hat b}\partial {\hat b}
\partial {\hat b}\, ,
\end{equation}

\noindent if we make the identification

\begin{equation}
\begin{array}{rcl}
\hat{\rho}^{(5)} & = & -\hat{\kappa}^{(5)} -5\Lambda^{(4)}\hat{b}
+30 (2\pi\alpha^{\prime}) \Lambda^{(2)}\hat{b} \partial \hat{b}  
-20  (2\pi\alpha^{\prime})^{2} \Lambda^{(0)}
\hat{b} \partial \hat{b}  \partial \hat{b}  \\
& & \\
& &
-20 (2\pi\alpha^{\prime})^{3} \hat{\rho}^{(0)} \partial\hat{c}^{(0)}
\partial \hat{b}  \partial \hat{b}  
+40 (2\pi\alpha^{\prime})^{2}\partial\hat{c}^{(0)} \lambda
\hat{b}\partial\hat{b} \\
& & \\
& &
+5\frac{m}{2} (2\pi\alpha^{\prime})^{3} \hat{\rho}^{(0)}
\hat{b}\partial\hat{b}\partial\hat{b}\, .\\
\end{array}
\end{equation}

In doing the reduction, we have to take into account that there are
background fields present in the 3-form field strength. The resulting
action in the quadratic approximation (linear in the equations of
motion) is

\begin{equation}
\label{eq:massivep5A-action}
\begin{array}{rcl}
\hat{S}[X^{\mu},\hat{c}^{(0)},\hat{b},\hat{a}^{(2)}] 
& = & \\
& & \\
& &
\hspace{-3.5cm}
-T_{M5}\int d^{6}\hat{\xi}\ e^{-2\phi}\ 
\sqrt {|g_{\hat{\imath}\hat{\jmath}}
-(2\pi\alpha^{\prime})^{2}\ e^{2\phi}\ \hat{\cal G}^{(1)}{}_{\hat{\imath}}
 \hat{\cal G}^{(1)}{}_{\hat{\jmath}} |}\
\times \\
& & \\
& & 
\hspace{-1cm}
\times \left\{1 
-{\textstyle\frac{1}{4\cdot 3!}} (2\pi\alpha^{\prime})^{2}\
e^{2\phi} \left( \hat{\cal H}^{(3)} \right)^{2} +\cdots \right\} \\
& & \\
& & 
\hspace{-3.5cm}
-(2\pi\alpha^{\prime})\frac{T_{M5}}{6!}\int d^{6}\hat{\xi}\
\hat{\epsilon}^{\hat{\imath}_{1}\cdots\hat{\imath}_{6}}
\hat{\tilde{\cal F}}_{\hat{\imath}_{1}\cdots\hat{\imath}_{6}}\, ,\\
\end{array}
\end{equation}

\noindent where the field strengths $\hat{\cal G}^{(1)},
\hat{\cal H}^{(3)}$ and $\hat{\tilde{\cal F}}$ are defined in
Section~\ref{sec-worldvolume}.


After getting the equations of motion from the action above one still
has to impose the anti-self-duality constraint, which in the linear
approximation used here reads:

\begin{equation}
\hat{\cal H}^{(3)} =-e^{\phi}\  {}^{\star}\hat{\cal H}^{(3)}\, .
\end{equation}

\noindent The factor $e^{-2\phi}$ in front of the kinetic term 
indicates that the physical mass is proportional to $g_{A}^{-2}$,
$g_{A}=e^{<\phi>}$ being the string coupling constant. This is the
right behavior for a solitonic object.


\subsubsection{\bf Double Dimensional Reduction: the Massive D-4-brane}

We take as our starting point the massive p-5A-brane action and
gauge-fix the g.c.t.~transformation in the ${\hat \xi}^5$-direction by
imposing the condition

\begin{equation}
{\hat c}^{(0)} = {\textstyle\frac{1}{2\pi \alpha^\prime}}
\hat{\xi}^{5}\, .
\end{equation}

\noindent To preserve this gauge condition we have to perform a 
compensating worldvolume g.c.t.~transformation in the $\hat{\xi}^{5}$
direction 

\begin{equation}
\label{compen}
\delta\hat{\xi}^{\hat{\imath}}=\delta^{\hat{\imath}5}
\left[-\Lambda^{(0)} +{\textstyle\frac{m}{2}}(2\pi\alpha^{\prime})
\hat{\rho}^{(0)} \right]\, ,
\end{equation}

\noindent leading to modifications in the $\Lambda^{(0)}$ and ${\hat
\rho}^{(0)}$ transformations of the reduced worldvolume fields. Now,
however, we cannot perform the  field redefinitions that we used in the
M-2--brane case. In this case we have target space fields transforming
under  $\Lambda^{(0)}$ and we do not want to modify their gauge
transformations with the introduction of $c^{(0)}$. This complicates
somewhat our work.

We  do the following identification:

\begin{equation} 
\begin{array}{rcl}
{\hat c}^{(6)}{}_{ijklm5} & = &
-5(2\pi\alpha^\prime) \partial_{[i}v^{(0)}
c^{(4)}{}_{jklm]}-5{\textstyle\frac{m}{2}}(2\pi\alpha^\prime)^{3}
v^{(0)} b_{[i}\partial_j b_k\partial_l b_{m]}\\
& & \\
& & 
+v^{(5)}{}_{ijklm}\, ,\\
\end{array}
\end{equation}

\noindent where $v^{(5)}$ is a vector field transforming according to

\begin{equation}
\left\{
\begin{array}{rcl}
\delta v^{(5)} & = & 5\partial \Delta^{(4)}\, ,\\
& & \\
\Delta^{(4)}{}_{ijkl} & = &
{\hat{\kappa}}^{(5)}{}_{ijkl5}+4(2\pi\alpha^\prime)v^{(0)}
\partial_{[i}\kappa^{(3)}{}_{jkl]}\\
& & \\
&&+\frac12 m(2\pi\alpha^\prime)^3
v^{(0)}\rho^{(0)}\partial_{[i}b_j\partial_k b_{l]}
+2m(2\pi\alpha^\prime)^2 v^{(0)}\lambda_{[i}b_j\partial_k
b_{l]}\, .\\
\end{array}
\right.
\end{equation}

The equation of motion for $c^{(4)}$ would imply that $v^{(0)}$ is a
constant and with this simplification we would proceed as in the
M-2-brane case.  This might be correct if we took into account
fermionic terms, but without fermions it is not clear to us whether we
can use the equation of motion of $c^{(4)}$ to set $v^{(0)}$ to a
constant. For the time being we will simply truncate the theory and we
will set $v^{(0)}=0$ by hand and then we will extend the calculation
to the non vanishing but constant value of $v^{(0)}$.  We will also
ignore total derivatives.  Thus, at this level we find the following
result

\begin{equation}
5\partial_{[i_{1}}{\hat {\tilde b}}{}_{i_{2}\cdots i_{5}]\ 5} 
+{\textstyle\frac{m}{2}}\hat{c}^{(6)}{}_{i_{1}\cdots i_{5}\ 5} 
 =  -5\partial_{[i_{1}}v^{(4)}{}_{i_{2}\cdots i_{5}]}
+{\textstyle\frac{m}{2}}v^{(5)}{}_{i_{1}\cdots i_{5}}\, , 
\end{equation} 

\noindent where $v^{(4)}$ is an auxiliary 4-form transforming by shifts

\begin{equation}
\delta v^{(4)}={\textstyle\frac{m}{2}}\Delta^{(4)}\, .
\end{equation}

\noindent This implies that the pair $v^{(4)},v^{(5)}$ constitutes a 
Stueckelberg pair, in which $v^{(4)}$ gets eaten by $v^{(5)}$ which
becomes massive. This is why we keep this total derivative. This part
of the WZ term is completely decoupled from the rest.

The rest of the worldvolume fields can be indentified as follows:

\begin{equation}
\begin{array}{rclrcl}
\hat{b}_{5} & = & v^{(0)}=0\, ,&
\hat{b}_{i} & = & b_{i}\, ,\hspace{1cm}\\
& & & & & \\
\hat{a}^{(2)}{}_{i5} & = & b_{i} +v^{(1)}{}_{i}\, ,&
\hat{a}^{(2)}{}_{ij} & = & a^{(2)\prime}{}_{ij}\, ,\\
\end{array}
\end{equation}

\noindent where $a^{(2)\prime}\neq a^{(2)}$ from the point of view of the
gauge transformations. 

Taking into account that the worldvolume metric is given by:

\footnotesize
\begin{equation}
g_{\hat{\imath}\hat{\jmath}} = e^{-2/3\phi} 
\left(
\begin{array}{cc}
g_{ij} - (2\pi\alpha^\prime)^2 e^{2\phi}{\cal G}^{(1)}_{i}
{\cal G}^{(1)}_{j} & 
-(2\pi\alpha^\prime)e^{2\phi}{\cal G}^{(1)}{}_{i}
\left[1-\frac{m}{2} (2\pi\alpha^\prime) v^{(0)} \right]   \\
& \\
-(2\pi\alpha^\prime)e^{2\phi}{\cal G}^{(1)}{}_{i}
\left[1-\frac{m}{2} (2\pi\alpha^\prime) v^{(0)} \right] 
\hspace{-.5cm}
& 
-e^{2\phi}\left[1-\frac{m}{2} (2\pi\alpha^\prime) 
v^{(0)} \right]^2 \\
\end{array}
\right)\, ,
\end{equation}
\normalsize

\noindent we find the action of the massive D-4-brane in the
``1-2-form'' formalism\footnote{Since it depends on the 1-form $b$ plus
the 2-form $a^{(2)\prime}$.}:

\begin{equation}
\begin{array}{rcl}
S\left[X^{\mu},a^{(2)\prime},b,v^{(1)},v^{(4)},v^{(5)}\right] & = & \\
& & \\
& & 
\hspace{-6cm}
-T_{M5}l\int d^{5}\xi\ e^{-\phi} \sqrt{|g_{ij}|}
\left\{ 1 -\frac{1}{4\cdot 3!}(2\pi\alpha^{\prime})^{2} 
\left({\cal R}^{(3)}\right)^{2}
+\frac{1}{4\cdot 2!}(2\pi\alpha^{\prime})^{2} 
\left({\cal F}+{\cal L}^{(2)}\right)^{2}
\right\} \\
& & \\
& & 
\hspace{-6cm}
+\frac{T_{M5}l}{5!}(2\pi\alpha^{\prime})\int d^{5}\xi\epsilon
\left\{
\frac{1}{2\pi\alpha^{\prime}}C^{(5)}
-5 C^{(3)}\left({\cal F}+{\cal L}^{(2)}\right)
+15\partial a^{(2)\prime}B 
\right.\\
& & \\
& & 
\hspace{-6cm}
-15\frac{m}{2}(2\pi\alpha^{\prime})b\partial b B
+30\frac{m}{2}(2\pi\alpha^{\prime})b\partial (b+v^{(1)}) B
+30\frac{m}{2}(2\pi\alpha^{\prime})b\partial b\partial (b+v^{(1)}) \\
& & \\
& & 
\hspace{-6cm}
\left.
-20\frac{m}{2}(2\pi\alpha^{\prime})^{2}b\partial b\partial b
-5\partial v^{(4)} +{\textstyle\frac{m}{2}}v^{5}
\right\}\, ,
\end{array}
\end{equation}

\noindent where the double-dimensionally reduced anti-self-duality condition 

\begin{equation}
{\cal R}^{(3)}= -e^{-\phi} {}^{\star}\left({\cal F}+{\cal L}^{(2)} \right)\, ,
\end{equation}

\noindent with

\begin{equation}
\begin{array}{rcl}
{\cal R}^{(3)} & = &
3\partial a^{(2)\prime} +{\textstyle\frac{1}{2\pi\alpha^{\prime}}}C^{(3)}
-3{\textstyle\frac{m}{2}}(2\pi\alpha^{\prime})b\partial b
-3{\textstyle\frac{m}{2}}(2\pi\alpha^{\prime})B b \\
& & \\
& &
-\left(C^{(1)} -{\textstyle\frac{m}{2}b}\right)
\left({\cal F}+{\cal L}^{(2)}\right)\, ,\\
\end{array}
\end{equation}

\noindent still needs to be imposed in the reduced action.

Either $b$ or $a^{(2)\prime}$ can be considered an auxiliary 
field and eliminated in the equations of motion after using the
constraint above, that relates their field strengths. This was the
procedure followed in Ref.~\cite{kn:BRO} to find the Born-Infeld
effective action from the massless M-5-brane.  Here we will follow an
alternative, simpler, procedure which consists in introducing the
constraint into the action by means of a Lagrange multiplier term.
Integrating out $a^{(2)\prime}$ yields the usual D-4-brane action and
integrating out $b$ we obtain the dual action in terms of the 2-form.
This action is the worldvolume dual of the usual massive D-4-brane.
We will not study this one here but we will concentrate on recovering
the standard D-4-brane action.

The Lagrange multiplier term, containing a Lagrange multiplier 1-form 
$\varrho$, that we add to the action is:

\begin{equation}
\label{lagrange}
+{\textstyle\frac{T_{M_5}l}{4!}}(2\pi\alpha^\prime)^2\int d^5\xi 
\epsilon (2\partial\varrho) (3\partial a^{(2)\prime})
\end{equation}

\noindent Integration over $\varrho$ enforces the Bianchi
identity for the two-form $a^{(2)\prime}$. On the other hand, integrating
out the 3-form ${\cal R}^{(3)}$ we can impose the anti-self-duality
constraint on-shell and obtain the action in the 1-form formalism.  In
particular, the equation of motion for ${\cal R}^{(3)}$ gives:

\begin{equation}
{\cal R}^{(3)}  = -e^{-\phi}{\cal F}^{\prime}\, ,
\hspace{1cm}
{\cal F}^{\prime} = 2\partial\varrho 
+{\textstyle\frac{1}{2\pi\alpha^{\prime}}}B\, .
\end{equation}

\noindent We can now substitute the above equation of motion to
eliminate $a^{(2)\prime}$. Now one can use the anti-self-duality
constraint which takes the form

\begin{equation}
{\cal F}^{\prime}={\cal F}+{\cal L}^{(2)}\, ,
\end{equation}

\noindent in the action. The result is

\begin{equation}
\begin{array}{rcl}
S\left[X^{\mu},b,v^{(1)},v^{(4)},v^{(5)}\right] & = & \\
& & \\
& & 
\hspace{-4cm}
-T_{M5}l\int d^{5}\xi\ e^{-\phi} \sqrt{|g_{ij}|}
\left\{ 1 +\frac{1}{2\cdot 2!}(2\pi\alpha^{\prime})^{2} 
\left({\cal F}+{\cal L}^{(2)}\right)^{2}
\right\} \\
& & \\
& & 
\hspace{-4cm}
+\frac{T_{M5}l}{5!}(2\pi\alpha^{\prime})\int d^{5}\xi\epsilon
\left\{
\frac{1}{2\pi\alpha^{\prime}}C^{(5)}
-10C^{(3)}\left({\cal F}+{\cal L}^{(2)}\right)
\right.\\
& & \\
& & 
\hspace{-4cm}
+15(2\pi\alpha^{\prime}) C^{(1)}\left({\cal F}+{\cal L}^{(2)}\right)^{2}
-60\frac{m}{2}(2\pi\alpha^{\prime})b\partial v^{(1)}\partial (b+v^{(1)})
\\
& & \\
& & 
\hspace{-4cm}
\left.
-20\frac{m}{2}(2\pi\alpha^{\prime})^{2}b\partial b\partial b
-5\partial v^{(4)} +{\textstyle\frac{m}{2}}v^{(5)}
\right\}\, .
\end{array}
\end{equation}

\noindent Now we make the field redefinition

\begin{equation}
b+v^{(1)}\rightarrow b\, ,
\end{equation}

\noindent and adding the total derivative $5\partial c^{(4)}$ to get a 
more compact expression we find the usual action of the D-4--brane
(``1-form formalism'') to quadratic order in ${\cal F}$ containing three
extra terms which are completely decoupled

\begin{equation}
\begin{array}{rcl}
S\left[X^{\mu},b,v^{(1)},v^{(4)},v^{(5)}\right] & = & \\
& & \\
& & 
\hspace{-5cm}
-T_{M5}l\int d^{5}\xi\ e^{-\phi} \sqrt{|g_{ij}|}
\left\{ 1 +\frac{1}{2\cdot 2!}(2\pi\alpha^{\prime})^{2} 
\left({\cal F}\right)^{2}
\right\} \\
& & \\
& & 
\hspace{-5cm}
+\frac{T_{M5}l}{5!}(2\pi\alpha^{\prime})\int d^{5}\xi\epsilon
\left\{ {\cal G}^{(5)}
-20\frac{m}{2}(2\pi\alpha^{\prime})^{2}v^{(1)}\partial v^{(1)}\partial v^{(1)}
-5\partial v^{(4)} +{\textstyle\frac{m}{2}}v^{5}
\right\}\, .
\end{array}
\end{equation}

Observe that $v^{(1)}$ appears only through a topological term.  The
consequences of the presence of a 5-form field in the WZ term of the
massive D-4-brane will be discussed in the next Section.

One can easily see that if we consider a constant but non-vanishing
$v^{(0)}$ the only change in the above result is that the tension has
to be replaced by

\begin{equation}
T = \left[1-{\textstyle\frac{m}{2}}
(2\pi\alpha^\prime) v^{(0)}\right]T_{M_5}l\, ,
\end{equation}

\noindent and the terms with $v$-forms get an extra factor
$\left[1-{\textstyle\frac{m}{2}} (2\pi\alpha^\prime)
v^{(0)}\right]^{-1}$ for each power of $v$.


\section{Conclusions} 
\label{sec-conclusion}

In this paper we have investigated the worldvolume effective theory of
massive branes, i.e.~branes moving in a background with a nonzero
cosmological constant, both in string theory and M-theory.  The
construction of the worldvolume actions requires new results on
the structure of both the target-space background fields as well as the
worldvolume fields.

As far as the target-space fields are concerned we encountered two
issues. First of all the coupling of massive IIA supergravity to the
type~IIA five-brane requires a dualization of the {\sl massive} NS-NS
target-space 2-form $B$.  Inspired by the work of \cite{kn:Q} we
succeeded in doing this and encountered an interesting mechanism
where, after dualization, the roles of the NS-NS and R-R fields get
interchanged.  Whereas in the usual formulation the R-R 1-form
$C^{(1)}$ is a Stueckelberg field and $B$ is a massive 2-form, we find
that, after dualization, the dual R-R 7-form $C^{(7)}$ becomes massive
while the dual NS-NS 6-form is a Stueckelberg field.

The second issue concerns the question of whether or not the massive
IIA supergravity theory has an eleven-dimensional origin. Inspired by
the worldvolume sigma model approach we made a proposal for a massive
eleven-dimensional supergravity theory. A word of caution is needed
here. We do not claim that a cosmological constant (containing a mass
parameter $m$) can be added to the usual (massless) eleven-dimensional
supergravity theory. Indeed, a no-go theorem for such a theory exists
\cite{kn:BDHS}. There is a simple argument for this obstruction.  One
can easily verify that any cosmological constant in eleven dimensions
of the form

\begin{equation}
{\cal L}_{{\rm d=11}} \sim \int d^{11}x\ m^2 \sqrt {|g|}
\end{equation}

\noindent leads, upon dimensional reduction to ten dimensions, 
to a cosmological constant with a dilaton coupling that differs from
the one in massive IIA supergravity.  What we find from the sigma
model approach is that it can be done only under the assumption that
the d=11 target space background fields have an isometry. This
assumption follows naturally from the fact that the underlying sigma
model turns out to be a {\sl gauged} sigma model whose construction
requires an isometry direction. Only in the massless case, when $m=0$,
this assumption can be deleted and we obtain the usual massless
supergravity theory with no isometry whatsoever. The obstruction
mentioned above is now avoided by considering a cosmological constant
of the form

\begin{equation}
{\cal L}_{massive\ d=11\ supergravity} \sim \int d^{11}x\ m^2 
|{\hat k}^2|^2\sqrt {|g|}\, ,
\end{equation}

\noindent involving the Killing vector ${\hat k}^\mu$. This term 
gives the correct dilaton coupling.

It is of interest to compare our proposal of massive
eleven-dimensional supergravity with the work of \cite{kn:HLW}
where a new d=11 supergravity theory is introduced that, upon
reduction to ten dimensions, leads to an alternative massive IIA
supergravity theory. In both cases the massive corrections to the
massless d=11 supergravity theory are triggered by modifying the
definition of the spin connection.  The specific form of the
modification, however, differs in the two cases. In our case (see
Eq.~(\ref{connection})) the extra term in the connection involves the
3-form $\hat C$ whereas the extra term in the connection of
\cite{kn:HLW} involves only the Killing vector, leading to a different
massive supergravity theory. In the case of \cite{kn:HLW} the modified
spin connection has a geometric interpretation in terms of a Weyl
superspace (see e.g.~\cite{Weyl})\footnote{This construction is based on the
fact that the equations of motion of d=11 supergravity are invariant under
scale transformations (see, e.g., \cite{BerJO}). It has recently been shown
that the new massive d=10 supergravity of \cite{kn:HLW}
can alternatively be understood as a generalized Scherk-Schwarz
reduction (making use of the scaling symmetry) of the ordinary
massless d=11 supergravity theory \cite{LLP}.}. 
It would be interesting to see whether the modified
connection we have introduced has a similar geometric interpretation.

Another question is whether the massive eleven-dimensional
supergravity theory considered in this work has something to do with
M-theory.  We remind that the relation between the usual massless d=11
supergravity theory and M-theory goes via the observation that
massless d=11 supergravity can be viewed as the decompactification
limit ($R_{11}\rightarrow \infty)$ of massless IIA supergravity and
that this decompactification limit is equivalent to the strong
coupling limit ($g_s \rightarrow \infty$) of type~IIA superstring
theory since $R_{11} = g_s^{2/3}$ \cite{W}. It is natural to consider
the isometry direction as an eleventh direction, since the massive
M-brane configurations we introduce in this work can be positioned in
this direction. Indeed, this feature enables us to construct two
different massive branes of string theory out of a single massive
M-brane. The two possibilities arise due to the fact that the isometry
direction can be transverse to the brane (direct dimensional
reduction) or along one of the worldvolume directions of the brane
(double dimensional reduction).  It would be interesting to consider
the massive eleven-dimensional supergravity theory of this work in the
context of the Ho{\v r}ava-Witten approach \cite{HorWit} where there
are two M-9-branes positioned at the end of spacetime.

As far as the worldvolume fields are concerned the construction of the
worldvolume actions for the massive branes of string theory requires
that we do not only introduce the usual embedding coordinates $X^\mu$
and BI 1-form $b$ but also a whole set of worldvolume p-form fields
$c^{(p)}$, which are in one to one correspondence to the target space
R-R fields $C^{(p+1)}$. Such p-form fields have been encountered
before in a scale-invariant formulation of branes \cite{kn:BLT} and in
a study of branes ending on branes \cite{kn:T2} (see also
\cite{kn:Sezgin}). They have also been recently used in the
construction of a manifestly $SL(2,\R)$ covariant formulation of IIB
brane worldvolume actions \cite{T,TC,CW}.  In the latter case all
$c^{(p)}$\ (p odd) occur in definite $SL(2,\R)$ representations: for
instance, there are doublets $ (b^{(1)},c^{(1)}), (b^{(5)}, c^{(5)})$
and singlets $c^{(3)}$, etc., where $b^{(1)} (b^{(5)})$ describes the
tension of a type~IIB string (type~IIB five-brane).  Finally, p-form
worldvolume fields also play an important role in the construction of
null super D-branes \cite{kn:BeTo}. We have shown in this paper that
the worldvolume p-form fields of IIA string theory can be obtained
from corresponding p-form worldvolume fields in M-theory via
reduction.  We plan to discuss these M-theory worldvolume fields in a
future publication \cite{kn:BJO3}.

With the required new knowledge on the structure of the target-space
and worldvolume fields described above, we proceeded to construct the
worldvolume actions for massive branes. Our strategy was to first
construct the action of a massive M-brane, i.e.  an M-brane moving in
a massive d=11 supergravity background, and next obtain, via direct
and double dimensional reduction, a pair of massive branes of string
theory.  We first discussed the massive M-0-brane which is a bit
special and only allows a direct reduction to the massive D-0-brane.
Next, we treated the massive M-2-brane \cite{kn:L,kn:O} and obtained
the massive fundamental string and the massive D-2-brane. The last
case we considered was the massive M-5-brane whose reduction leads to
the massive type~IIA five-brane and the massive D-4-brane, although we
must impose a truncation in order to make contact with the massive
D-4-brane action given in the literature.

One interesting outcome of our work is that we find that the massive
type~IIA 5-brane has an additional coupling to a worldvolume 6-form
$c^{(6)}$, describing the tension of a D-6-brane, with the strength of
the coupling proportional to $m$:

\begin{equation}
\label{term2}
S_{\rm massive\  NS5-brane} \sim \int d^6\xi\ m\  \epsilon^{i_1\cdots i_6}
c^{(6)}{}_{i_1\cdots i_6}\, .
\end{equation}
A similar coupling occurs in the massive D-0-brane action:

\begin{equation}
\label{term1}
S_{\rm massive\  D0-brane} \sim \int d\xi\ m\ b_\xi\, ,
\end{equation}

\noindent where $b_\xi$ describes the tension of a fundamental string.
In \cite{kn:Kl,kn:DaFe} it was pointed out, for the massive D-0-brane,
that this new coupling has implications for the anomalous creation of
branes \cite{HaWi,BDG} (for a more recent discussion see
\cite{kn:GrBa}). To be precise, if a D-0-particle crosses a D-8-brane
a stretched fundamental string is created in the single direction
transverse to the D-8-brane.  This process is, via duality, related to
the creation of a stretched D3-brane if a D-5-brane crosses a
NS5-brane \cite{HaWi}. In the latter case the intersecting
configuration is given by

\begin{equation}
\label{conf1}
 \begin{array}{c|c}
{\rm D5}:\ \ \          \x & \x    \x   -   -   -  \x  \x   \x   - \\
{\rm NS5}:\          \x & \x   \x  \x  \x  \x  -  -   -   -    \\
{\rm D3}: \ \ \       \x & \x \x - - - - - - \x  
                         \end{array} 
\end{equation}

\noindent where we have used the notation of \cite{kn:Groningen}
\footnote{Each horizontal line indicates the 10 directions $0,1,\cdots
  9$ in spacetime.  A $\x (-)$ means that the corresponding direction
  is in the worldvolume of (transverse to) the brane.}. The
intersecting configuration of \cite{kn:Kl,kn:DaFe} is obtained by
first applying T-duality in the directions 1 and 2, next applying an
S-duality and, finally, applying a T-duality in the directions 6,7 and
8 \cite{kn:Kl,kn:DaFe}:
 
\begin{equation}
\label{conf2}
 \begin{array}{c|c}
{\rm D0}:\ \ \          \x & -    -   -   -   -  -  - -    - \\
{\rm D8}:\ \ \          \x & \x   \x  \x  \x  \x  \x \x \x   -    \\
{\rm F1}: \ \ \       \x & - - - - - - - - \x  
                         \end{array} 
\end{equation}
 
One would expect that the new coupling we find for the massive 5-brane
has similar implications\footnote{We thank C.~Bachas for an
  illuminating discussion on this point.}.  To be precise, one would
expect that if a NS5-brane crosses a D-8-brane a D-6-brane stretched
between them is created. This process would be the dual of the one
considered in \cite{kn:Kl,kn:DaFe}.  Indeed, this is exactly the
process which has been considered in \cite{Hanany} where it was used
to construct $N=1$ supersymmetric gauge theories in four dimensions
with chiral matter\footnote{We thank P.~Townsend for pointing
  reference \cite{Hanany} out to us.}.

The intersecting configuration corresponding to this process is
obtained by applying T-duality in the directions 3,4 and 5 on the
configuration (\ref{conf1}) \cite{Hanany}:

\begin{equation}
\label{conf3}
 \begin{array}{c|c}
{\rm D8}:\ \ \          \x & \x    \x   \x \x \x  \x  \x   \x   - \\
{\rm NS5}:\          \x & \x   \x  \x  \x  \x  -  -   -   -    \\
{\rm D6}: \ \ \       \x & \x \x \x \x \x  - - - \x  
                         \end{array} 
\end{equation}

\noindent A special feature of this configuration is that when the 
NS5-brane passes through the D-8-brane it is completely embedded
within the D-8-brane. After passing through the D-8-brane a D-6-brane
is created that has 5 of its directions on the worldvolume of the
D-8-brane and in the remaining directions is stretched in the single
direction transverse to the D-8-brane.

We may consider the above process from an M-theory perspective. The
intersecting M-brane configuration that, via reduction (over the 10
direction), yields the intersecting configuration given in
(\ref{conf3}) is given by\footnote{This configuration, alternatively
  denoted by $(5|M9,KK)$, is known to preserve $1/4$ of the
  supersymmetry \cite{deRoo}.}:

\begin{equation}
\label{conf4}
 \begin{array}{c|c}
{\rm M9}:\ \ \          \x & \x    \x   \x \x \x   \x  \x   \x   - \x \\
{\rm M5}:\ \ \          \x & \x   \x  \x  \x  \x  -  -   -   -   -  \\
{\rm KK11}: \       \x & \x \x \x \x \x  - - - \x z 
                         \end{array} 
\end{equation}

\noindent where $z$ refers to the $U(1)$ isometry direction in the
Taub-NUT space of the monopole. This intersecting configuration
suggests that when an M-5-brane passes through a (hypothetical)
M-9-brane a stretched 11-dimensional KK monopole is created. The
anomalous creation of this KK monopole is related to the following
term we found in the massive M5-brane action:

\begin{equation}
S_{\rm massive\  M5-brane} \sim \int d^6\hat \xi\ m\ \epsilon^{\hat 
\imath_1\cdots 
\hat \imath_6}
\hat{\omega}^{(6)}{}_{\hat \imath_1\cdots \hat \imath_6}\, .
\end{equation}

\noindent This suggests that, in the same way as $c^{(6)}$ 
describes the tension of a D-6-brane, the worldvolume 6-form ${\hat
  \omega}^{(6)} $ is related to the tension of an 11-dimensional KK
monopole.  A gauged version of this 6-form can be used to construct
the WZ term in the KK11-monopole worldvolume action \cite{kn:BJO3}.

We have found in the action of the massive D-4-brane a term of the form

\begin{equation}
\label{term4}
S_{\rm massive\ D4-brane} \sim \int d^5\xi\ m\ v^{(5)}\, ,
\end{equation}

\noindent where $v^{(5)}$ is related to the tension of a KK10-monopole.
This coupling suggests that whenever a D4-brane
passes through a D8-brane a stretched KK10-monopole is created with 4 of
its worldvolume directions inside the D8-brane. Indeed, 
performing first a T-duality in the 9 direction and next a T-duality
in the 3 and 4 directions on the configuration (\ref{conf1}) we obtain
the intersecting brane configuration corresponding to this process:

\begin{equation}
\label{conf5}
 \begin{array}{c|c}
{\rm D8}:\ \ \ \          \x & \x    \x   \x   \x   -  \x  \x   \x   \x \\
{\rm KK10}:\          \x & \x   \x  \x  \x  \x  -  -   -   z    \\
{\rm D4}: \ \ \ \       \x & \x \x \x \x - - - - -  
                         \end{array} 
\end{equation}
The same intersecting configuration is obtained if one reduces the
intersecting M-brane configuration (\ref{conf4}) over the
5-direction.

There is one more argument in favor of the above process.
It turns out that, of all branes in string theory, only
$D0,\ NS5,\ D4$ (and a wave) 
can be embedded in a D8-brane and the intersections are given by

\begin{eqnarray}
&&(0|D0,D8)\, ,\nonumber\\
&&(5|NS5,D8)\, ,\\
&&(4|D4,D8)\, ,\nonumber
\end{eqnarray}
respectively.
The first 2 cases are required for the creation of a fundamental string
(D-6-brane) whenever a D-0-brane (NS5-brane) passes through a D-8-brane. 
The third and only other possible embedding of a brane into a D-8-brane
is that of a D-4-brane and this would correspond to the process 
(\ref{conf5}) where an anomalous KK10-monopole is created.

There is one more potential role to be played by the worldvolume
6-form field $\hat{\omega}^{(6)}$ we found in the massive M-5-brane
action. It has been shown that the selfdual 2-form
$\hat{\omega}^{(2)}$ occurring in the same action can be used to
construct a 1-brane soliton on the M-5-brane worldvolume
\cite{kn:HLW2}.  This 1-brane soliton corresponds to the occurrence of
a 1-form central charge in the 6-dimensional (2,0) worldvolume
supersymmetry algebra.  The complete (2,0) algebra also suggests a
3-brane and a 5-brane soliton.  The 3-brane soliton has been recently
constructed by dualizing one of the embedding scalars into a
worldvolume 4-form \cite{kn:HLW2}.  The 5-brane soliton should
describe the embedding of the M-5-brane into something else and a
natural possibility is an M-9-brane.  It would be interesting to see
whether such a 5-brane soliton can indeed be constructed and whether
the 6-form $\hat{\omega}^{(6)}$ (which does not describe a dynamical
degree of freedom) could describe the charge of such a 5-brane
soliton.

In this work we have shown that the massive branes of M-theory,
i.e.~the M-0-brane, M-2-brane and M-5-brane, are described by a {\sl
  gauged} sigma model.  It would be interesting to see whether the
gauged sigma model approach can also be applied to describe a massive
KK11 monopole and/or the conjectured M9-brane. Concerning the
KK11-monopole, a new feature in this case is that the gauged sigma
model is already needed to describe the dynamics of the KK11-monopole
in a {\sl massless} background \cite{kn:BJO2}\footnote{Note that the
  gauge transformation rule of the 6-form ${\hat \omega}^{(6)}$ contains
  $m$-dependent terms. This shows that the WZ term of a massive
  KK11-monopole differs from the WZ term of a massless
  KK11-monopole.}.

It would be interesting to see whether our results shed new light on
the evasive 11-dimensional M-9-brane\footnote{For other discussions of
  the M-9-brane, see
  e.g.~\cite{HorWit,kn:BRGPT,HoweSezgin,PapaTo,Polchinski,Duff,Hull,Se}.}.
A standard argument against a {\sl freely moving} M-9-brane is that
the corresponding 10-dimensional worldvolume field theory does not
allow multiplets containing a single scalar to indicate the position
of the M9-brane\footnote{Such a scalar would not be needed if the
  M9-brane was positioned at the end of spacetime, like in
  \cite{HorWit}.}.  A way out of this is to assume that the M-9-brane
is really an 8-brane with an extra isometry in one of the 2 transverse
directions, leading to a gauged sigma-model, like it happened in the
case of the KK11-monopole. Now, we are dealing with a nine-dimensional
field theory which naturally contains a vector multiplet with a single
scalar.  We will not pursue this line of thought further here.

We would like to end with the following remark. Although this paper is
dealing with {\sl massive} branes, our results also have implications,
via {\sl massive} T-duality, for the worldvolume theories of {\sl
  massless} branes.  To be specific, it is known that the term
(\ref{term1}), via the T-duality rule \cite{kn:BR}

\begin{equation}
C^{(0)} = m x
\end{equation}

\noindent in the direction $x$, leads to the following 
gauge-invariant term in the D-1-brane action:

\begin{equation}
S_{\rm D1-brane} \sim \int d^2\xi\ C^{(0)}  \epsilon^{ij}
{\cal F}_{ij}\, .
\end{equation}

\noindent This term is needed for an $SL(2,\R)$-covariant formulation of
(p,q)-strings \cite{T}. Similarly, the new term (\ref{term2}) we have
found in the worldvolume action of the p-5A-brane leads, via massive
T-duality, to a gauge-invariant term in the worldvolume action of the
type~IIB KK-monopole and the NS5B-brane containing the RR-scalar
$C^{(0)}$. The latter case suggests the following gauge-invariant term
in the NS5B-brane action:

\begin{equation}
S_{\rm NS5B-brane} \sim \int d^6\xi\ C^{(0)} {\cal G}^{(6)}\, ,
\end{equation}

\noindent with ${\cal G}^{(6)}$ the curvature of the D-5-brane tension.
Via $S$-duality this implies that the D-5-brane action contains the
term

\begin{equation}
S_{\rm D5-brane} \sim \int d^6\xi\ C^{(0)}  {\cal F}_{IIB}\, ,
\end{equation}

\noindent with ${\cal F}_{IIB}$ the curvature of the NS5B-brane tension.
It would be interesting to see whether the above modifications are
indeed present, in which case they are likely to play a role in the
construction of an $SL(2,\R)$-covariant formulation of
$(p,q)$-5-branes.  In order to verify this one needs to explore the
T-duality properties of the different worldvolume p-forms introduced
in this work.


\section*{Acknowledgments}

E.B.~would like to thank C.~Bachas, M.B.~Green, J.P.~van der Schaar,
J.~Schwarz, P.K.~Townsend and, especially E.~Eyras, for useful
discussions.  Y.L.~would also like to thank J.~Gauntlett and C.M.~Hull
for useful discussions, and the Theory Division at CERN for its
hospitality while part of this work was done.  T.O.~would like to
thank E.~\'Alvarez for many useful conversations.  He is also indebted
to M.M.~Fern\'andez for her support.

The work of E.B.~is supported by the European Commission TMR program
ERBFMRX-CT96-0045, in which E.B.~is associated to the University of
Utrecht. The work of Y.L.~is supported by a TMR fellowship from the
European Commission.  The work of T.O.~is supported by the European
Union TMR program FMRX-CT96-0012 {\sl Integrability, Non-perturbative
  Effects, and Symmetry in Quantum Field Theory} and by the Spanish
grant AEN96-1655.

\appendix

\section{Duality of Massive $k$-Form Fields}
\label{sec-massdual}

In this Appendix we focus on the study of duality for massive
$d$-dimensional $k$-form fields in general to then try to apply the
ideas to massive type~IIA supergravity.  Following Ref.~\cite{kn:Q} we
start by briefly discussing duality of massless $k$-form fields in the
presence of magnetic sources.

Let us consider a $(p+1)$-form potential $A_{(p+1)}$ with field
strength

\begin{equation}
F_{(p+2)}=(p+2)\partial A_{(p+1)}\, ,
\end{equation}

\noindent and action

\begin{equation}
\label{eq:masslessp}
S[A_{(p+1)}] =\int d^{d}x\ \sqrt{|g|}\
\left[ {\textstyle\frac{(-1)^{(p+1)}}{2\cdot (p+2)!}} 
{\textstyle\frac{1}{e_{(p)}^{2}}} F_{(p+2)}^{2}\right]\, .
\end{equation}

If $A_{(p+1)}$ is defined everywhere, then the field strength
satisfies the Bianchi identity\footnote{Here $\tilde{p}=d-p-4$ 
so the Hodge dual of 
a $(p+2)$-form is a $(\tilde{p}+2)$-form.}

\begin{equation}
\nabla_{\nu}\left({}^{\star}
F_{(p+2)}\right)^{\nu\mu_{1}\ldots\mu_{\hat{p}+1}} = 0\, .
\end{equation}

\noindent If we consider the presence of (non-dynamical) magnetic 
sources then the Bianchi identity becomes

\begin{equation}
\nabla_{\nu}\left({}^{\star}
F_{(p+2)}\right)^{\nu\mu_{1}\ldots\mu_{\tilde{p}+1}} 
=\tilde{J}^{\mu_{1}\ldots\mu_{\tilde{p}+1}}\, ,
\end{equation} 

\noindent which means that $F_{(p+2)}$ cannot be $(p+2)\partial A_{(p+1)}$ 
everywhere (i.e.~it is not an exact form anymore) but is singular at
the Dirac singularity. In general we can write

\begin{equation}
F_{(p+2)}=(p+2)\partial A_{(p+1)}+B_{(p+2)}\, ,  
\end{equation}

\noindent where $A_{(p+1)}$ is  defined everywhere and

\begin{equation}
\nabla_{\nu}\left({}^{\star}
B_{(p+2)}\right)^{\nu\mu_{1}\ldots\mu_{\tilde{p}+1}} 
=\tilde{J}^{\mu_{1}\ldots\mu_{\tilde{p}+1}}\, .
\end{equation}

With this modification the action is still the same
(\ref{eq:masslessp}). This modification makes sense as long as we deal
with a number of (non-dynamical) magnetic objects. If we allow the
magnetic objects to condense then we have to promote the field that
describes them ($B_{(p+2)}$) to a dynamical one by the simple addition
of a kinetic term for it:

\begin{equation}
\label{eq:masslesspcondensed}
S[A_{(p+1)},B_{(p+2)}] =\int d^{d}x\ \sqrt{|g|}\ \left[
{\textstyle\frac{(-1)^{(p+1)}}{2\cdot (p+2)!} \frac{1}{e_{(p)}^{2}}}
F_{(p+2)}^{2} 
+{\textstyle\frac{(-1)^{(p+2)}}{2\cdot (p+3)!} \frac{1}{e_{(p+1)}^{2}}}
H_{(p+3)}^{2} \right]\, ,
\end{equation}

\noindent where

\begin{equation}
H_{(p+3)} = (p+3)\partial B_{(p+2)}\, . 
\end{equation}

The action above is invariant under the usual gauge transformations of
the $A_{(p+1)}$ potential

\begin{equation}
\delta A_{(p+1)} = (p+1)\partial\chi_{(p)}\, ,
\end{equation}

\noindent and under the {\it massive gauge transformations}

\begin{equation}
\left\{
\begin{array}{rcl}
\delta B_{(p+2)} & = & -(p+2) \partial\lambda_{(p+1)}\, ,\\
& & \\
\delta A_{(p+1)} & = & \lambda_{(p+1)}\, ,\\
\end{array}
\right.
\end{equation}

\noindent which allow us to eliminate $A_{(p+1)}$ by setting it to zero.
Doing this, we get the following action for a massive $(p+2)$-form
potential\footnote{It is often said that the $(p+2)$-form has
  ``eaten'' the $(p+1)$-form, thus getting a mass.}:

\begin{equation}
\label{eq:massivep}
S[B_{(p+2)}] =\int d^{d}x\ \sqrt{|g|}\ \left[
{\textstyle\frac{(-1)^{(p+2)}}{2\cdot (p+3)!} \frac{1}{e_{(p+1)}^{2}}}
H_{(p+3)}^{2}
+{\textstyle\frac{(-1)^{(p+1)}}{2\cdot (p+2)!} \frac{1}{e_{(p)}^{2}}}
B_{(p+2)}^{2}  \right]\, ,
\end{equation}

\noindent which is the one studied in Ref.~\cite{kn:Q}. 
This action is also a truncation of the massive type~IIA supergravity
action for $p=0$\footnote{This does not necessarily mean that the
  mechanism through which $B$ gets mass in the type~IIA theory must be
  the condensation of 6-branes. The origin of the mass in the type~IIA
  theory is the presence of 8-branes \cite{kn:PW,kn:BRGPT}. Still, the
  reason why it is precisely $B$ the field that becomes massive is not
  clear to us, and other mechanisms for giving mass to different
  fields of the type~IIA theory should not be excluded. We thank
  J.L.F.~Barb\'on for discussions on this point.}  with $e_{(1)}=1$
and $e_{(0)}=1/m$.

In Ref.~\cite{kn:Q} it is mentioned that the above action enjoys no
evident global symmetry, but that in spite of this it can be dualized.
What we have just seen is that the action for the massive $(p+2)$-form
potential is equivalent to the action with $A_{(p+1)}$ (which plays
the role of Stueckelberg field here) with a massive gauge symmetry,
and that in this formulation the dualization can be done systematically,
which may mean that a global symmetry (which we have not identified)
is present.

In any case, in Ref.~\cite{kn:Q} the dual action is found via an
intermediate first-order action with an auxiliary $(\tilde{p}+1)$-form
field. Integration of the auxiliary field gives (\ref{eq:massivep})
and integration of the field strength $H_{(p+3)}$ gives the dual
action. Instead of using this intermediate action, which in principle
one has to construct in a case by case basis, we are going to follow a
different, more systematic, procedure using the usual recipe for
Poincar\'e duality in the action with the Stueckelberg
field\footnote{The usual recipe also involves the construction of an
  intermediate action, but our construction is systematic.}. This
recipe cannot be applied to the dualization of $B_{(p+2)}$ because it
occurs explicitly in the action instead of appearing only through
$H_{(p+3)}$ (hence the need of an intermediate action) but it can be
applied to the dualization of the Stueckelberg field $A_{(p+1)}$ as a
first step. Thus, we consider now $F_{(p+2)}$ as a fundamental field
and we add to the action (\ref{eq:masslesspcondensed}) a Lagrange
multiplier term to enforce its Bianchi identity. The resulting action
is, after integration by parts

\begin{equation}
\label{eq:masslesspcondensedintermediate}
\begin{array}{rcl}
S[F_{(p+2)},B_{(p+2)},\tilde{A}_{(\tilde{p}+1)}] =\hspace{-2cm}
& &\\
& & \\
& & 
 \int d^{d}x\ \sqrt{|g|}\ \left\{
{\textstyle\frac{(-1)^{(p+1)}}{2\cdot (p+2)!}} \frac{1}{e_{(p)}^{2}}
F_{(p+2)}^{2} 
+{\textstyle\frac{(-1)^{(p+2)}}{2\cdot (p+3)!}} \frac{1}{e_{(p+1)}^{2}}
 H_{(p+3)}^{2} 
\right.
\\
& & \\
& &
\left.
+{\textstyle\frac{(-1)^{p}}{(p+2)!(\tilde{p}+1)!}} 
\frac{\epsilon}{\sqrt{|g|}} \left[F_{(p+2)}-B_{(p+2)}\right]
\partial\tilde{A}_{(\tilde{p}+1)}\right\}\, .
\end{array}
\end{equation}

Substituting into this action the equation of motion for $F_{(p+2)}$

\begin{equation}
F_{(p+2)}=e_{(p)}^{2}\ {}^{\star}\tilde{F}_{(\tilde{p}+2)}\, ,
\hspace{1cm}
\tilde{F}_{(\tilde{p}+2)}=(\tilde{p}+2)\partial \tilde{A}_{(\tilde{p}+1)}\, ,
\end{equation}

\noindent we find (after integration by parts) the dual action

\begin{equation}
\label{eq:masslesspcondenseddual1}
\begin{array}{rcl}
S[B_{(p+2)},\tilde{A}_{(\tilde{p}+1)}] =\hspace{-1.5cm}
&  & \\ 
& & \\
& & \int d^{d}x\ \sqrt{|g|}\ \left\{
{\textstyle\frac{(-1)^{(p+2)}}{2\cdot (p+3)!}} \frac{1}{e_{(p+1)}^{2}}
H_{(p+3)}^{2} 
+{\textstyle\frac{(-1)^{(\tilde{p}+1)}}{2\cdot (\tilde{p}+2)!}} 
e_{(p)}^{2} \tilde{F}_{(\tilde{p}+2)}^{2} 
\right.
\\
& & \\
& &
\left.
+{\textstyle\frac{1}{(p+3)!(\tilde{p}+1)!}} 
\frac{\epsilon}{\sqrt{|g|}} H_{(p+3)}\tilde{A}_{(\tilde{p}+1)}\right\}\, .
\end{array}
\end{equation}

We have substituted a massive $(p+2)$-form potential by a massless
$(p+2)$-form potential and a massless $(\tilde{p}+1)$-form potential.
None of these fields is auxiliary nor it can be eliminated by a gauge
transformation of some kind. Using that the number of degrees of
freedom of a massive $d$-dimensional $k$-form potential is

\begin{equation}
N(d,k)_{\rm massive}=\frac{(d-1)!}{k!(d-1-k)!}\, ,  
\end{equation}

\noindent and that the number of degrees of freedom of a massless
$d$-dimensional $k$-form potential is

\begin{equation}
N(d,k)_{\rm massless}=\frac{(d-2)!}{k!(d-2-k)!}\, ,  
\end{equation}

\noindent one can check that the number of degrees of freedom of the
theory (\ref{eq:masslesspcondenseddual1}) is the same as that of the
original (\ref{eq:masslesspcondensed})\footnote{The Higgs mechanism by
  which the massless $(p+2)$-form ``eats'' the $(p+1)$-form becoming a
  massive $(p+2)$-form is based in it.}:

\begin{equation}
\label{eq:massive-massless}
N(d,p+2)_{\rm massive} = N(d,p+2)_{\rm massless} 
+N(d,\tilde{p}+1)_{\rm massless}\, .
\end{equation}

The usual electric-magnetic duality for massless $d$-dimensional
$(p+1)$- and $(\tilde{p}+1)$-form potentials is based on the identity

\begin{equation}
N(d,p+1)_{\rm massless}=N(d,\tilde{p}+1)_{\rm massless}\, ,  
\end{equation}

\noindent while the electric-magnetic duality between massive
$d$-dimensional $(p+2)$- and $(\tilde{p}+1)$-form potentials (which we
are going to see next) is based on

\begin{equation}
N(d,p+2)_{\rm massive}=N(d,\tilde{p}+1)_{\rm massive}\, .  
\end{equation}

Now, the action (\ref{eq:masslesspcondenseddual1}) does not depend
explicitly on $B_{(p+2)}$ anymore and we can follow again the usual
prescription for dualization of this field. We find the intermediate
action (after integration by parts)

\begin{equation}
\label{eq:masslesspcondenseddual1intermediate}
\begin{array}{rcl}
S[H_{(p+3)},\tilde{A}_{(\tilde{p}+1)},\tilde{B}_{(\tilde{p})}] =\hspace{-2cm}
&  & \\ 
& & \\
& & \int d^{d}x\ \sqrt{|g|}\ \left\{
{\textstyle\frac{(-1)^{(p+2)}}{2\cdot (p+3)!}} \frac{1}{e_{(p+1)}^{2}}
H_{(p+3)}^{2} 
+{\textstyle\frac{(-1)^{(\tilde{p}+1)}}{2\cdot (\tilde{p}+2)!}} 
e_{(p)}^{2} \tilde{F}_{(\tilde{p}+2)}^{2} 
\right.
\\
& & \\
& &
\left.
+{\textstyle\frac{1}{(p+3)!}}
 H_{(p+3)}{}^{\star}\tilde{H}_{(\tilde{p}+1)}\right\}\, ,
\end{array}
\end{equation}

\noindent where

\begin{equation}
\tilde{H}_{(\tilde{p}+1)} = (\tilde{p}+1)\partial\tilde{B}_{(\tilde{p})} 
+\tilde{A}_{(\tilde{p}+1)}\, .  
\end{equation}

Substituting the equation of motion of $H_{(p+3)}$

\begin{equation}
H_{(p+3)}=(-1)^{(p+1)}e^{2}_{(p+1)}\ {}^{\star}\tilde{H}_{(\tilde{p}+1)}\, ,
\end{equation}

\noindent we arrive at the action

\begin{equation}
\label{eq:masslesspcondenseddual2}
S[\tilde{B}_{(\tilde{p})},\tilde{A}_{(\tilde{p}+1)}] =
\int d^{d}x\ \sqrt{|g|}\ \left\{
{\textstyle\frac{(-1)^{(\tilde{p}+1)}}{2\cdot (\tilde{p}+2)!}} 
e_{(p)}^{2} \tilde{F}_{(\tilde{p}+2)}^{2} 
+{\textstyle\frac{(-1)^{\tilde{p}}}{2\cdot (\tilde{p}+1)!}} e_{(p+1)}^{2}
\tilde{H}_{(\tilde{p}+1)}^{2}\right\} 
\end{equation}

This action is now invariant under the usual gauge transformations of
$\tilde{B}_{(\tilde{p})}$

\begin{equation}
\delta\tilde{B}_{(\tilde{p})} = 
\tilde{p}\partial\tilde{\chi}_{(\tilde{p}-1)}\, ,
\end{equation}

\noindent and the following massive gauge
transformations ($\tilde{B}_{(\tilde{p})}$ playing now the role of
Stueckelberg field):

\begin{equation}
\left\{
\begin{array}{rcl}
\delta \tilde{A}_{(\tilde{p}+1)} & = & 
-(\tilde{p}+1) \partial\tilde{\lambda}_{(\tilde{p})}\, ,\\
& & \\
\delta \tilde{B}_{(\tilde{p})} & = & \tilde{\lambda}_{(\tilde{p})}\, .\\
\end{array}
\right.
\end{equation}

These transformations allow us to eliminate $\tilde{B}_{(\tilde{p})}$ by
setting it to zero. We then get the following action 

\begin{equation}
\label{eq:massivepdual}
S[\tilde{A}_{(\tilde{p}+1)}] =\int d^{d}x\ \sqrt{|g|}\ \left[
{\textstyle\frac{(-1)^{(\tilde{p}+1)}}{2\cdot (\tilde{p}+2)!}} 
e_{(p+1)}^{2}\tilde{F}_{(\tilde{p}+2)}^{2}
+{\textstyle\frac{(-1)^{\tilde{p}}}{2\cdot (\tilde{p}+1)!}}e_{(p)}^{2}
\tilde{A}_{(\tilde{p}+1)}^{2}  \right]\, ,
\end{equation}

\noindent which is the one given in Ref.~\cite{kn:Q}.
Had we set $p=0$, $e_{(1)}=1$ and $e_{(0)}=1/m$ at the beginning,
rescaling now $\tilde{A}_{(\tilde{p}+1)}$ we would have found that the
dual mass parameter is identical to the original one $\tilde{m}=m$. We
would also have found that the dual NS/NS 6-form $\tilde{B}$ is the
Stueckelberg field for the R-R 7-form $C^{(7)}$.
  
It seems strange that one can replace a massive field by a pair of
massless fields. Observe that one of them (the $(\tilde{p}+1)$-form
potential) corresponds to the $\tilde{p}$-extended objects whose
condensation gives mass to the $(p+2)$-form potential which was
previously massless. Then, the action makes sense as a slightly more
general action which can describe the situation in which there is no
condensation of $\tilde{p}$-branes and the $(p+1)$-branes are
massless, and also the opposite situation. The WZ interaction term is
crucial for this.


\section{Target Space Fields} 
\label{sec-fields}

In this Appendix we collect the field strengths and the gauge
transformations of the different fields that appear throughout the
paper. We also give the relation with 11-dimensional fields.

Let us start with some conventions: when indices are not shown
explicitly and partial derivatives are used, we assume that all
indices are completely antisymmetrized in the obvious order.  For
instance:

\begin{equation}
G^{(4)} = 4\left(\partial C^{(3)} -3\partial B C^{(1)} 
+{\textstyle\frac{3}{8}}m B^{2}\right)\, ,
\end{equation}

\noindent means

\begin{equation}
G^{(4)}{}_{\mu\nu\rho\sigma}
=4\left[\partial_{[\mu}C^{(3)}{}_{\nu\rho\sigma]} 
-3\left(\partial_{[\mu}B_{\nu\rho}\right) C^{(1)}{}_{\sigma]} 
+{\textstyle\frac{3}{8}}m B_{[\mu\nu}B_{\rho\sigma]}\right]\, .
\end{equation}

Differential form components are defined as follows:

\begin{equation}
A_{(k)}={\textstyle\frac{1}{k!}} A_{(k)\mu_{1}\ldots\mu_{k}} dx^{\mu_{1}}
\wedge
\ldots\wedge dx^{\mu_{k}}\, ,
\end{equation}

\noindent but we will often omit the $\wedge$ symbol.

The 11-dimensional 3-form and dual 6-form fields are denoted by
$\hat{C}$ and $\hat{\tilde{C}}$ respectively. In the massless theory
their field strengths $\hat{G}$ and $\hat{\tilde{G}}$ are:

\begin{equation}
\left\{
\begin{array}{rcl}
\hat{G} & = & d\hat{C}\, ,\\
& & \\\
\hat{\tilde{G}} & = & d\hat{\tilde{C}} +\frac{1}{2}\hat{C}d\hat{C}
={}^{\star}\hat{G}\, ,\\
\end{array}
\right.
\end{equation}

\noindent and their gauge transformation laws are

\begin{equation}
\left\{
\begin{array}{rcl}
\delta\hat{C} & = & d\hat{\chi}\, ,\\
& & \\\
\delta\hat{\tilde{C}} & = & d\hat{\tilde{\chi}} 
+\frac{1}{2}d\hat{\chi}\hat{C}\, .\\
\end{array}
\right.
\end{equation}

The action of 11-dimensional supergravity in differential forms
language is

\begin{equation}
S = {\textstyle\frac{1}{16\pi G_{N}^{(11)}}}
 \int \left\{\Omega_{(11)} \hat{R}
-{\textstyle\frac{1}{2}} \hat{G}\wedge {}^{\star}\hat{G}
+{\textstyle\frac{1}{6}} d\hat{C}\wedge d\hat{C}\wedge \hat{C} \right\}\, .
\end{equation}

Ten-dimensional R-R $k$-form fields are denoted by $C^{(k)}$ and their
field strengths by $G^{(k+1)}$. $B$ is the NS/NS 2-form and
$\tilde{B}$ its dual NS/NS 6-form, and their field strengths are $H$
and $\tilde{H}$.  The $(k-1)$-form parameters of the R-R gauge
transformations are denoted by $\Lambda^{(k-1)}$, and those of the
NS/NS fields $B,\tilde{B}$ are denoted by $\Lambda,\tilde{\Lambda}$.
The parameters of the massive gauge transformations are
$\lambda,\tilde{\lambda}$.

The definitions of the field strengths of all these fields
are\footnote{The gauge parameters used in Ref.~\cite{kn:GHT} are
  related to ours by $\Lambda=\Lambda_{GHT}e^{B}$. Of course, other
  differences in conventions need also be taken into account.}
\cite{kn:GHT,kn:BCT}

\begin{equation}
\begin{array}{rcl}
G^{(2)} & = & d C^{(1)} +\frac{m}{2}B\, , \\
& & \\
H & = & dB\, ,\\
& & \\
G^{(4)} & = & d C^{(3)} -dB C^{(1)}+\frac{1}{2!} \frac{m}{2}B^{2}\, , \\
& & \\
G^{(6)} & = & d C^{(5)} -dB C^{(3)}+\frac{1}{3!} \frac{m}{2}B^{3}\, , \\
& & \\
\tilde{H} & = & d\tilde{B} +G^{(6)} C^{(1)} 
-\frac{1}{2}C^{(3)}dC^{(3)} \\
& & \\
& & 
-\frac{m}{2}\left[C^{(7)} -C^{(5)}B +\frac{1}{2}C^{(3)}B^{2}\right]\, ,\\
& & \\
G^{(8)} & = & d C^{(7)} 
-dB C^{(5)}+\frac{1}{4!} \frac{m}{2}B^{4}\, . \\
\end{array}
\end{equation}

The field strengths of the massless theory are obtained by setting
$m=0$. In this case
the gauge transformations of the fields are:

\begin{equation}
\begin{array}{rcl}
\delta C^{(1)} & = & d \Lambda^{(0)}\, ,\\
& & \\
\delta B & = & d\Lambda\, ,\\
& & \\
\delta C^{(3)} & = & d \Lambda^{(2)} +d\Lambda^{(0)} B\, ,\\
& & \\
\delta C^{(5)} & = & d \Lambda^{(4)} +d \Lambda^{(2)}B 
+\frac{1}{2!}d\Lambda^{(0)} B^{2}\, ,\\
& & \\
\delta\tilde{B} & = & d\tilde{\Lambda} 
-\frac{1}{2} d\Lambda^{(2)} C^{(3)}
+d\Lambda^{(0)} \left( C^{(5)} -\frac{1}{2}C^{(3)}B \right)\, ,\\
& & \\
\delta C^{(7)} & = & d\Lambda^{(6)} +d \Lambda^{(4)}B 
+\frac{1}{2}d \Lambda^{(2)}B^{2}
+\frac{1}{3!}d\Lambda^{(0)} B^{3}\, ,\\
\end{array}
\end{equation}

\noindent and those of the massive theory are obtained by the
replacements

\begin{equation}
\left\{
\begin{array}{rcl}
d\Lambda^{(0)}  & \rightarrow & d\Lambda^{(0)} +m\lambda\, ,\\
& & \\
\Lambda & \rightarrow  & -2\lambda\, ,\\
& & \\
d \tilde{\Lambda} & \rightarrow &  d \tilde{\Lambda} 
+m \tilde{\lambda}\, , \\
& & \\
\Lambda^{(6)} & \rightarrow & 2 \tilde{\lambda}\, ,\\
\end{array}
\right.
\end{equation}

\noindent that is

\begin{equation}
\begin{array}{rcl}
\delta C^{(1)} & = & d \Lambda^{(0)}+m\lambda\, ,\\
& & \\
\delta B & = & -2d\lambda\, ,\\
& & \\
\delta C^{(3)} & = & d \Lambda^{(2)} +\left( d\Lambda^{(0)} 
+m\lambda\right)B\, ,\\
& & \\
\delta C^{(5)} & = & d \Lambda^{(4)} +d \Lambda^{(2)}B 
+\frac{1}{2!}\left(d\Lambda^{(0)} +m\lambda \right)B^{2}\, ,\\
& & \\
\delta\tilde{B} & = & \left(d\tilde{\Lambda} +m\tilde{\lambda}\right) 
-\frac{1}{2} d\Lambda^{(2)} C^{(3)}
+\left(d\Lambda^{(0)} +m\lambda\right)
\left( C^{(5)} -\frac{1}{2}C^{(3)}B \right)\, ,\\
& & \\
\delta C^{(7)} & = & 2d\tilde{\lambda} +d \Lambda^{(4)}B 
+\frac{1}{2}d \Lambda^{(2)}B^{2}
+\frac{1}{3!}\left(d\Lambda^{(0)} +m\lambda\right)B^{3}\, .\\
\end{array}
\end{equation}

Defining, as in Ref.~\cite{kn:GHT} the Grassmann algebra objects

\begin{equation}
\begin{array}{rcl}
C & = & C^{(0)} + C^{(1)} + C^{(2)} +\ldots \\
& & \\
G & = & G^{(0)} + G^{(1)} + G^{(2)} +\ldots \\
& & \\
\Lambda^{(\cdot)}  & = & \Lambda^{(0)} +\Lambda^{(1)} + \ldots\, , \\
\end{array}
\end{equation}

\noindent the gauge transformations and field strengths of the R-R fields
can be written in the more compact form

\begin{equation}
\begin{array}{rcl}
\delta C & = & \left(\Lambda^{(\cdot)} +m\lambda\right)e^{B}\, ,\\
& & \\
G & = & d C - dB C + \frac{m}{2}e^{B}\, .\\
& & \\
\end{array}
\end{equation}

The action of massive ten dimensional IIA supergravity can be written
in differential forms language as follows:

\begin{equation}
\begin{array}{rcl}
S & = & \frac{1}{16\pi G_{N}^{(10)}} \int \left\{
e^{-2\phi} \left[ \Omega_{(10)} R
-4d\phi\wedge{}^{\star}d\phi +\frac{1}{2}H\wedge {}^{\star}H \right]
\right.\\
& & \\
& & 
+\frac{1}{2}G^{(2)}\wedge {}^{\star}G^{(2)} 
+\frac{1}{2}G^{(4)}\wedge {}^{\star}G^{(4)}
+\frac{1}{2}\left(\frac{m}{2}\right)\wedge {}^{\star}
\left(\frac{m}{2}\right) \\
& & \\
& & 
\left.
+\frac{1}{2}dC^{(3)}\wedge dC^{(3)}\wedge B
+\frac{1}{3!}\frac{m}{2}dC^{(3)}\wedge B^{3}
+\frac{3}{5!}\left(\frac{m}{2}\right)^{2}B^{5} 
\right\}\, .
\end{array}
\end{equation}

%
%
%
%
%
%
%

\section{Worldvolume Fields}
\label{sec-wvfields}

Below we collect some useful gauge transformation rules of worldvolume
fields needed in the text. The compact form notation of these rules
was given in Section~\ref{sec-worldvolume}. The gauge transformation
rules of the $c^{(p)}$\ (p=0,2,4,6) are given by

\begin{equation}
\label{transwor}
\left\{
\begin{array}{rcl}
\delta c^{(0)} & = & -\frac{1}{2\pi\alpha^\prime} \Lambda^{(0)} 
+\frac{m}{2}\rho^{(0)}\, ,\\
& & \\
\delta c^{(2)} & = & 2\partial\kappa^{(1)} -{1\over 2\pi\alpha^\prime}
\Lambda^{(2)} + 2\Lambda^{(0)}\partial b -\frac{m}{2}(2\pi\alpha^\prime)
\rho^{(0)}\partial b -m\lambda b\, ,\\
& & \\
\delta c^{(4)} & = & 4\partial \kappa^{(3)} -
{1\over 2\pi\alpha^\prime}\Lambda^{(4)} +12\Lambda^{(2)}\partial b
-12(2\pi\alpha^\prime)\Lambda^{(0)}\partial b\partial b\\
& & \\
&&
+2m(2\pi\alpha^\prime)^2\rho^{(0)}\partial b\partial b 
+8m(2\pi\alpha^\prime)\lambda b\partial b\, ,\\
& & \\
\delta c^{(6)} & = & 6\partial\kappa^{(5)} 
-{1\over 2\pi\alpha^\prime} 2{\tilde
\lambda} + 30\Lambda^{(4)}\partial b -180 (2\pi\alpha^\prime)
\Lambda^{(2)}\partial b\partial b\\
& & \\
&&
+120 (2\pi\alpha^\prime)^2\Lambda^{(0)}\partial b\partial b\partial b
-15m(2\pi\alpha^\prime)^3\rho^{(0)}\partial b\partial b\partial b\\
& & \\
& &
-90m (2\pi\alpha^\prime)^2\lambda b\partial b\partial b\, .\\
\end{array}
\right.
\end{equation}

The corresponding field strengths are

\begin{equation}
\left\{
\begin{array}{rcl}
{\cal G}^{(1)} & = & \partial c^{(0)} 
+\frac{1}{2\pi\alpha^{\prime}} C^{(1)} 
-{\textstyle\frac{m}{2}} b, , \\
& & \\
{\cal G}^{(3)} & = & 3\partial c^{(2)}
+{\textstyle\frac{1}{2\pi\alpha^{\prime}}}
C^{(3)} -3 C^{(1)}{\cal F} +3 {\textstyle\frac{m}{2}}\ 
(2\pi\alpha^{\prime})\ b \partial b\, ,\\
& & \\
{\cal G}^{(5)} & = &
5\partial c^{(4)} +\frac{1}{2\pi\alpha^{\prime}}C^{(5)}
-10 C^{(3)}{\cal F} +15(2\pi\alpha^{\prime})C^{(1)}{\cal F}{\cal F}\\
& & \\
& &
-20\frac{m}{2}(2\pi\alpha^{\prime})^{2}b\partial b\partial b\, . \\
\end{array}
\right.
\end{equation}

The following gauge transformations of the $a^{(p)}$ fields are also
used in the paper:

\begin{equation}
\label{aes}
\left\{
\begin{array}{rcl}
\delta a^{(0)} & = & \delta c^{(0)}\\
& & \\
& = & -\frac{1}{2\pi\alpha^\prime}
\Lambda^{(0)}+\frac{m}{2}\rho^{(0)} \, ,\\
& & \\
\delta a^{(2)} & = & 2\partial \mu^{(1)}
-\frac{1}{2\pi\alpha^\prime}\Lambda^{(2)}
-4\partial a^{(0)}\lambda+\frac{m}{2} (2\pi\alpha^\prime)
\rho^{(0)}\partial b-m\lambda b\, , \\
& & \\
\delta a^{(4)} & = & 4\partial\mu^{(3)}
-\frac{1}{2\pi\alpha^\prime}\Lambda^{(4)}
-24\partial a^{(2)}\lambda
+2(2\pi\alpha^\prime)^2 m\rho^{(0)}(\partial b)^2 \\
& & \\
&&
-4(2\pi\alpha^\prime)m\lambda b\partial b\, .\\
\end{array}
\right.
\end{equation}

The corresponding field strengths are

\begin{equation}
\left\{
\begin{array}{rcl}
{\cal H}^{(1)} & = & {\cal G}^{(1)}\\
& & \\
& = & \partial a^{(0)} 
+\frac{1}{2\pi\alpha^{\prime}} C^{(1)} 
-{\textstyle\frac{m}{2}} b, , \\
& & \\
{\cal H}^{(3)} & = & 3\partial a^{(2)} +3\partial a^{(0)}B
+\frac{1}{2\pi\alpha^{\prime}}C^{(3)}
-3\ \frac{m}{2}\ (2\pi\alpha^{\prime})b\partial b
-3\ \frac{m}{2}\ B b\, ,\\
& & \\
{\cal H}^{(5)} & = & 5\partial a^{(4)}+30\partial^{(2)}B
+15\partial a^{(0)}B^2+\frac{1}{2\pi\alpha^\prime}C^{(5)}
-20\frac{m}{2}(2\pi\alpha^\prime)^2 b(\partial b)^2\\
&& \\
&& -30m(2\pi\alpha^\prime)Bb\partial b-15\frac{m}{2}bB^2\,.
\\
\end{array}
\right.
\end{equation}

For the fields that compensate the total derivatives in p-brane WZ
terms we have

\begin{equation}
\left\{
\begin{array}{rcl}
\delta b & = & \frac{1}{2\pi\alpha^{\prime}}2\lambda 
+\partial\rho^{(0)}\, ,\\
& & \\
\delta \tilde{b} & = & -5\partial\rho^{(4)}
-\frac{1}{2\pi\alpha^{\prime}}\tilde{\Lambda} 
+\frac{m}{2}\rho^{(5)} +5\Lambda^{(4)}\partial c^{(0)}\\
& & \\
& &
+15(2\pi\alpha^{\prime})\partial a^{(2)}
\left(\delta a^{(2)}-2\partial\mu^{(1)}\right)\, ,\\
\end{array}
\right.
\end{equation}

\noindent and their respective field strengths are

\begin{equation}
\left\{
\begin{array}{rcl}
{\cal F} & = & 2\partial b + \frac{1}{2\pi\alpha^{\prime}}B\, ,\\
& & \\
\tilde{\cal F} & = & 6\partial\tilde{b} +\frac{m}{2}c^{(6)} 
+\frac{1}{2\pi\alpha^{\prime}}\tilde{B}
-6\left(C^{(5)} -5C^{(3)}B\right)\left(\partial c^{(0)} 
-\frac{m}{2}b\right)\\
& & \\
& &
-30(2\pi\alpha^{\prime})\left[\partial a^{(2)} 
-\frac{m}{2}(2\pi\alpha^{\prime})b\partial b\right]
\left({\cal H}^{(3)} -3\partial a^{(2)}\right)\\
& & \\
& & 
-120 \frac{m}{2}(2\pi\alpha^{\prime})^3\partial c^{(0)}
b\partial b\partial b\, .\\
\end{array}
\right.
\end{equation}


\end{document}